# Technical Review on RF-Amplifiers for Quantum Computer Circuits: New Architectures of Josephson Parametric Amplifier

Ahmad Salmanogli[1,3], Hesam Zandi[2,3,6], Mahdi Esmaeili, Abolfazl Eskandari[3], Mohsen Akbari[5]
[1]Department of Electrical and Electronics, Ankara Yildirim Beyazit University, Turkey
[2]Faculty of Electrical Engineering, K. N. Toosi University of Technology, Tehran, Iran
[3]Iranian Quantum Technologies Research Center (IQTEC), Tehran, Iran
[5]Quantum optics lab, Department of Physics, Kharazmi University, Tehran, Iran
[6]Electronic Materials Laboratory, K. N. Toosi University of Technology, Tehran, Iran

**Abstract:**
Josephson Parametric Amplifiers are fundamental components in quantum information processing and computing due to their quantum-limited noise performance, allowing amplification of weak quantum signals with minimal added noise. This capability is particularly essential in applications such as qubit readout, quantum sensing, and communication, where signal fidelity and preservation of quantum coherence are critical. Unlike conventional CMOS and HEMT amplifiers, which are widely used in RF applications, the Josephson parametric amplifiers are specifically optimized for cryogenic operation at mK temperatures. While CMOS amplifiers offer excellent scalability and integration potential, they suffer from higher noise figures and degraded performance at ultra-low temperatures. HEMT amplifiers provide low noise and high gain but are power-hungry and less compatible with millikelvin environments. In contrast, Josephson parametric amplifiers deliver ultra-low noise figures (quantum limit), minimal power consumption, and exceptional compatibility with cryogenic conditions, making them ideal for quantum systems. The first part of this work specifically centers on the comparison among RF amplifiers and why cryogenic applications (mK) prefers the Josephson parametric amplifier. The second part of this work focuses on the design and analysis of the parametric amplifiers based on both single Josephson junctions and junction arrays. A single-junction amplifiers utilizes the nonlinear inductance of a Josephson junction to achieve amplification via parametric conversion. However, such amplifiers are prone to gain compression at high input powers and are sensitive to fabrication and bias variations, which limits their dynamic range and operational stability. To address these limitations, amplifiers using arrays of Josephson junctions are explored. By distributing the nonlinear response, the array configuration enhances linearity, increases power handling, broadens dynamic range, and provides better impedance tunability and gain control. Furthermore, arrays reduce phase noise and improve coherence. Additionally, several novel parametric amplifier architectures are designed, modeled, and thoroughly analyzed. Their performance is then compared with that of conventional array-based parametric amplifiers to evaluate potential improvements and trade-offs. This study uses quantum theory and CAD simulations to model and analyze the performance improvements enabled by these structural optimizations.

**Introduction**:

The Josephson Parametric Amplifier (JPA) [1-5] is a superconducting device based on the Josephson effect, which leverages nonlinear properties of superconducting circuits to quantum amplify signals with minimal noise. The JPA relies on the unique behavior of superconductors under specific conditions (using nonlinearity) to achieve amplification that is critical for quantum computing, radio astronomy, and sensitive electromagnetic measurements [1-14]. The JPA is considered a key component in quantum measurement systems due to its ability to achieve near-quantum-limited noise performance, making it ideal for applications that require highly sensitive detection of weak signals.

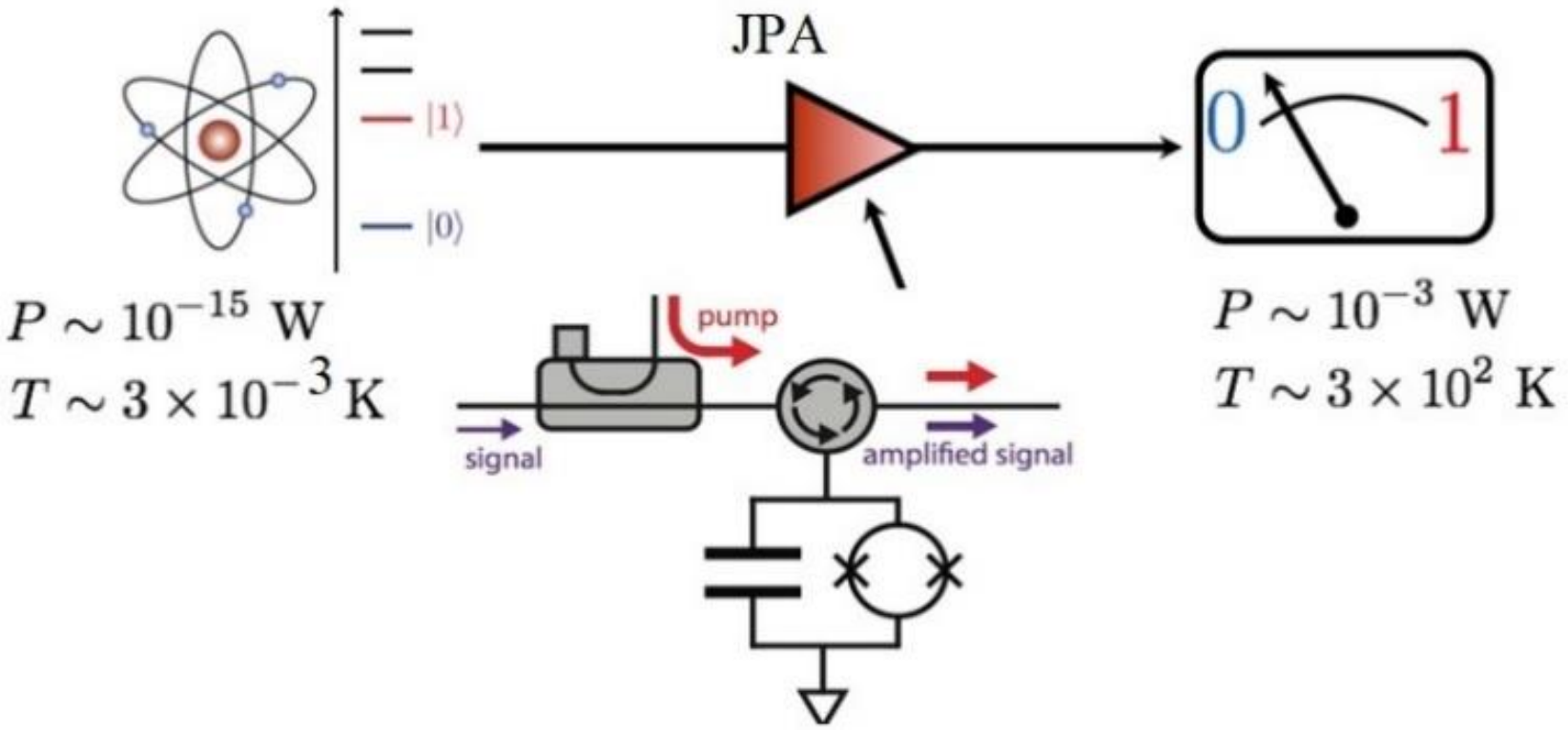

Fig. 1 Quantum signal Amplification using a JPA in quantum computing circuits; showing quantum signal power and temperature before and after amplifications.

JPAs are a type of parametric amplifier, a class of amplifiers that exploit the modulation of a parameter within an electrical circuit, such as capacitance or inductance, to transfer energy and achieve signal amplification. Unlike conventional amplifiers, parametric amplifiers like JPAs use this modulation to create an amplified output without relying on resistive components, thereby minimizing thermal noise. The core element of the JPA is the Josephson junction (JJ), a superconducting device that enables energy transfer through a nonlinear inductance, which can be modulated to enhance weak input signals. Indeed, a JPA is used to amplify the weak signals generated by a quantum chip illustrated in Fig. 1. This diagram illustrates a quantum computing setup where weak signals from a qubit system are amplified using a JPA. The qubit, represented by two energy states |0⟩ and |1⟩, operates at extremely low power (~$10^{-15}$ W) and temperature (~$10^{-3}$ W) and at a higher temperature (~300 K) for reliable measurement. This amplification step is essential for preserving quantum information while boosting it for readout, a critical component of scalable quantum computing architectures.

There are several configurations and types of JPAs, which vary based on their design and operational requirements: Non-degenerate JPAs: These JPAs use a single-frequency signal, or "pump," to achieve amplification. In non-degenerate mode, the input signal and the pump are at different frequencies, and this type is generally simpler in design but may be less efficient for certain applications [5, 6, 9]. Degenerate JPAs: Degenerate JPAs, or phase-sensitive JPAs, operate by mixing the input signal with a pump signal at the same frequency. This type of JPA offers phase-sensitive amplification, meaning it amplifies only one quadrature of the input signal while de-amplifying the orthogonal quadrature, which can be beneficial in noise-sensitive applications [5, 6, 9, 13]. Bifluxon JPAs: This type of JPA is designed to operate with a bistable state, where it can switch between two flux states, allowing it to have both amplification and memory capabilities. Bifluxon JPAs are of interest for quantum information processing, as they can serve as memory elements in addition to providing amplification [15, 16]. Josephson Traveling-Wave Parametric

Amplifiers (JTWPAs) (Fig. 2): Unlike standard JPAs that operate in a lumped-element configuration, JTWPAs use a distributed network of JJ, creating a traveling wave structure that supports broader bandwidth amplification. JTWPAs are essential for applications requiring wide bandwidths, such as broadband microwave detection in quantum information systems [5, 6, 9, 13].

Each type of JPA offers unique advantages and is suited for different application scenarios, depending on the bandwidth, gain, and noise requirements of the system. The JPA has several distinctive advantages over conventional amplifiers, particularly in the context of low-noise applications and quantum systems [6, 9]: Quantum-Limited Noise Performance: JPAs can achieve quantum-limited amplification, meaning they add the minimum possible noise allowed by quantum mechanics. This is essential for quantum computing and other applications that require precise detection of weak signals without introducing additional noise. High Sensitivity: The nonlinear inductance of the JJ in JPAs enables extremely sensitive amplification. JPAs can detect signals on the order of a single photon, making them highly valuable in applications where signal strength is limited. Low Power Consumption: JPAs, as superconducting devices, operate at cryogenic temperatures with minimal power consumption, which is beneficial in environments like quantum computing, where heat can interfere with quantum states. Phase-Sensitive Amplification: Certain types of JPAs, such as degenerate JPAs, offer phase-sensitive amplification, which allows them to amplify one quadrature component of the signal while suppressing the other. This capability is advantageous in scenarios where phase information is crucial, such as in quantum information systems. Non-destructive Readout: In quantum applications, JPAs enable non-destructive readout of qubit states, allowing repeated measurements without disturbing the quantum state of the system. However, to get fully know about the JPA properties we need a powerful theory than the classical ones that it has been normally used to analyze the amplifier; this theory is the full quantum theory using which some new degree of freedom are introduced to cover all parts of the device under studied. Indeed, quantum theory plays a pivotal role in analyzing quantum devices such as JPAs and Purcell filters [13, 14a, 17, 18], which are integral to quantum information processing and sensing applications. The first step in this analysis involves constructing a Hamiltonian that describes the system's quantum states. For JPAs, this Hamiltonian typically includes terms representing the Josephson junction's nonlinearity and inductive and capacitive elements. Using this Hamiltonian, the equations of motion can be derived, often through Heisenberg's equation or the Master equation. Solving these equations provides insight into the dynamic behavior of the JPA, such as how energy states evolve over time under different external drive conditions. In this review article, we attempt to fully analyze some unique JPA structures using quantum theory and discuss about their features.

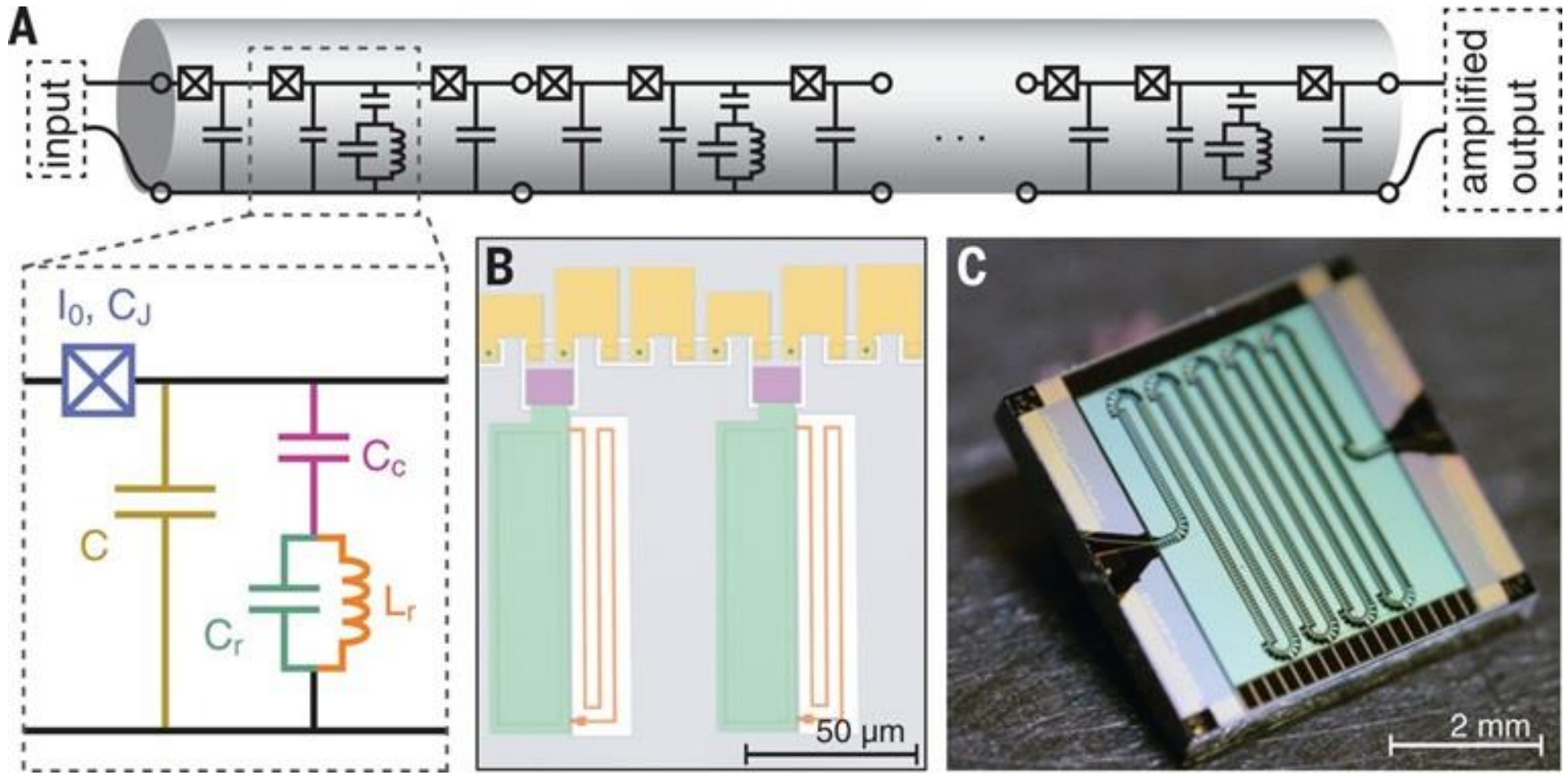

Fig. 2 Josephson traveling-wave parametric amplifier. (A) Schematic diagram: The JTWPA is designed as a nonlinear transmission line made up of lumped elements. Each unit cell includes a Josephson junction with a critical current of μA and an intrinsic capacitance of fF, along with a capacitive shunt to ground of fF. (B) False-color optical micrograph: The colors correspond to the inset in (A), with the lower metal layer shown in gray. (C) Image of a 2037-junction JTWPA: The line is meandered multiple times on the 5 mm by 5 mm chip to reach the desired amplification level [5b].

However, when considering alternatives to JPAs, high-electron-mobility transistors (HEMTs) [20-36] and complementary metal-oxide-semiconductor (CMOS)-based amplifiers are often discussed [37-49]. Accordingly, JPAs offer several advantages that make them more suitable for certain applications, especially in quantum information and precision measurement systems: Lower Noise than HEMT Amplifiers: HEMTs are popular in cryogenic applications due to their low noise properties, but they still cannot achieve the quantum-limited noise performance of JPAs. HEMTs introduce noise due to thermal effects and other factors, which can limit their effectiveness in ultra-sensitive applications. JPAs, by contrast, are capable of near-quantum-limited noise, making them ideal for quantum measurements and other applications requiring extreme sensitivity. Lower Power Dissipation: CMOS amplifiers, while widely used in traditional electronics, are not ideal for cryogenic environments because they consume more power and generate heat that can interfere with superconducting systems. JPAs, as superconducting devices, are inherently low-power, which allows them to operate in cryogenic environments with minimal thermal impact on nearby quantum systems. Higher Linearity and Stability: CMOS amplifiers can suffer from issues like gain compression and non-linear distortions, especially at high frequencies or in high-gain applications. JPAs, thanks to their superconducting properties, provide more stable and linear amplification, which is crucial in systems that rely on precise signal integrity. Compatibility with Quantum Systems: Quantum systems, especially those used in computing and sensing, require components that operate at very low temperatures and have minimal interference with quantum states. JPAs, designed to operate at cryogenic temperatures, fit seamlessly into these systems and offer quantum-limited noise, which is essential for maintaining the coherence of quantum information. Due to these advantages, JPAs are generally preferred over HEMT and CMOS amplifiers in quantum applications and other fields where ultra-low-noise amplification is necessary. However, HEMT and CMOS circuits have been widely used in the quantum computing circuits at temperature above 4.2 K and also in the room temperature. In the following, we attempt to focus on this point and show that how much these circuits are effectively utilized in quantum computers. In Fig. 3a a complete circuit relating to a typical quantum analog-mixing circuit shown.

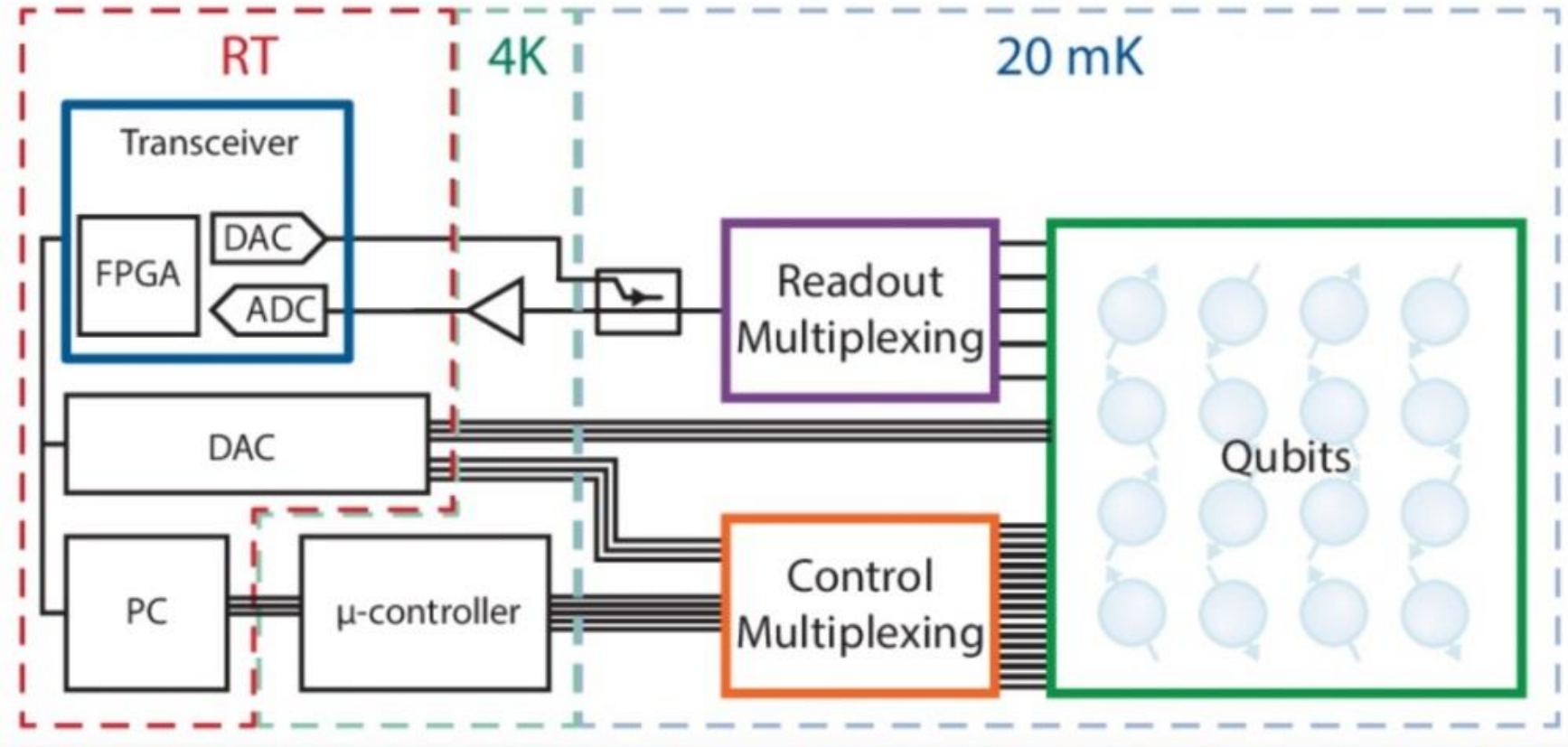

Fig. 3a Quantum computing readout and control circuit with HEMT and CMOS amplifiers containing sub-cryogenic: quantum chip, cryogenic: HEMT and micro-controllers, and room temperature electronics: mixed-analog electronics and FPGA [14b].

This circuit diagram represents a quantum computing readout and control system divided into three temperature regions: room temperature, 4 K, and 20 mK. At room temperature, CMOS-based components, such as the FPGA, digital-to-analog converters (DACs), and an analog-to-digital converter (ADC), manage data processing and initial signal generation. The CMOS components are advantageous here due to their low power consumption and scalability in digital electronics. At 4 K, the HEMT amplifier is crucial for low-noise signal amplification. HEMT amplifiers are ideal in cryogenic environments around 4 K due to their superior noise performance, which is essential in preserving quantum signals coming from qubits at ultra-low temperatures. By amplifying weak quantum signals at 4 K, the HEMT prevents excessive noise from degrading the signal before further processing at room temperature. At 20 mK, qubits operate in an environment isolated from thermal noise, ensuring stable quantum coherence. Signals generated by qubits-transmission line (TL) coupling are transmitted through control and readout multiplexing structures, which regulate the interactions between the qubits and the external control and readout electronics. This configuration allows for accurate measurement and manipulation of qubits, a crucial aspect for quantum information processing. Aside from, one can find an extra detail relating to the relationship among the analog, digital, and quantum circuits in Fig. 3b. This figure shows the experimental setup for state measurement in a quantum computing environment, highlighting the essential components used to generate, filter, and amplify signals for precise qubit control and measurement. The setup operates across three temperature stages, each with specific components designed to optimize performance at different stages of the cooling process. At the room temperature stage, an Arbitrary Waveform Generator (AWG) generates pulses that are mixed with a Local Oscillator (LO) to create gigahertz-frequency signals. These signals are split into two components, labeled I and Q, which are then directed towards the qubit system. An ADC captures and digitizes the reflected or transmitted signals for further processing. The diagram includes a series of microwave generators, mixers, differential amplifiers, and buffer amplifiers to prepare and condition the signals before they move down to lower temperature stages. At the 4K stage, cold attenuators and filters, including infrared filters and copper powder filters, help suppress noise and thermal radiation, preventing unwanted disturbances from reaching the qubits. This stage also houses a HEMT amplifier, which boosts weak signals without introducing significant noise. The setup utilizes RC filters and a custom voltage source to generate a stable flux bias for the parametric amplifier, ensuring precise control over the amplification process. The 40mK stage, the coldest point in the setup, houses the qubit chip and parametric amplifiers to generate and intensify the quantum signals. A six-port switch allows the system to alternate between multiple samples, noise references, and parametric amplifier, improving flexibility and enabling accurate measurement of the qubit states. This switching capability is crucial for experiments requiring low-noise measurements, as it allows for easy calibration and comparison between the signal and noise references.

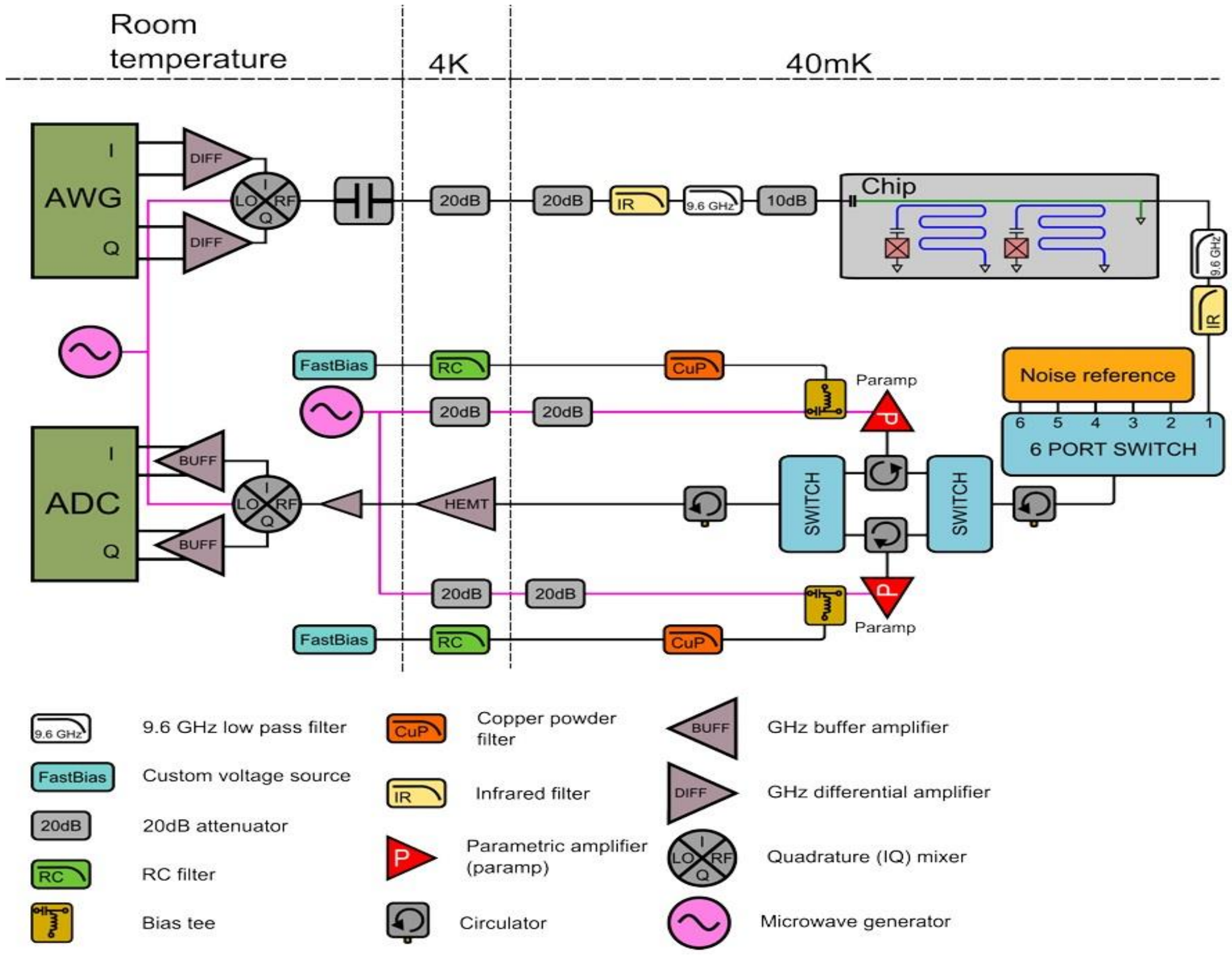

Fig. 3b Experimental setup for qubit state measurement. Pulses generated by the AWG are mixed to gigahertz frequencies and directed to the qubit chip at the 40mK stage, where state measurements occur. Noise and thermal radiation are suppressed by a series of attenuators and filters, including infrared and copper powder filters, at different temperature stages. The signal path includes switches that enable selection between parametric amplifiers and multiple samples for comparison. Amplification stages, including HEMT and room-temperature amplifiers, ensure signal integrity before digitization by the ADC [14C].

It is shown above while HEMT- and CMOS-based amplifiers are indeed applicable in quantum circuits, a key question remains: can these technologies be advanced to the point where they could effectively replace JPAs? If so, what specific parameters must be optimized to achieve comparable performance in quantum applications? The following materials attempts to answer these questions.

## JPA as an amplifier for quantum applications: why not CMOS or HEMT at sub-cryogenic temperature (< 30 mK)

### *CMOS Technology*

CMOS technology is a foundational technology in modern electronics, known for its efficiency in power consumption and its widespread use in digital logic circuits. CMOS circuits consist of complementary and symmetrical pairs of p-type and n-type MOSFETs, which allow them to operate with high noise immunity and low static power consumption. In the realm of RF (Radio Frequency) applications, CMOS technology has been adapted to design RF amplifiers that can operate at various frequency bands, including the C-band (4-8 GHz) and X-band (8-12 GHz). Despite being traditionally considered less suitable for high-frequency applications due to inherent limitations in performance, advances in CMOS technology have allowed for significant improvements. The development of CMOS RF amplifiers focuses on integrating analog and digital circuitry on a single chip, leveraging the cost-effectiveness and scalability of CMOS processes [37-42]. This integration is particularly beneficial in RF applications, where compactness and energy efficiency are crucial. Following, a few interesting and relevant works are studied in details. Work cited in [37] introduces a cryogenic CMOS receiver designed for multi-qubit gate-based RF readout in silicon spin qubit systems. Fabricated in 40nm CMOS, the receiver operates at 4 K and supports the 6–8 GHz band with 2 GHz bandwidth, a gain of 58 dB, and a noise figure of ~0.6 dB, offering performance close to that of room-temperature racks. The power consumption is 170 μW/qubit, and the system supports frequency division multiplexing (FDMA) with 5 MHz channel spacing, enabling scalable readout of ~400 qubits. A cascoded common-source low noise amplifier (LNA), quadrature mixer, and interstage image filtering ensure high linearity ($P_{1dB}$ (1 dB compression point) = –58.4 dBm, $IIP_3$ (third-order intercept point) = –50.8 dBm). This receiver enables single-shot readout with sub-μs integration time and is optimized for dilution refrigerator integration. It demonstrates a promising approach for scalable, low-power quantum-classical interfacing.
The other work [38] reports a cryo-CMOS XY gate modulator in 28nm CMOS for transmon qubits, operating at 4–8 GHz with sub-2 mW power consumption. This IC replaces traditional AWGs with on-chip vector modulation and digital-to-analog converter (DAC) based envelope generation, enabling coherent qubit control at 3 K. The system supports 16 waveform storage, amplitude and phase tuning, and symmetric cosine-based pulse shaping, essential for minimizing spectral leakage and gate error. Experimental validation with Rabi and Bloch sphere control experiments confirms the IC's ability to execute high-fidelity rotations with <12% RMS control error, comparable to standard setups. The IC integrates passive upconversion mixers, current-mode DACs, and programmable LO drivers, optimizing power–performance tradeoffs. It represents a compact, efficient solution for scalable cryogenic quantum control, essential for transitioning from room-temperature racks to cryogenic integrated control layers in large-scale quantum processors.

In addition, Ref [39] presents a scalable cryogenic controller fabricated in 22nm FinFET CMOS, designed to operate over a 2–20 GHz range and support up to 432 frequency-multiplexed qubits. The architecture integrates arbitrary waveform generation, Z-corrections, and digital I/Q modulation, enabling fine-grain control with SFDR >44 dB, $IM_3$>50 dB, and SNR ~ 48 dB (25 MHz BW). A digitally intensive backend stores 40960 waveform points, executed via an instruction set optimized for low-latency triggering and envelope reuse. The controller supports both spin qubits and transmons with flexible power scaling and FDMA-compatible multiplexing. Measured at 3 K, the transmitter achieves –17.5 dBm output power, with integrated LO/image leakage calibration. The high-band (15–20 GHz) support uses 3rd harmonic LO upconversion, demonstrating a wideband, multi-qubit solution. This work marks a major advance in building compact, cryogenic-scale quantum control stacks for large qubit arrays. Eventually, CMOS technology's compatibility with standard semiconductor manufacturing processes makes it highly desirable for mass production, enabling the development of complex RF systems that can be produced at scale with uniform performance into a chip (e.g CMOS RF circuit design for quantum application illustrated in Fig. 4). The figure shows a typical pad frame layout of a fully integrated System-on-Chip (SoC) receiver [49]. The layout includes several key components essential for RF signal processing: RF LNAs, Voltage-Controlled Oscillator (VCO), mixer, and Intermediate Frequency (IF) amplifier. The LNA, specifically designed to operate at 5.2 GHz, is crucial for amplifying weak RF signals with minimal noise addition, which improves the receiver's sensitivity. The mixer follows the LNA, where it down-converts the high-frequency RF signal to a lower IF signal. This process simplifies subsequent signal processing. The VCO, an integral part of the local oscillator section, generates a stable oscillation frequency, which is essential for frequency translation in the mixer. Inset diagrams highlight some intricate sections of the layout, focusing on the LNA, mixer, VCO and the IF amplifiers. The compact layout, measuring 0.6 mm by 0.4 mm, demonstrates the SoC receiver's small footprint, which is ideal for applications requiring miniaturized RF front-end solutions in portable or embedded devices. The design's integration of multiple RF functions within a small area showcases advancements in SoC design, allowing high performance in compact RF systems. Finally, the ability of CMOS technology to integrate with other electronic components also facilitates the creation of sophisticated systems for RF signal processing, where both analog and digital functions are necessary [37-49]. In addition to the technical aspects previously discussed, it is important to explore the role of CMOS technology within the quantum domain. The following section highlights key areas in quantum science and engineering where CMOS has been effectively applied.

***CMOS Applications in the Quantum Realm*:**

CMOS technology has become increasingly relevant in quantum computing, particularly in the development of scalable quantum processors and readout systems. The use of CMOS in quantum systems offers several advantages: I) Integration with Classical Electronics: CMOS technology allows for the seamless integration of quantum circuits with classical control and readout electronics on the same chip. This integration is crucial for scaling up quantum processors, as it minimizes the need for external connections and enhances the overall system's coherence and stability. II) Low-Power RF Amplification: CMOS-based RF amplifiers are utilized in quantum systems for signal readout, where low-power operation is essential to minimize thermal noise and maintain the integrity of quantum signals. These amplifiers can be designed to operate in the C- and X-bands, providing the necessary bandwidth and gain for detecting weak quantum signals. III) Cryogenic Operation: Advances in CMOS technology have enabled the development of cryogenic RF amplifiers that can operate at temperatures close to absolute zero, where quantum computers typically function. These amplifiers are used in qubit readout systems to amplify quantum signals without introducing significant noise, which is critical for accurate quantum state measurement. IV) Scalability and Manufacturability: CMOS technology's mature fabrication processes make it highly scalable and manufacturable, allowing for the production of large-scale quantum processors with high yield and uniformity. This scalability is essential for the future of quantum computing, where large arrays of qubits and associated electronics will be required. As a brief conclusion, in the quantum realm, CMOS technology is uniquely positioned to enable the development of scalable and integrated quantum systems [37-45]. Its ability to operate at cryogenic temperatures and integrate with classical electronics makes it a key technology for quantum processors, particularly in applications where high levels of integration and low power consumption are required. CMOS-based RF amplifiers are also used in quantum sensing and metrology, where precise measurement of weak signals is critical.

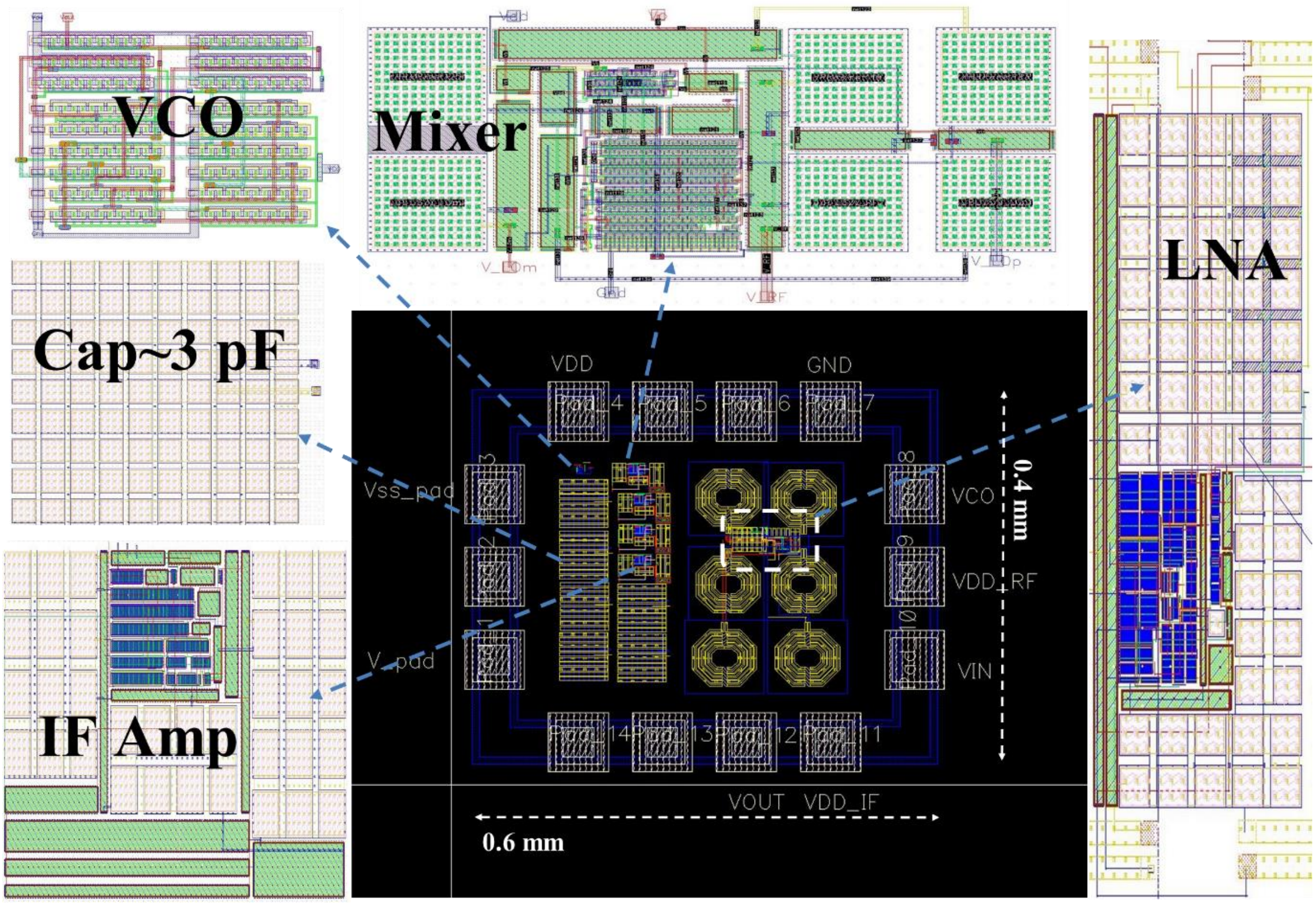

Fig. 4 Floor plane of the fully integrated SoC designed for C-band (4.1–8.2 GHz) quantum-RF applications. The SoC includes a LNA, Mixer, VCO, IF amplifier, coupling capacitors (~3 pF), and RF filters. The total chip size is approximately 0.6×0.4 mm². The design emphasizes compactness by minimizing the use of inductors (only in LNA), employing active components for space efficiency, implementing noise cancellation and feedback techniques, and using optimized matching networks to reduce the noise figure. Post-layout simulations confirm that layout parasitics have a negligible impact on performance [49].

***HEMT Technology***

HEMT technology is a specialized field-effect transistor (FET) known for its exceptional high-frequency performance and low noise characteristics. HEMTs are constructed using a heterojunction, typically between materials like Gallium Arsenide (GaAs) and Aluminum Gallium Arsenide (AlGaAs) or Gallium Nitride (GaN) and Aluminum Gallium Nitride (AlGaN). This heterostructure creates a two-dimensional electron gas (2DEG) at the interface, where electrons can move with very high mobility due to the lack of impurity scattering, resulting in fast switching speeds and excellent frequency response [20-36]. Following, a few interesting works are cited. Utilizing HEMT technology, Ref [20] presents two wideband monolithic microwave integrated circuit (MMIC) LNAs utilizing 100 nm InP HEMTs designed for cryogenic environments (~4 K). Operating over 0.3–14 GHz and 16–28 GHz, the LNAs achieve record-low noise temperatures—down to 2.2 K at 6 GHz and 4.8 K at 20.8 GHz—while maintaining high gain (41.6 dB and 32.3 dB, respectively). The design benefits from aggressive scaling of the gate length and barrier thickness, which increases transconductance and cutoff frequencies (fT = 240 GHz at 5 K), while minimizing parasitic resistances and gate leakage. External and integrated matching networks, respectively, were optimized for low-frequency and high-frequency bands. The work sets a new standard in cryogenic broadband LNA design for radio astronomy and deep-space communication, with a strong focus on noise optimization and gain flatness across a wide bandwidth.
Another work cited in [21] introduces a 100-nm InP HEMT optimized for ultra-low-power cryogenic LNA operation, targeting sub-milliwatt dissipation. By reducing the gate-channel distance to 5 nm via a Pt-gated design and employing an InP etch-stop layer, the device achieves superior subthreshold characteristics (SS = 14 mV/dec at 5 K) and low gate leakage (~0.1 μA/mm). A 4–8 GHz hybrid LNA demonstrates remarkable performance at only 112 μW total power: average noise temperature of 4.1 K and 20 dB gain. At 300 μW, performance improves to 2.8 K noise temperature and 27 dB gain. The minimal increase in $T_{min}$ across different bias points and high transconductance-to-drain current ratio confirm the suitability of this design for quantum processor readout near the 1 K stage. This work represents a significant advancement in cryogenic amplifier design by enabling efficient operation at ultra-low power levels without sacrificing noise or gain performance.
The other interesting work cited in [23] investigates the stability of two-finger 100-nm InP HEMTs under cryogenic conditions for use in ultra-low-noise Ka- and Q-band MMIC LNAs (24–40 GHz and 28–52 GHz). Devices with gate widths of 30–50 μm exhibited cryogenic instabilities, such as current jumps and transconductance discontinuities, attributed to high-frequency oscillations and asymmetrical gate layout. Three stabilization techniques were proposed: source air bridges, gate-end interconnects, and increased gate resistance. These methods effectively suppressed anomalies, ensuring stable, reproducible operation. LNAs incorporating these improvements achieved minimum noise temperatures of 6.7–7 K and average gains of 29–34 dB at 5.5 K, setting benchmarks for InP-based cryogenic amplifiers. This paper highlights the importance of transistor layout and gate parasitics in determining high-frequency stability at cryogenic temperatures, and provides a reliable design path for scalable, high-frequency, low-noise receiver systems.
It is known that HEMTs in line with CMOS are particularly valued in RF applications, including amplifiers operating in different bands (Fig. 5) and more importantly it can be used to generate nonclassicality (Fig. 6). Figs. 5 to 7 emphasize the critical need to minimize the noise figure for quantum applications. They demonstrate how various layout and design strategies can be utilized to achieve this, and more importantly, explore the potential for attaining nonclassical behavior. Fig. 5a shows the PCB layout of a cryogenic LNA specifically engineered to deliver an ultra-low noise figure of approximately 0.01 dB (Fig. 5b), making it well-suited for quantum applications. This design employs negative capacitive feedback and source degeneration in the transistors to enhance impedance matching. Fig. 5c displays the results of an EM simulation, where adjustments to the PCB layout have been made to achieve optimal matching between $\Gamma_s$ and $\Gamma_{out}$, where the parameters mentioned are the input and output reflection coefficients. The primary objective of Fig. 6 is to develop a high-performance LNA that achieves an ultra-low NF, high gain, and high linearity, allowing it to be somewhat comparable to a JPA. Previous research has demonstrated [50] that by carefully designing the LNA with an emphasis on impedance matching (specifically the reflection coefficient), the NF can be reduced to as low as 0.04 dB around 1.392 GHz, corresponding to a noise temperature of about 2.68 K. Although this is higher than the 0.4 K noise temperatures achievable by large arrays of JPAs, it is still quite competitive. Furthermore, using a cryogenic Ultra-LNA with such a minimized noise figure can enhance the likelihood of generating entangled microwave photons, as predicted by quantum theory. Additionally, the gain of LNA shown in Figure 5c, reaches an average of 22 dB across the targeted bandwidth, which is notably higher than the gain of approximately 20 dB provided by JPA. This high gain, combined with low noise performance, makes the designed LNA a valuable component in quantum and low-temperature applications. Along with, Fig.7 illustrates that in the region where the NF is minimized, the entanglement metric rises. However, when the transconductance ($g_m$) is low, the likelihood of generating entangled microwave photons also increases. This effect (the entanglement metric rising) is significantly attributed to the system's nonlinear Hamiltonian [50]. Theare are some critical terms discussed in [50], which play a crucial role in modifying the interaction between the two oscillators, leading to a substantial change in the phase-sensitive properties of the system. Additional details highlighted in the figures within dashed circles provide further insights. They show that in regions with exceptionally low NF, the probability of entanglement creation becomes significantly higher. This analysis underlines the importance of the system's design parameters, such as coupling and transconductance, in optimizing entanglement generation, which is essential for applications in quantum information processing [50].
Additionally, HEMTs exhibit very low noise figures, which is critical in applications like radar, satellite communications, and high-frequency data transmission, where signal integrity is paramount. The ability of HEMTs to operate efficiently at cryogenic temperatures further extends their application in sensitive environments, such as quantum computing, where low noise and high performance are essential. HEMT technology also supports high power operation, which is beneficial in applications requiring significant amplification of weak signals, such as in long-range communication systems [20-36].

***HEMT Applications in the Quantum Realm***:
HEMT technology plays a significant role in quantum computing and quantum communication systems, particularly in applications requiring high-performance RF amplification at cryogenic temperatures: I) Low Noise Amplification: HEMTs are renowned for their extremely low noise figures, which are critical in quantum systems where the detection of ultra-weak signals is essential. In quantum computing, HEMT-based amplifiers are often used in qubit readout chains to amplify quantum signals without adding significant noise, ensuring accurate quantum state measurements. II) Cryogenic Operation: HEMT amplifiers are capable of operating at cryogenic temperatures around 4.2 K, making them suitable for use in quantum systems that require cooling to minimize thermal noise and decoherence. These amplifiers are often deployed in dilution refrigerators, where they serve as the first stage of amplification for signals coming from superconducting qubits. III) Quantum-Limited Amplification: HEMTs can achieve near quantum-limited performance, meaning they add minimal noise close to the quantum limit, making them ideal for quantum measurement applications. This property is crucial for experiments in quantum optics and quantum information processing, where preserving the quantum nature of signals is paramount. IV) High-Frequency Applications: HEMTs' ability to operate at high frequencies makes them suitable for quantum communication systems, particularly in the C- and X-bands. They are used in the development of quantum radars, quantum key distribution systems, and other applications that require secure and efficient transmission of quantum information over long distances. HEMT technology is uniquely suited for quantum applications that demand low noise, high-frequency operation, and cryogenic compatibility. Its role in quantum computing is particularly prominent in the readout and control of qubits, where precision and noise reduction are critical. HEMTs are also used in quantum sensing and metrology, where they enable the detection and amplification of weak quantum signals with minimal disturbance [20-26].

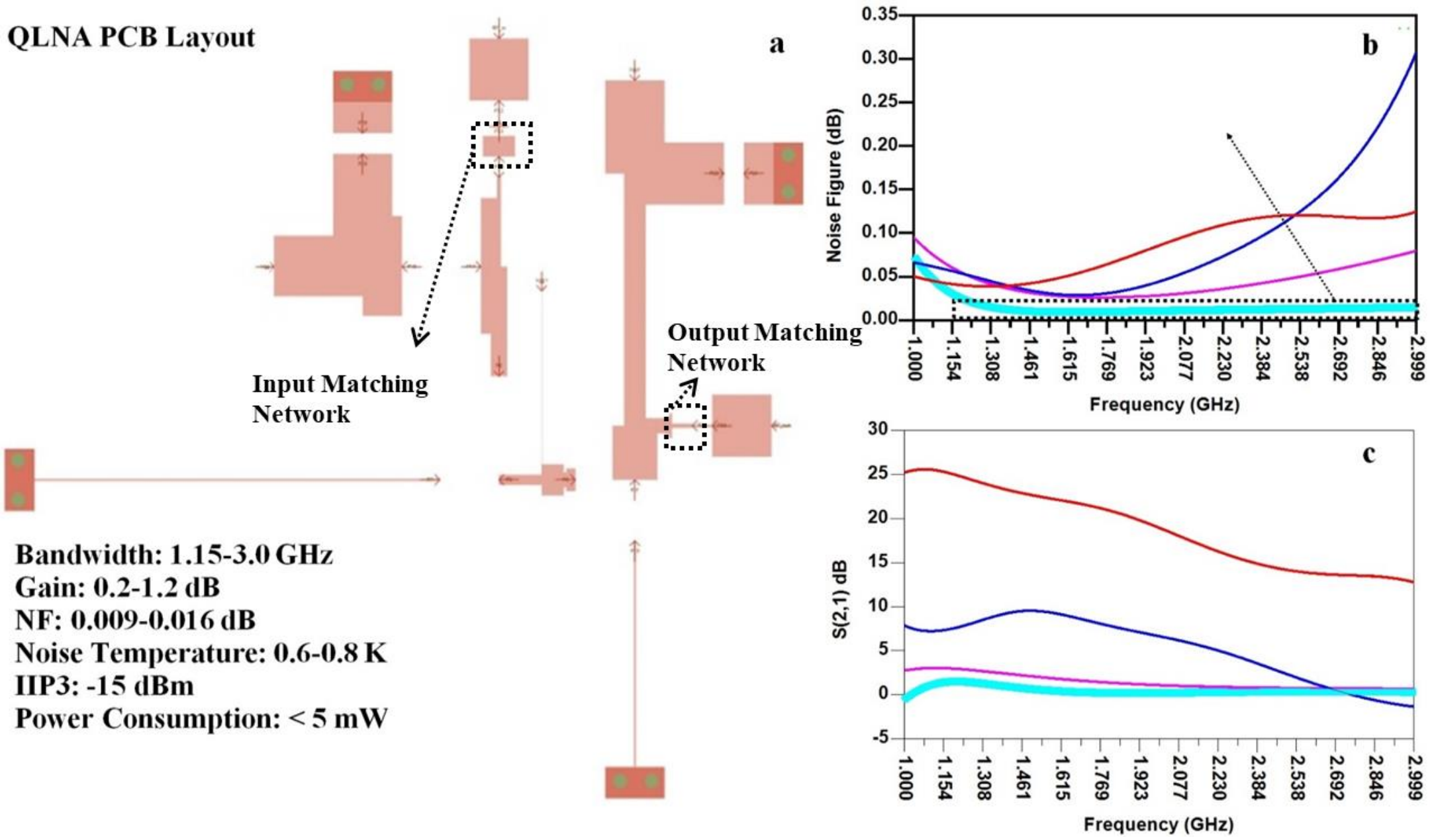

Fig. 5a) Schematic of the PCB layout of the LNA circuit utilizing HEMT designed to be operated in the quantum applications, b) Noise figure of four different designed LNA, c) the comparison between the gains of the four additional LNA [33].

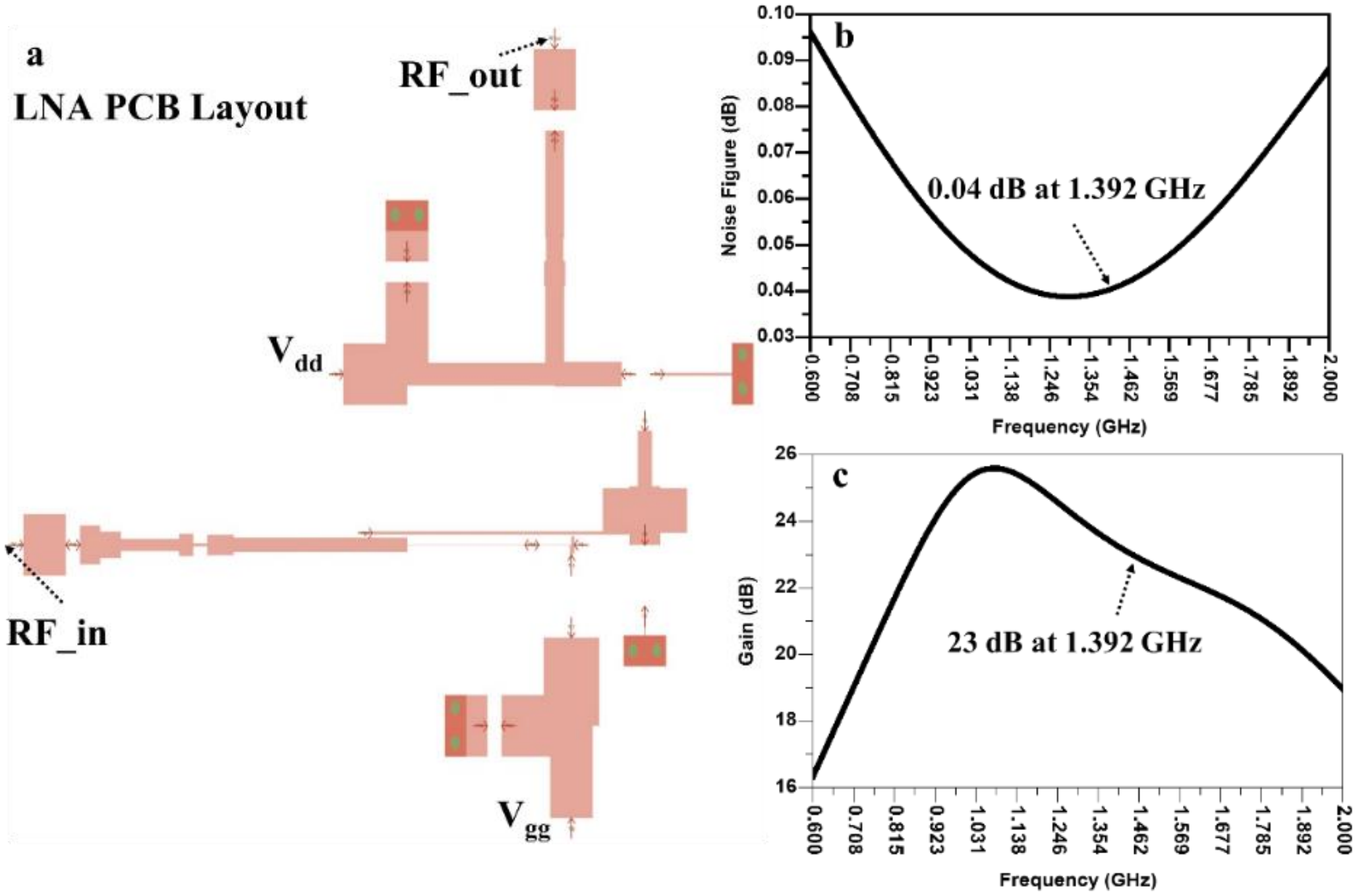

Fig 6. a) PCB layout of the designed LNA using HEMT, b) Noise figure (dB) vs. Frequency (GHz), c) Gain (dB) vs. Frequency (GHz) [50].

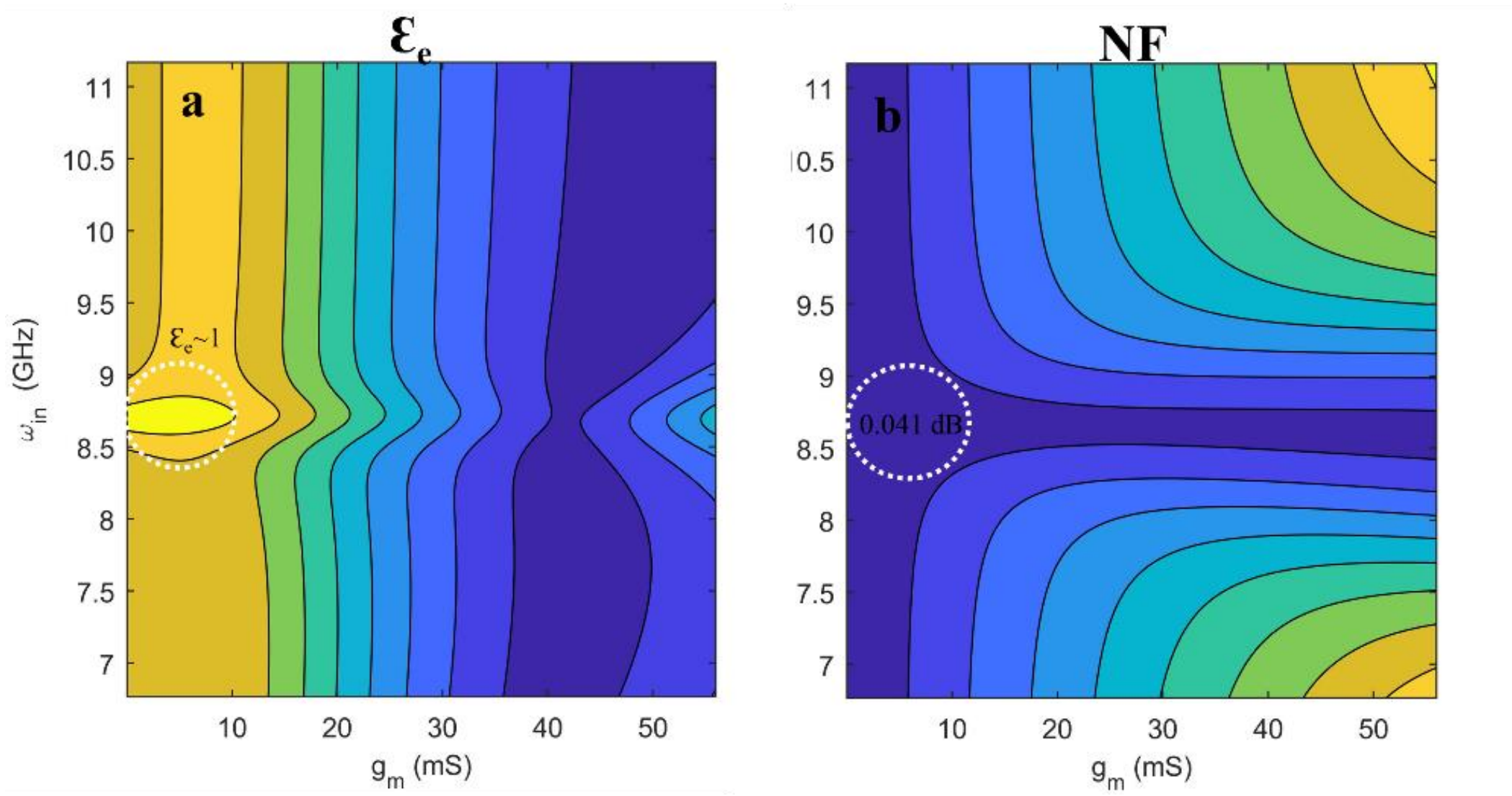

Fig 7. Entanglement profile of LNA depicted in Fig. 6, b) Noise figure (dB), vs. RF source angular frequency (GHz) and transistor intrinsic transconductance $g_m$ (mS) [50].

***Josephson Parametric Amplifier***

A JPA is a superconducting device that utilizes the unique properties of Josephson junctions to achieve ultra-low-noise amplification at microwave frequencies. A Josephson junction is formed by sandwiching a thin insulating layer between two superconducting materials, allowing for the tunneling of Cooper pairs without resistance [1-5]. The nonlinear inductance of the Josephson junction is key to the parametric amplification process, where an external pump signal modulates the junction's inductance, enabling the amplification of weak input signals with minimal added noise. JPAs are renowned for their quantum-limited performance, meaning they add noise very close to the theoretical minimum set by quantum mechanics. This characteristic makes JPAs indispensable in quantum computing and quantum optics, where preserving the quantum nature of signals is critical. JPAs operate at frequencies typically in the microwave range, including the C-band and X-band, and are used in the readout of qubits, where they amplify the minute signals generated by quantum states without disturbing their coherence. The ability to achieve high gain and low noise makes JPAs ideal for quantum measurement applications, where detecting and amplifying weak signals is necessary for accurate quantum state determination. In contrast with the CMOS and HEMT, JPAs can generate and manipulate squeezed states of light, which are important for precision measurements and quantum communication systems. Similar to the approach taken with other amplifier technologies, we begin with a literature survey that outlines the fundamental design features and performance characteristics of JJ-based JPAs. In the second part of this article, we then turn our attention to the theoretical analysis of single-JJ and arrayed-JJ JPAs, as well as emerging JPA architectures, and compare our simulation and modeling results with existing findings in the literature.

The review article [70] outlines the critical role of JPAs in superconducting qubit readout, highlighting both degenerate and nondegenerate configurations. Modern JPAs, often built with flux-pumped SQUIDs or Josephson junction arrays, offer near quantum-limited noise performance, achieving added noise as low as 1–2 times the Standard Quantum Limit (SQL). Typical gain values reach 20–30 dB, but bandwidths are constrained to <10–50 MHz due to the gain-bandwidth trade-off. The $P_{1dB}$ compression point lies near –110 dBm, limited by junction critical currents. A major advancement is the Field-Programmable Josephson Amplifier, featuring three tunable resonances and nonreciprocal operation. The work also discusses

JTWPA, achieving >4 GHz bandwidth and higher power handling, albeit with slightly degraded noise (~4× SQL). Key challenges remain in scalability, power handling, and eliminating bulky cryogenic circulators via on-chip nonreciprocity schemes.
The other work cited in [71] presents a high-performance JPA based on a λ/4 CPW resonator terminated by a flux-tunable dc-SQUID. The study emphasizes novel fabrication techniques, including optimized surface treatments and argon ion milling, which enable unprecedented internal quality factors ($Q_{int}$) exceeding $1.2\times10^5$ at the single-photon level. The JPA demonstrates wide frequency tunability (>750 MHz) and supports the generation of squeezed vacuum states with up to 11.75 dB squeezing and purity ~99%, validating low internal loss. The gain-bandwidth product is indirectly inferred via quality factor tuning, though explicit gain or $P_{1dB}$ data isn't detailed. The novelty lies in achieving near-record $Q_{int}$ through Al/Nb hybrid fabrication and interface passivation, which minimize intrinsic loss channels. This device marks a significant advancement in low-noise superconducting circuits, supporting scalable quantum technologies.
One of the interesting works in this area [72] demonstrates a gate-tunable JPA using a graphene-based Josephson junction (gJJ) embedded in a microwave resonator. The amplifier achieves a gain of 22 dB, gain-bandwidth product of 33 MHz, and exhibits frequency tunability over 1 GHz via a side gate voltage. The $P_{1dB}$ is measured at –123 dBm, comparable to conventional JPAs. Critically, the device adds noise close to the SQL, with intrinsic added noise estimated at ~0.68 photons. Fabrication uses a van der Waals heterostructure and h-BN encapsulation to minimize dissipation and enhance carrier mobility. Unlike flux-tuned devices, the electric-field control allows for localized, scalable tuning with reduced crosstalk. This graphene JPA offers a promising route toward compact, high-performance parametric amplifiers for quantum technologies, with potential for both four-wave and three-wave mixing amplification mechanisms.
Other interesting work cited in [73] present a broadband lumped-element JPA fabricated using a single-step e-beam lithography process, significantly simplifying device fabrication. The JPA operates in the three-wave mixing regime, using a flux-pumped SQUID shunted by an interdigital capacitor. The amplifier achieves a gain of 20 dB, a bandwidth of 95 MHz around 5 GHz, and exhibits tunable operation over >1 GHz by adjusting the pump frequency. The design favors wideband operation over high dynamic range, resulting in a P1dB of approximately –125 dBm. The amplifier demonstrates near-quantum-limited noise performance, making it suitable for qubit readout, shot noise spectroscopy, and other low-noise quantum applications. Key novelties include its ease of fabrication, low-Q architecture, and rapid tunability, which support frequency multiplexing schemes. Despite the trade-off in saturation power, this JPA is highly scalable and can be further optimized with matching structures or multiple SQUID stages.
In addition, [74] present a flux-pumped, impedance-engineered JPA designed for broadband quantum-limited operation. The amplifier employs a lumped-element SQUID resonator integrated with an on-chip impedance transformer, which enables efficient broadband impedance matching. The JPA operates in three-wave mixing mode, achieving a gain >20 dB over a 300 MHz bandwidth, and exhibits a $P_{1dB}$ compression point of –116 dBm, with near-quantum-limited noise performance. A key novelty is the fully monolithic integration of the JPA with a compact design that avoids complex transmission-line structures seen in TWPAs. The architecture also supports four-wave mixing operation with 200 MHz bandwidth. Their engineering approach—optimizing flux-coupling geometry and using pumpistor-based simulations—enables low-loss, scalable fabrication, making this JPA suitable for multiplexed qubit readout. The separation of pump and signal paths simplifies integration in cryogenic systems, positioning this amplifier as a robust solution for high-fidelity, broadband quantum measurements.
In line with, the study of [75] present a non-degenerate JPA based on a Josephson ring modulator (JRM) intersected by two half-wave microstrip resonators supporting signal and idler modes. Operating in a three-wave mixing configuration, the amplifier achieves up to 20 dB gain, bandwidths of 9–11 MHz, and center frequency tunability over 60 MHz, with signal and idler frequencies near 8.1 GHz and 6.4 GHz, respectively. The 1 dB compression point is measured around –125 dBm, suitable for single-photon-level qubit readout. Key novelties include the use of a Wheatstone-bridge-style JRM, enabling balanced, broadband, phase-preserving amplification. The authors also report near-quantum-limited noise performance (~1 photon added), making it a reliable preamplifier for quantum measurements. The design offers enhanced fabrication simplicity and system integration, showing potential for scalable superconducting quantum circuits with minimal noise and high fidelity.

The other article cited in [76] presents the single-shot readout of a superconducting flux qubit using a flux-driven JPA. This JPA, distinctively pumped by AC flux rather than current, operates in a degenerate three-photon mode—offering a practical benefit by avoiding the need for extra cancellation tones. The device exhibits a gain of up to 29 dB, with bandwidth tunable between 2.8 and 4.8 MHz, and a 1 dB compression point around –53.5 dBm, corresponding to a dynamic range suitable for single-photon-level signals. The system achieved a 74% single-shot readout contrast, primarily limited by JPA bandwidth and qubit relaxation ($T_1 \approx 600$ ns). Additionally, quantum jumps were observed in continuous monitoring, affirming the nondestructive nature of the readout. Notably, the flux-driven JPA's broad tunability and compatibility with dispersive readout position it as a robust preamplifier for quantum measurement applications.

The article cited in [77] presents a comprehensive study of a flux-driven JPA with a strong focus on vacuum squeezing performance and noise properties. This flux-driven JPA features a large pump-signal isolation, eliminating the need for pump cancellation circuitry. Operating both in degenerate and non-degenerate modes, it achieves a maximum gain of 25.5 dB, bandwidths up to 3.95 MHz, and a 1 dB-compression point at –133 dBm, corresponding to ~10 photons in a 1 MHz cavity—making it suitable for quantum-limited signal detection. Using dual-path cross-correlation techniques, the JPA demonstrated 4.9 ± 0.2 dB squeezing below vacuum at 10 dB gain. Importantly, in degenerate mode, it achieves a noise temperature below the standard quantum limit, confirming its capability as a phase-sensitive amplifier. These results highlight the JPA's strong potential for use in quantum measurement, entanglement generation, and continuous-variable quantum information processing.

Another article cited in [78] introduces a strongly environmentally coupled JPA, termed the Impedance-Transformed Parametric Amplifier (IMPA). This JPA operates in a non-degenerate (phase-preserving) mode and leverages a tapered transmission line to lower the environmental impedance from 50 Ω to ~15 Ω, thereby reducing Q, enhancing both bandwidth (up to 700 MHz) and input saturation power (as high as –103 dBm at 15 dB gain). Unlike traditional JPAs, this device exhibits multi-peaked gain profiles, explained by an extended "pumpistor" model, which incorporates the impact of environmental admittance at signal and idler frequencies. The study validates that gain profile complexity can be tuned by adjusting the impedance environment, offering flexibility without compromising near-quantum-limited noise performance. The work demonstrates a tenfold improvement in bandwidth and enhanced robustness—key advantages for multi-qubit readout and scalable quantum systems.

***Applications in the Quantum Realm***:

JPAs have become a cornerstone in the field of quantum information processing, where they are used for the following applications: I) Qubit Readout: In quantum computing, JPAs are commonly used in the readout chains of superconducting qubits. They amplify the signals generated by qubits during measurement, allowing for high-fidelity readout with minimal noise addition. The quantum-limited nature of JPAs ensures that the quantum information encoded in the qubit states is preserved during amplification.

II) Quantum Sensing: JPAs are employed in quantum sensing applications, where they enable the detection of extremely weak signals, such as those generated in dark matter searches or gravitational wave detection. Their low noise figure and high gain make them ideal for enhancing the sensitivity of quantum sensors. III) Squeezed State Generation: JPAs are also used to generate squeezed states of light, which are essential for quantum-enhanced measurements. Squeezed states have reduced noise in one quadrature at the expense of increased noise in the orthogonal quadrature, allowing for precision measurements beyond the standard quantum limit. IV) Quantum Communication: In quantum communication systems, JPAs can be used to amplify quantum signals transmitted over long distances, such as in quantum key distribution (QKD). Their ability to amplify signals without significantly degrading the quantum properties of the signal makes them valuable in maintaining secure communication channels.

As a brief conclusion, JPAs are uniquely valuable in quantum systems where the preservation of quantum coherence and minimization of noise are paramount. Their role in qubit readout, quantum sensing, and quantum communication highlights their importance in the development of quantum technologies. The ability to generate and manipulate squeezed states also positions JPAs as critical components in quantum metrology [1-10].

## Technical Comparison of RF Amplifiers in Quantum Technology

When comparing CMOS [37-49], HEMT [20-36], and JPA [3-10] for quantum technology applications, particularly considering their operation at cryogenic temperatures (4.2 K for CMOS and HEMT, and 10 mK for JPA), it is essential to evaluate key performance parameters. These include gain, bandwidth, $P_{1dB}$, $IIP_3$, noise figure, power consumption, component size, and scalability. Below is a technical comparison of these parameters: **Gain**: 1) CMOS: CMOS amplifiers typically offer moderate gain, ranging from 10-20 dB at cryogenic temperatures. However, achieving high gain at such low temperatures can be challenging due to reduced transconductance. 2) HEMT: HEMT amplifiers excel in high-gain performance, with typical gains ranging from 25-45 dB. The high electron mobility in HEMTs enables superior gain even at cryogenic temperatures. 3) JPA: JPAs provide quantum-limited gain, typically in the range of 20-25 dB. This high gain is crucial for amplifying weak quantum signals without adding significant noise. **Bandwidth**: 1) CMOS: The bandwidth of CMOS amplifiers is generally limited, with typical values around 1-4 GHz. The bandwidth may further narrow at cryogenic temperatures due to increased parasitic effects. 2) HEMT: HEMT amplifiers offer wide bandwidth, often exceeding 4 GHz, making them suitable for applications that require high-frequency operation over a broad spectrum. 3) JPA: JPAs typically have narrower bandwidths, usually in the range of 10-50 MHz, which is a trade-off for achieving quantum-limited noise performance. **$P_{1dB}$:** 1) CMOS: The $P_{1dB}$ of CMOS amplifiers is generally lower, around -20 to -10 dBm, indicating limited power handling capabilities. 2) HEMT: HEMTs have a higher $P_{1dB}$, typically around 0 to 10 dBm, due to their ability to handle higher power levels. 3) JPA: JPAs have very low $P_{1dB}$ values, often in the range of -125 to -110 dBm, reflecting their focus on amplifying extremely weak signals. **$IIP_3$:** 1) CMOS: CMOS amplifiers have a moderate $IIP_3$, usually around 0 to 10 dBm, with non-linearities becoming more pronounced at higher frequencies. 2) HEMT: HEMTs exhibit a higher $IIP_3$, typically in the range of 10 to 20 dBm, indicating better linearity and lower distortion. 3) JPA: JPAs, being designed for quantum-limited amplification, generally do not prioritize $IIP_3$, and their values are much lower, as they are optimized for very small-signal operation. **Noise Figure**: 1) CMOS: The noise figure for CMOS amplifiers at 4.2 K is typically around 0.6~0.9 dB. While this is lower than room temperature, it still lags behind HEMT and JPA technologies. 2) HEMT: HEMTs are known for their extremely low noise figures, often below 0.04 dB, even at cryogenic temperatures, making them highly suitable for low-noise quantum applications. 3) JPA:

JPAs achieve near quantum-limited noise figures, typically below 0.0065 dB, which is essential for preserving the integrity of quantum signals. **Power Consumption**: 1) CMOS: CMOS amplifiers are highly energy-efficient, with power consumption typically in the milliwatt-microwatt range, even at cryogenic temperatures. 2) HEMT: HEMTs consume more power, usually in the range of tens of milliwatts, which is still manageable but higher than CMOS. 3) JPA: JPAs consume very little power, typically in the microwatt range, as they rely on superconducting circuits operating at millikelvin temperatures. **Size of Components**: 1) CMOS: CMOS amplifiers are compact and can be integrated into small chips, making them suitable for large-scale integration in quantum processors. 2) HEMT: HEMTs are generally larger than CMOS components but can still be integrated into compact designs, though not as easily as CMOS. 3) JPA: JPAs are bulkier due to the need for superconducting materials and cryogenic setups, limiting their integration density compared to CMOS and HEMT. **Scalability**: 1) CMOS: CMOS technology is highly scalable, benefiting from well-established semiconductor fabrication processes, allowing for mass production and integration into large quantum systems. 2) HEMT: HEMT technology is less scalable than CMOS but still allows for moderate integration levels. The fabrication processes are more complex and costly. 3) JPA: JPA scalability is limited due to the need for superconducting materials and operation at mK temperatures, making large-scale integration challenging.

**Table 1.** Comparison of amplifier parameters at cryogenic temperatures. CMOS and HEMT amplifiers at 4.2 K, along with JPAs at 10 mK, are evaluated in terms of gain, bandwidth, $P_{1dB}$, noise figure, power consumption, size, and scalability.

| **Parameter** | **CMOS (4.2 K) [37-49]** | **HEMT (4.2 K) [20-36]** | **JPA (10 mK) [3-10]** |
|---|---|---|---|
| **Gain** | 10-20 dB | 20-30 dB | 20-25 dB |
| **Bandwidth** | ~1-4 GHz | ~4 GHz | 10-40 MHz |
| **$P_{1dB}$** | -20 to -10 dBm | 0 to 10 dBm | -125 to -108 dBm |
| **Noise Figure** | 0.6~0.9 dB | ~0.02 dB | ~0.0065 dB |
| **Power Consumption** | Milliwatts | Tens of milliwatts | Microwatts |
| **Size** | Very compact | Moderate | Larger, bulkier |
| **Scalability** | Highly scalable | Moderately scalable | Limited scalability |

As an interesting result from the table introduced, JPAs demonstrate exceptional noise performance but are less scalable, whereas CMOS and HEMT amplifiers balance scalability and performance with higher power consumption and larger bandwidths.

The latter section presents a comprehensive comparison of RF amplifiers based on CMOS, HEMT, and JPA technologies, with a focus on their application in quantum systems operating at cryogenic temperatures. It tries to answer the latter questions regarding why JPA is preferred rather than CMOS-and HEMT-based amplifier in quantum circuits. CMOS and HEMT amplifiers are evaluated for their performance at 4.2 K, while JPAs are assessed at 10 mK, the typical operating temperature for quantum computing environments. Key parameters such as gain, bandwidth, $P_{1dB}$, $IIP_3$, noise figure, power consumption, component size, and scalability are analyzed to determine their suitability for quantum applications. The results indicate that while CMOS and HEMT technologies offer advantages in terms of scalability, integration, gain, $P_{1dB}$, and wider bandwidth, they are limited by higher noise figures, greater power consumption, and reduced compatibility with the ultra-low temperatures required for quantum systems. In contrast, JPAs, with their quantum-limited noise performance, low power consumption, moderate gain, and operation at mK temperatures are uniquely suited to the stringent demands of quantum applications. Despite their narrower bandwidth and challenges in scalability, JPAs provide unparalleled precision in amplifying weak quantum signals, making them the preferred choice for maintaining quantum coherence and ensuring accurate quantum state readout. This comparison highlights the critical role of JPAs in advancing quantum technology, particularly in applications where minimizing noise and maintaining low-temperature conditions are essential. Building on the emphasis placed on JPAs, we conduct a comprehensive quantum-theoretical analysis of a typical JPA and introduce several novel structural variations aimed at addressing key limitations of single JPAs, such as limited $P_{1dB}$ and bandwidth. Finally, we compare the results obtained from quantum theory and CAD simulations with available experimental data.

## Design of a single Josephson Junction JPA and its technical features:

A single JJ can function as a parametric amplifier by leveraging its intrinsic nonlinearity to enable energy transfer between electromagnetic modes. Fundamentally, a JJ consists of two superconducting electrodes separated by a thin insulating barrier, allowing Cooper pairs to tunnel through via the Josephson effect [6-13]. The junction exhibits a nonlinear inductance, which is the key ingredient for parametric amplification. When driven by an external pump tone at frequency $F_{pump}$, the nonlinear inductance modulates periodically, allowing for energy exchange between different frequency components. This parametric process can lead to either degenerate or non-degenerate amplification [6-10]. In degenerate mode, the signal and idler share the same frequency but have a phase relationship that governs the gain. In non-degenerate mode, the idler frequency is different from the signal, leading to broader operational flexibility. This process is governed by the quantum Langevin equation and input-output formalism, where the interaction Hamiltonian describes the coupling between the pump, signal, and idler fields. The strength of amplification is dictated by the junction's critical current $I_c$, its shunt capacitance, and the external impedance environment [67, 68]. These parameters influence the bandwidth, gain, and stability of the amplifier. The gain mechanism fundamentally relies on parametric conversion, where the pump injects energy into the signal and idler fields while maintaining overall energy conservation. Following, the article center on the theoretical analysis of single-JJ and arrayed-JJ JPAs, as well as emerging JPA architectures, and compare our simulation and modeling results with existing findings in the literature.

### *Theory behind the design of a single JJ JPA: JPA gain calculation using quantum theory*

In this section, a single JJ JPA is quantum mechanically analyzed and it attempts to theoretically derive amplifier critical parameters such as gain. The approach begins with the theoretical derivation of the system's Hamiltonian, followed by an analysis of the dynamics using the quantum Langevin equation. The Hamiltonian of lumped element JPA (that is a capacitively shunted JJ coupled to a transmission line [67-68]) is expressed as (setting $\hbar = 1$ for reminder of the article):

$$H_{JPA} = \frac{\hat{Q}^2}{2C} - E_J \cos(\hat{\varphi}) \longrightarrow \approx \omega_0 \hat{a}^+ \hat{a} + \frac{\Lambda}{6}\left(\hat{a}^+ + \hat{a}\right)^4 \quad (1)$$

where C, $E_J$, $Q$, and $\varphi$ are the capacitance, Josephson energy, charge, the phase across the junction, respectively. The other side of the arrow is concluded by expanding the cosine up to forth order and introducing annihilation and creation ($a$, $a^+$) operators. In this equation, the Kerr coefficient $\Lambda = -E_c/2$ and the bare frequency of the resonator, $\omega = \sqrt{8E_cE_J}$, where $E_c = e^2/2C$ is the JJ charging energy. Using the point mentioned, the Hamiltonian of a single JJ JPA drive by a monochromatic current pump [67, 68] is expressed as

$$H_{CP} = H_{JPA} + \varepsilon e^{(-i\omega_p t)}\hat{a}^+ + \varepsilon^* e^{(i\omega_p t)}\hat{a} \tag{2}$$

where $\mathcal{E}$ and $\omega_p$ are the drive pump amplitude and its frequency, respectively. To simplify the successive algebra, it becomes versatile to eliminate the pump Hamiltonian using a displacement operator as $a = \alpha + \delta a$, where $\alpha$ and $\delta a$ are the classic field and quantum fluctuation, respectively. In the displaced frame mentioned, the Hamiltonian of the system becomes:

$$H_{CP} = \Delta_0 \delta\hat{a}^+\delta\hat{a} + \frac{\lambda_1}{2}\delta\hat{a}^{+2} + \frac{\lambda_1^*}{2}\delta\hat{a}^2 + H_{nc} \tag{3}$$

In this equation the shifted detuning is $\Delta_0 = \omega_0 + 4|\alpha|^2\Lambda - \omega_p$, and the effective parametric pump strength is $\lambda_1 = 2\alpha^2\Lambda$, and also the classic field can be solved using the quantum Langevin equation [67, 68]. Finally, the last term in the equation, $H_{nc}$, is the nonlinear correction Hamiltonian arisen from the displacement of the Kerr nonlinearity and can be figured out as:

$$H_{nc} = \mu_0 \delta\hat{a}^{+2}\delta\hat{a} + \mu_0^* \delta\hat{a}^+\delta\hat{a}^2 + \Lambda\delta\hat{a}^{+2}\delta\hat{a}^2 \tag{4}$$

where $\mu_0 = 2\alpha^2\Lambda$ is the cubic term coefficient. However, to obtain a linear equation one can ignore $H_{nc}$ for the small quantum fluctuation, which is valid in low gain and low Kerr nonlinearity regime [67]. It is notable to indicate that the output state of the JPA pumped with a monochromatic current regarding Eq. 3 and Eq.4 is a displaced squeezed state. In the following, we just focus on Eq. 3 by ignoring $H_{nc}$, and using the input-output formula ($\delta a_{out} = \sqrt{\kappa}\delta a - \delta a_{in}$) [67, 68] attempt to analytically calculate the gain of the system discussed. Generally, the behavior of the intracavity mode may be calculated by master equation method; still one can use quantum Langevin equation which is for single mode cavity as:

$$\frac{d\delta\hat{a}}{dt} = -i\left[\delta\hat{a},(H_{CP} - H_{nc})\right] - \frac{\kappa}{2}\delta\hat{a} + \sqrt{\kappa}\delta\hat{a}_{in} \tag{5}$$

where $\kappa$ and $\delta a_{in}$ are the JPA damping constant created due to the any mismatching between the JPA and environment embedded, quantum fluctuation of external field applied to the system, respectively. Eq. 5 can be expanded for *δa and* its conjugate as:

$$\begin{aligned} \frac{d\delta\hat{a}}{dt} &= -\left(i\Delta_0 + \frac{\kappa}{2}\right)\delta\hat{a} - i\lambda_1\delta\hat{a}^+ + \sqrt{\kappa}\delta\hat{a}_{in} \\ \frac{d\delta\hat{a}^+}{dt} &= -\left(-i\Delta_0 + \frac{\kappa}{2}\right)\delta\hat{a}^+ + i\lambda_1^*\delta\hat{a} + \sqrt{\kappa}\delta\hat{a}_{in}^{\ +} \end{aligned} \tag{6}$$

Eq. 6 is a linear and in terms of frequency components ($a(\omega) = 1/\sqrt{2\pi}\int a(t)e^{i\omega t}dt$) it becomes:

$$\begin{aligned} -i\omega\delta\hat{a}(\omega) &= -\left(i\Delta_0 + \frac{\kappa}{2}\right)\delta\hat{a}(\omega) - i\lambda_1\delta\hat{a}^+(-\omega) + \sqrt{\kappa}\delta\hat{a}_{in}(\omega) \\ i\omega\delta\hat{a}^+(-\omega) &= -\left(-i\Delta_0 + \frac{\kappa}{2}\right)\delta\hat{a}^+(-\omega) + i\lambda_1^*\delta\hat{a}(\omega) + \sqrt{\kappa}\delta\hat{a}_{in}^{\ +}(-\omega) \end{aligned} \tag{7}$$

The last step to calculate the gain is to establish the relating scattering matrix, then employing the input-output formula, the gain of the signal and idler can be calculated [7], [19], [20]. It is shown as follows:

$$\begin{cases} \left\{-i(\omega+\Delta_0)+\frac{\kappa}{2}\right\}\delta\hat{a}(\omega)+i\lambda_1\delta\hat{a}^+(-\omega)=\sqrt{\kappa}\delta\hat{a}_{in}(\omega) \\ \left\{i(\omega-\Delta_0)+\frac{\kappa}{2}\right\}\delta\hat{a}^+(-\omega)-i\lambda_1^*\delta\hat{a}(\omega)=\sqrt{\kappa}\delta\hat{a}_{in}^{\ +}(-\omega) \end{cases} \tag{8a}$$

$$\Rightarrow\begin{bmatrix}\delta\hat{a}_{out}(\omega)\\ \delta\hat{a}_{out}^{+}(-\omega)\end{bmatrix}=\left\{\kappa\begin{bmatrix}\left\{-i(\omega+\Delta_0)+\frac{\kappa}{2}\right\} & i\lambda \\ -i\lambda_1^* & \left\{i(\omega-\Delta_0)+\frac{\kappa}{2}\right\}\end{bmatrix}^{(-1)}-\begin{pmatrix}1&0\\0&1\end{pmatrix}\right\}\begin{bmatrix}\delta\hat{a}_{in}(\omega)\\ \delta\hat{a}_{in}^{\ +}(-\omega)\end{bmatrix}$$

Finally, the gain matrix is presented as:

$$\begin{bmatrix}\delta\hat{a}_{out}(\omega)\\ \delta\hat{a}_{out}^{+}(-\omega)\end{bmatrix}=\begin{bmatrix}\dfrac{\kappa\left\{i(\omega-\Delta_0)+\frac{\kappa}{2}\right\}}{\frac{\kappa^2}{4}-i\kappa\Delta_0+\omega^2-\Delta_0^{\ 2}}-1 & \dfrac{-i\lambda\kappa}{\frac{\kappa^2}{4}-i\kappa\Delta_0+\omega^2-\Delta_0^{\ 2}} \\ \dfrac{i\lambda_1^*\kappa}{\frac{\kappa^2}{4}-i\kappa\Delta_0+\omega^2-\Delta_0^{\ 2}} & \dfrac{\kappa\left\{-i(\omega+\Delta_0)+\frac{\kappa}{2}\right\}}{\frac{\kappa^2}{4}-i\kappa\Delta_0+\omega^2-\Delta_0^{\ 2}}-1\end{bmatrix}\begin{bmatrix}\delta\hat{a}_{in}(\omega)\\ \delta\hat{a}_{in}^{\ +}(-\omega)\end{bmatrix} \tag{8b}$$

where the matrix (1,1) element determines the signal gain and (1,2) element signifies the idler gain. It should be noted that the effective parametric pump strength affects the idler gain not signal gain; in contrast the signal gain depends on the quantum system decay and frequency detuning. The gain of the single JJ JPA is defined as:

$$G_{\sin gle_JJ_JPA}=\left\{\frac{\kappa\left\{i(\omega-\Delta_0)+\frac{\kappa}{2}\right\}}{\frac{\kappa^2}{4}-i\kappa\Delta_0+\omega^2-\Delta_0^{\ 2}}-1\right\}-\left\{\frac{-i\lambda\kappa}{\frac{\kappa^2}{4}-i\kappa\Delta_0+\omega^2-\Delta_0^{\ 2}}\right\} \tag{9}$$

where the first part signifies the signal gain and the second part determines the idler gain. In the next section, a single JJ JPA will be simulated using a CAD software and we study all factors that affect the important parameters of a typical JPA such as gain, bandwidth, and linearity.

### *Simulation and modeling of a single JJ JPA*

The schematic of the circuit used as a JPA is shown in Fig. 8. In this configuration, a single JJ serves as the nonlinear medium, where both the signal and pump tones are applied to enable energy transfer from the pump to the signal through the junction's intrinsic nonlinearity. A circulator is included to direct the amplified signal toward the output while isolating and blocking any reflected signals. To attain information regarding the spectrum of the circuit illustrated, the output frequency response of the JPA is displayed in Fig. 9. The spectral graph reveals a prominent peak close to 5 GHz, indicating that the device is successfully amplifying the quantum signal around the frequency mentioned. The measured spectrum provides insights into the gain profile, the extent of amplification, and the presence of harmonics or spurious signals generated during the nonlinear amplification process. This figure is critical for assessing the purity and efficiency of the JPA's operation, especially in quantum applications where low noise is paramount. Furthermore, the spectral width of the amplified signal offers an indication of the amplifier's effective bandwidth, which is dramatically affected by the JPA energy decaying.

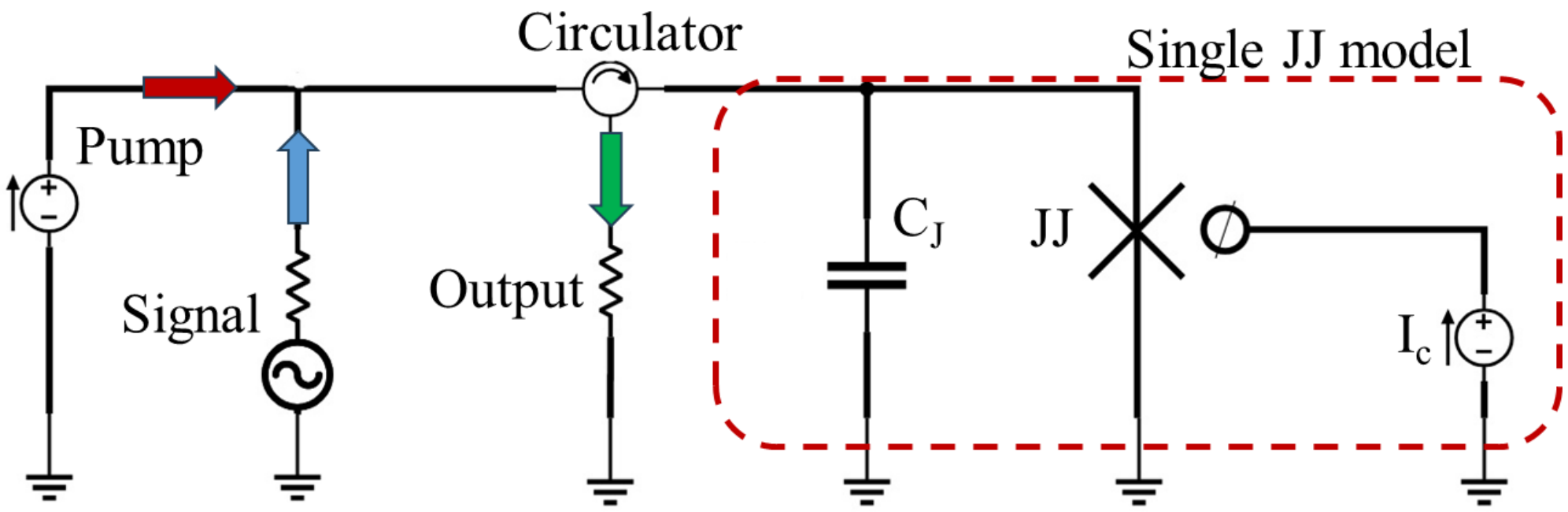

Fig. 8 Schematic of single JJ JPA; herein single JJ is used as a nonlinear medium, and signal and pump is applied to transfer the pump energy to signal via the nonlinearity of the medium; finally circulator is used to direction the signals and blocks the back coming signals.

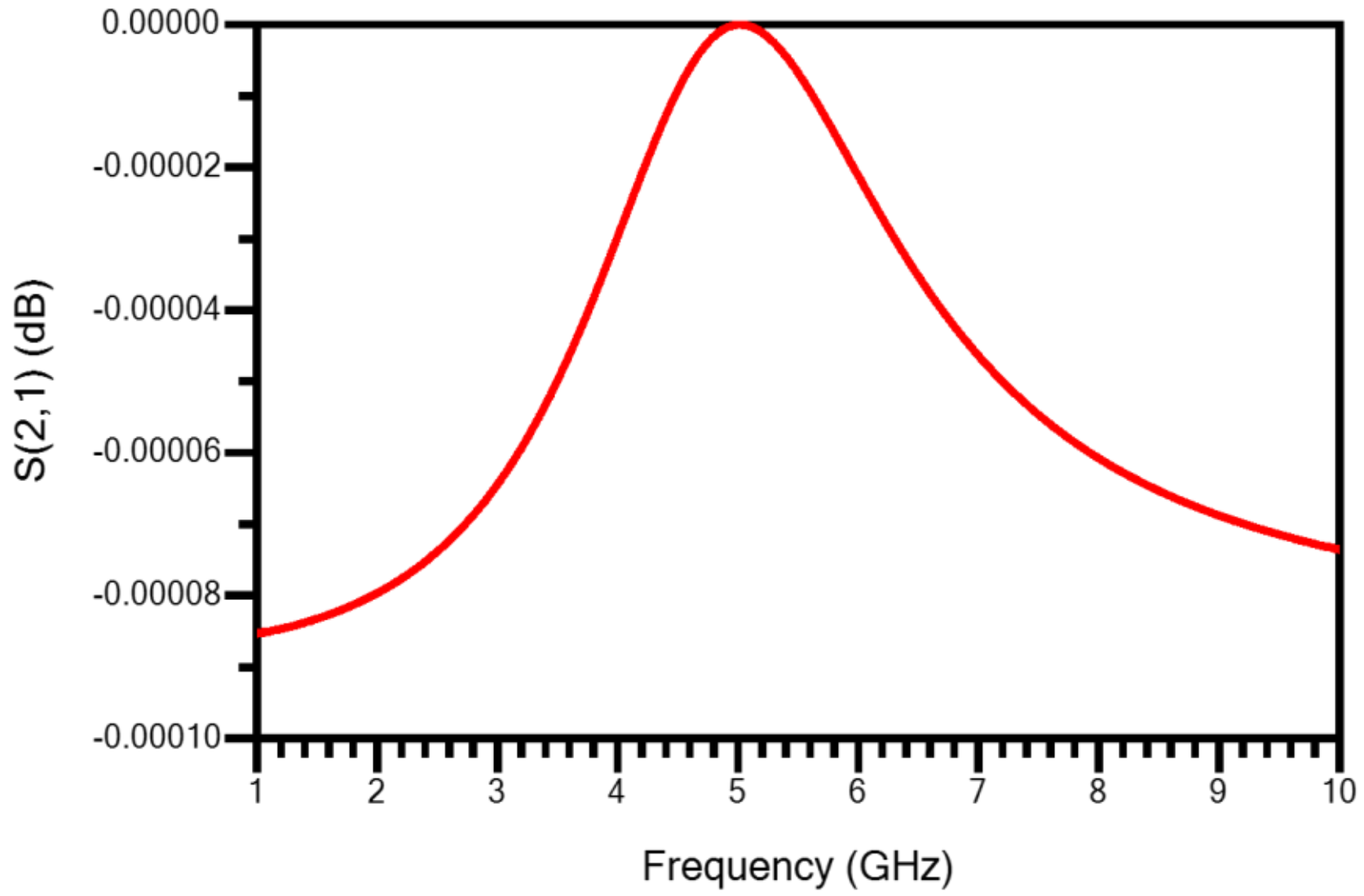

Fig. 9 Spectrum of single JJ JPA; $F_{sig}$ = 5.0 GHz, $F_{pump}$ = 10 GHz.

As a routine way, we study how the single-JJ JPA's amplification factor changes as a function of varying pump bias currents while operating with a fixed signal frequency of 5.0 GHz and pump frequency of 10 GHz; the results illustrated in Fig. 10. The graph shows multiple gain curves corresponding to different bias current levels. At lower bias currents, the device exhibits moderate gain over a range of frequencies, indicating that the nonlinear inductance of the junction is activated without reaching the onset of strong nonlinearities. As the bias current increases, the gain of the amplifier initially increases because the nonlinearity becomes more pronounced, facilitating efficient energy transfer from the pump to the signal. This enhancement in gain is beneficial for applications requiring high sensitivity, as the device can amplify extremely weak quantum signals to measurable levels. The plotted gain profiles also help in identifying the optimal bias current region where the amplifier provides maximum gain without introducing excessive distortion or instability. However, the graph also reveals an important phenomenon: the eventual decrease in gain when the bias current is increased beyond a certain point. When the bias current nears or exceeds the critical current of the JJ, the device enters a regime where the junction's nonlinearity becomes too strong, leading to early saturation and a reduction in the effective gain. This saturation effect is directly related to $P_{1dB}$, which is observed to drop as the bias current increases further. The reduction in $P_{1dB}$ indicates that the amplifier can handle a lower maximum input signal before experiencing significant compression of the gain. Essentially, once the bias current exceeds the critical point, the junction's superconducting properties are disrupted, and the increased energy dissipation leads to a decline in the parametric amplification efficiency. This behavior underscores the necessity of precisely tuning the bias current in a JPA. It is a delicate balance: increasing the bias current enhances the nonlinearity up to an optimal point, but pushing it further results in excessive nonlinearity, reduced gain, and lower dynamic range, thereby limiting the overall performance of the amplifier.

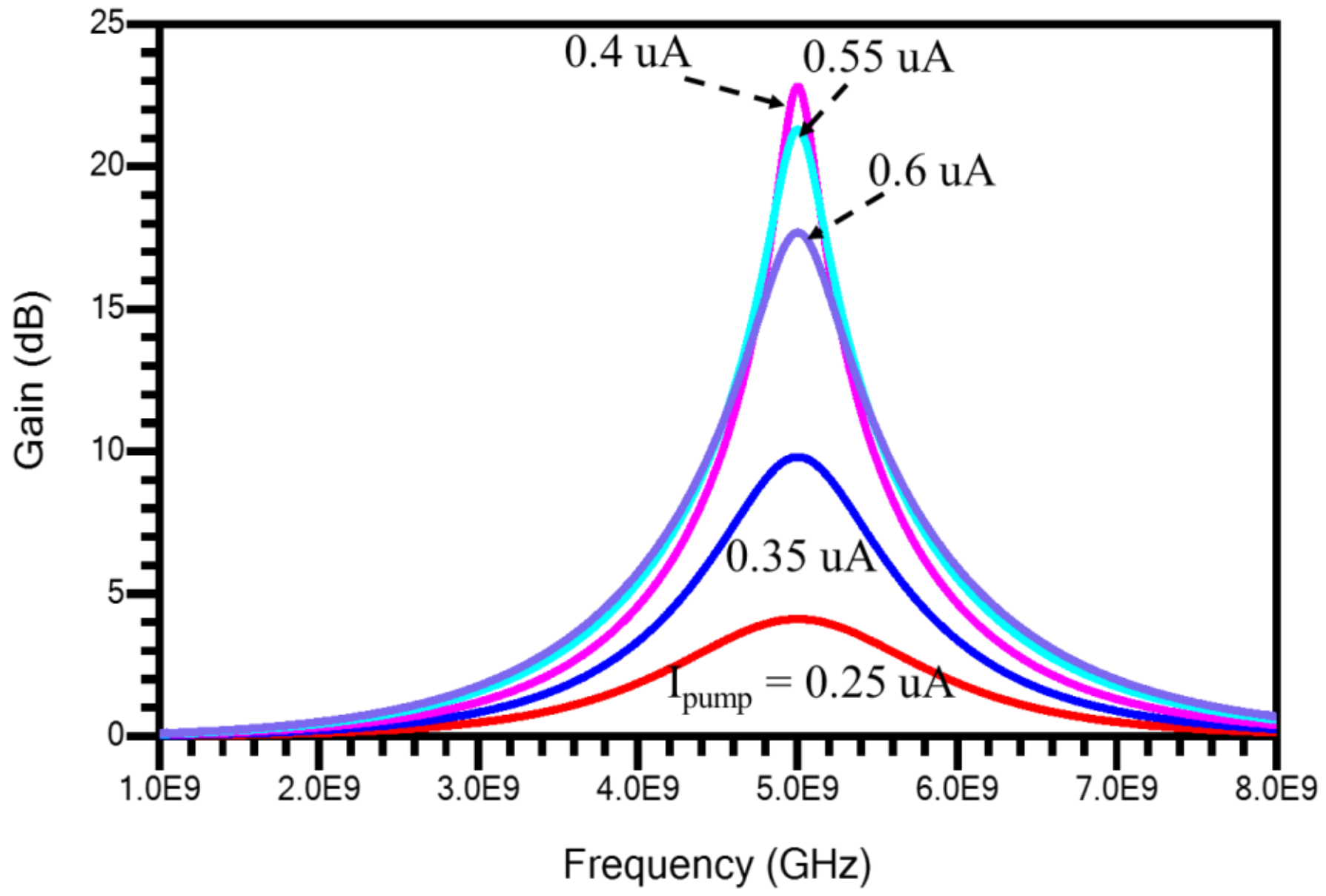

Fig. 10 the spectrum of single JJ JPA gains for different pump current, $F_{sig}$ = 5.0 GHz, $F_{pump}$ = 10 GHz.

In the line with the latter results, Fig. 11 provides a detailed analysis of the 1-dB compression point of the single-JJ JPA as a function of the pump bias current. This figure is crucial for understanding the dynamic range of the amplifier and its linearity. The $P_{1dB}$ represents the input power level at which the gain of the amplifier drops by 1 dB from its small-signal value. In this graph, as the bias current is varied, one can observe that $P_{1dB}$ decreases with an increase in the bias current. At lower bias currents, the amplifier operates in a relatively linear regime, where the gain remains stable over a broader range of input powers. This is indicative of a regime where the junction's nonlinearity is moderate, and the system can sustain a higher input power without significant gain compression. The graph hence demonstrates the advantage of operating at lower bias currents when a wide dynamic range is needed, such as in applications where signal strength can vary substantially. On the other hand, the figure clearly shows that when the bias current is increased toward or beyond the critical current, $P_{1dB}$ is markedly reduced. This decrease is due to the excessive nonlinearity that sets in as the bias current increases. In the high bias regime, the nonlinear inductance of the JJ becomes so pronounced that even a slight increase in the input signal results in a disproportionate increase in the nonlinear response, leading to earlier gain compression. Essentially, the junction's effective impedance and its reactive properties are altered by the high bias, causing the energy transfer process to become inefficient. This saturation effect means that the amplifier can no longer handle higher input signals without sacrificing gain, thus reducing the dynamic range. Moreover, the excessive bias current can also induce phase instabilities and increased noise levels, further degrading performance.

This behavior underscores the importance of optimizing the bias current: while a higher bias can increase the parametric gain up to a point, exceeding the critical current drastically reduces the 1-dB compression point, thereby limiting the overall performance and linearity of the JPA. Balancing these factors is critical in designing a JPA that can achieve high sensitivity while maintaining robust performance in the face of varying signal conditions.

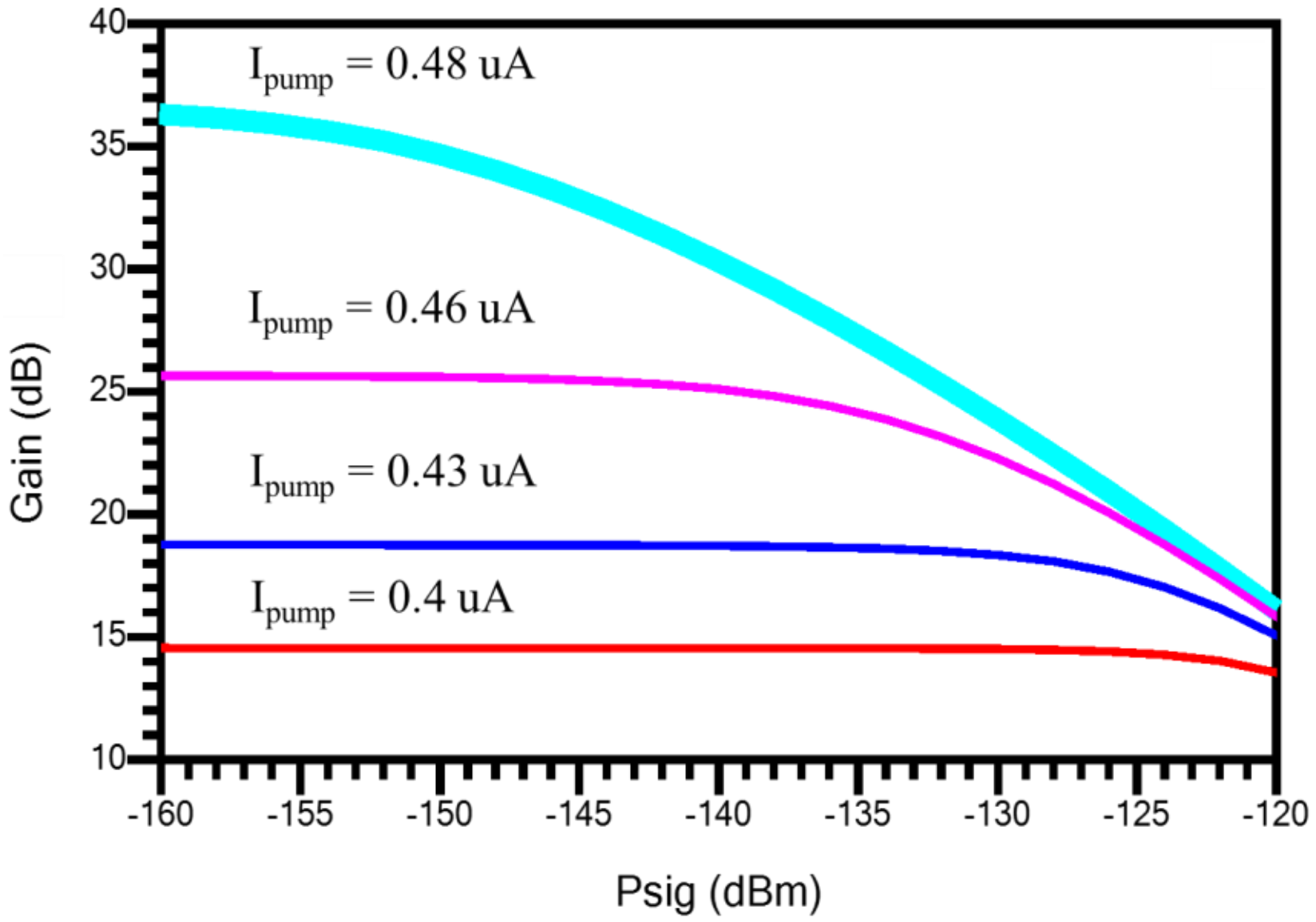

Fig. 11 Single JJ JPA $P_{1dB}$ vs signal power (dBm) for different pump current, $F_{sig}$ = 5.0 GHz, $F_{pump}$ = 10 GHz.

While a single JJ provides the necessary nonlinearity for parametric amplification, it comes with inherent limitations specifically a very limited $P_{1dB}$ or dynamics range that constrain its practical use. Due to the strong nonlinearity of a single JJ, its gain rapidly compresses at high input powers, leading to distortion and a limited dynamic range. Furthermore, the quantum noise performance of a single-JJ JPA is strongly dependent on phase fluctuations in the pump tone. Because parametric amplification relies on coherent energy transfer, any noise or instability in the pump signal can degrade the amplifier's performance, leading to excess phase noise and gain variations. Additionally, the gain-bandwidth product is inherently limited in a single-JJ system, requiring trade-offs between achieving high gain and maintaining a sufficiently broad operational bandwidth. Despite these limitations, single-JJ JPAs remain a fundamental building block in superconducting quantum technologies, often serving as prototypes for more advanced multi-junction designs that address these challenges while preserving the underlying parametric amplification mechanism. Nonetheless, it is necessary to improve some JPA critical properties such as linearity and bandwidth [79-87]. For this, one of the versatile approaches is to employ the array of JJ rather than a single JJ. The subsequent section focuses on this point and discuss how array of JJ can improve the linearity of the JPA and alongside how it is possible to array configuration enhance the JPA critical properties. Before delving into the theoretical analysis, we first present a literature survey emphasizing the key design features and performance characteristics of JJ-array-based JPAs. This sets the stage for a detailed examination of our theoretical model, followed by a comparison of our simulation results with relevant findings from the literature.

The reference [79] presents a distributed JPA using a 1000-junction series array embedded in a coplanar waveguide. Operated in the doubly degenerate mode at 19 GHz and 1.7 K, the amplifier achieves a peak gain of 16 dB with an instantaneous bandwidth of 125 MHz. The double sideband noise temperature referred to the input is approximately 0.56 ± 0.1 K, close to the theoretical Nyquist noise limit. Linearity tests yield a 3 dB compression point at −90.5 dBm and a dynamic range of 38 dB. This amplifier avoids impedance-matching challenges by using the series array architecture, forming a distributed transmission line with a nonlinear inductance. The design is optimized for wide bandwidth without resonant matching structures, and the low noise performance suggests suitability for sensitive microwave measurements, including quantum applications.

The other study [80] explores Josephson tunnel junction arrays (10–40 junctions) configured for parametric amplification at 10 GHz. These amplifiers achieved up to 24 dB of signal gain using −50 to −70 dBm of pump power. Coherence among 50–90% of the junctions contributed to effective amplification. A 20 dB gain bandwidth was observed to be 15 MHz, with power width (3 dB range around peak gain) around 0.25 dB. The amplifiers were tunable via external magnetic fields, enabling optimization of gain and phase coherence. The authors confirmed theoretical predictions from the SUPARAMP model and emphasized practical utility in the millimeter-wave regime, noting high potential for future quantum and cryogenic electronics due to the robustness and tunability of tunnel-junction arrays.

Reference [81] proposes a flux-driven JTWPA based on a linear, non-Kerr architecture. The amplifier consists of symmetric dc SQUIDs forming the signal line, coupled to a separate pump line that drives a traveling flux wave. This configuration supports three-wave mixing ($\omega_s + \omega_i = \omega_p$) while minimizing self- and cross-modulation, typically problematic in four-wave-mixing JTWPAs. With proper phase matching, this design offers broad bandwidth, high dynamic range, and reduced pump depletion. Simulated gains of 20 dB over a wide bandwidth ($\sim 0.5\omega_p$) are achieved with e-folding lengths determined by coupling strength and pump power. Unlike conventional designs, signal and pump ports are separated, simplifying system integration. The architecture is well-suited for scalable quantum systems due to its linearity, reduced nonlinearity constraints, and independent signal/pump paths.

The work [82] introduces a flux-modulated JPA built from a lumped-element LC resonator with the inductance formed by a series of 8 SQUIDs. The amplifier operates using three-wave mixing, with parametric modulation applied at twice the resonator frequency (≈5.97 GHz). Key achievements include 31 dB maximum gain, gain-bandwidth product >60 MHz, and a $P_{1dB}$ of –123 dBm at 20 dB gain. The array of SQUIDs reduces overall nonlinearity, enhancing dynamic range by ~18 dB over single-SQUID designs. Operation in both degenerate and nondegenerate modes is demonstrated, with phase-sensitive amplification reaching 37 dB gain. This architecture supports high-fidelity readout and squeezing while avoiding strong pump leakage through signal ports.

The other paper [83] proposes a JTWPA based on one-junction SQUIDs embedded in a transmission line, enabling controlled balance of quadratic and cubic nonlinearities via external magnetic flux. This enables three-wave mixing ($\omega_s + \omega_i = \omega_p$) operation with minimal phase mismatch and no pump leakage into the signal band. With N = 300 sections and a 12 GHz pump, the amplifier achieves an estimated flat gain of 20 dB across 5.6 GHz bandwidth. The design reduces pump power requirements and supports high dynamic range. The architecture separates gain and phase matching control through distinct nonlinearities ($\beta$ for gain, $\gamma$ for dispersion), simplifying optimization. This three-wave JTWPA architecture is particularly suitable for wideband, low-noise applications such as multiplexed qubit readout.

The other interesting study [84] presents detailed numerical modeling of a JTWPA with 990 rf-SQUID-based cells. Unlike simplified models, it captures full nonlinear dynamics through coupled differential equations, revealing gain behavior, pump depletion effects, and signal distortion. The modeled device exhibits gain up to ~11 dB, bandwidth ~6 GHz centered at 7 GHz, and noise near the quantum limit. Operating regimes include both four-wave mixing (4WM) and three-wave mixing (3WM). The study also shows that excessive pump power or high-frequency pumping induces chaotic behavior, degrading linearity and causing broadband noise. The work underscores the importance of tuning pump conditions to avoid harmonic distortion and optimize signal fidelity—critical for multi-qubit readout or axion detection systems.

Study [85] presents a JTWPA leveraging 3WM in a noncentrosymmetric Josephson metamaterial composed of an array of one-junction SQUIDs. By biasing the SQUIDs with a DC flux to achieve a phase drop of $\pm\pi/2$, the system introduces a quadratic nonlinearity while suppressing unwanted Kerr-type effects, allowing phase matching without complex dispersion engineering. The amplifier achieves a measured gain of ~11 dB over a broad 3 GHz bandwidth (4.8–7.8 GHz) with estimated saturation power around –95 dBm and near quantum-limited performance. This design minimizes cubic nonlinearity, leading to smoother gain profiles and better dynamic range. The architecture is scalable and tunable, with exponential gain dependent on the length of the nonlinear transmission line, marking it a promising candidate for low-noise, broadband quantum signal amplification.

The paper cited in [86] introduces a lumped-element JPA using a SQUID-array-based quarter-wave resonator whose resonance frequency is tuned via magnetic flux from 4 to 7.8 GHz, enabling center frequency reconfiguration while maintaining strong performance. The amplifier achieves a maximum gain of 28 dB, with 3 dB bandwidths up to 300 kHz, and adds less than twice the vacuum noise, verifying its near quantum-limited noise performance. Despite being narrowband, this JPA is ideal for quantum applications requiring precise frequency targeting, such as qubit readout. The nonlinear Josephson inductance creates a Kerr nonlinearity used for degenerate and non-degenerate amplification. This device offers excellent tunability, but its limited dynamic range ($P_{1dB} < -125$ dBm) restricts its use to low-power applications unless integrated with additional amplification stages.

As an interesting work referred in [87] develop a Josephson junction-based Traveling-Wave Parametric Amplifier (TWPA) that employs minimal resonator phase matching using periodically inserted $\lambda/4$ resonators between nonlinear sections. This design achieves an average gain of 12 dB across a 4 GHz bandwidth, with a $P_{1dB}$ near –92 dBm, and quantum-limited noise performance. Unlike standard JPA designs, this TWPA offers directional amplification and does not require high-Q resonators, thereby enabling high dynamic range and scalability. The use of low-loss amorphous silicon capacitors and discrete resonator phase shifters permits efficient phase correction and suppresses gain ripples across the operating band. The architecture is especially advantageous for frequency-multiplexed quantum readout, with significantly lower pump power requirements than NbTiN-based TWPAs and a more compact chip footprint.

To compare our simulated and modeling results, it is initially necessary to analyze the arrayed JJ JPA with quantum theory in detail and theoretically derive all critical parameters.

## Array of JJ as a JPA for further engineering

Basically, JPAs rely on the intrinsic nonlinearity of a JJ to enable parametric amplification. However, using a single JJ poses significant limitations, particularly in terms of gain, $P_{1dB}$, bandwidth, and power handling. A single JJ provides strong nonlinearity, but it also introduces high impedance mismatches and limited tunability, making it challenging to achieve stable, high-gain amplification across a broad frequency range (limited dynamics range). Additionally, a single JJ is highly susceptible to fabrication variations, leading to inconsistent performance [60, 61]. By incorporating an array of JJs, engineers can distribute the nonlinear response across multiple elements, effectively lowering the impedance and improving coupling to external circuitry. The collective behavior of an array provides a more robust amplification mechanism, crucial for applications in quantum computing and ultra-low-noise microwave signal processing [62-66].

One of the most critical challenges in single-JJ amplifiers is the limited dynamic range, as they tend to saturate quickly at higher input power levels. A single JJ exhibits strong gain compression around -125 dBm at $I_{pump}$ = 0.4 uA, meaning that its amplification capability degrades rapidly when the signal power exceeds a threshold, leading to significant nonlinearity and distortion. This issue is particularly problematic in quantum systems, where precise and noise-free signal amplification is necessary. Arrays of JJs distribute the overall gain across multiple junctions, reducing the burden on any single element and thereby increasing $P_{1dB}$ [61-66]. This enables the amplifier to handle stronger signals before reaching saturation. Furthermore, a single JJ is prone to excess phase noise due to its highly nonlinear response [59, 60]. By using an array, the noise contributions of individual junctions are averaged out, leading to improved phase stability and reduced excess noise in the amplification process.

The primary advantage of using an array of JJs in a JPA is the ability to engineer a highly tunable, broadband, and stable amplifier with superior power handling capabilities. The impedance of an individual JJ is typically very high, making impedance matching to external circuits difficult, while it arranged in an array, the effective impedance is significantly reduced, improving coupling efficiency with microwave transmission lines. Additionally, arrays allow for tunable dispersion engineering, enabling designers to optimize the gain-bandwidth trade-off. In applications such as superconducting qubit readout, where high-fidelity signal detection is essential, JJ arrays help maintain gain uniformity and spectral purity. Furthermore, the distributed nonlinearity in an array reduces the impact of local defects or fabrication imperfections, making JPAs with JJ arrays more reproducible and reliable for large-scale quantum computing architectures. The enhanced tunability of the arrayed structure also allows for more efficient gain control via external magnetic flux or bias current modulation [3-10]. Despite their advantages, the use of JJ arrays in JPAs introduces certain trade-offs and challenges. One major drawback is the increased complexity in fabrication and design [64, 66]. Maintaining uniformity across a large array of JJs requires precise control of junction parameters such as critical current, capacitance, and resistance, which can be difficult to achieve with high accuracy. Additionally, while the impedance reduction improves coupling, it can also introduce unwanted parasitic capacitances and resonances, potentially limiting the amplifier's bandwidth. Another challenge is phase coherence across the array; although averaging reduces noise, variations in junction properties can lead to phase dispersion, affecting signal integrity. Furthermore, larger arrays require higher pump power for parametric amplification [13]. Despite these challenges, the advantages of using JJ arrays in JPAs make them an essential component for achieving high-performance, quantum-limited amplification in modern quantum information systems [2-10, 62-66].

### *Theoretical background of array JJ JPA: JPA gain calculation using quantum theory*

Before focusing on the simulation of the arrayed JPA and analyzing the resulting contributions, it is essential to first address the critical question raised at the end of the previous section using quantum mechanics theory. The key questions we aim to explore are: how an array of JJ can improve the linearity of the JPA, and how the array different configurations can enhance the JPA's critical properties. A typical JPA containing array of JJ as a quantum circuit is illustrated in Fig. 12 in which information related to the parasitic capacitance and also the resonator connected the array are given. In the same way, to examine the quantum circuit dynamics and its quantum features one needs to initially derive the Lagrangian of the system expressed as:

$$L_t = \frac{C_g}{2}\sum_{k=0}^{N-1}\dot{\phi}_k^{\,2} + \frac{C_{out}}{2}\dot{\phi}_N^{\,2} + \frac{C}{2}\sum_{k=0}^{N-1}\left(\dot{\phi}_k - \dot{\phi}_{k+1}\right)^2 - \frac{1}{2L_J}\sum_{k=0}^{N-1}\left(\phi_k - \phi_{k+1}\right)^2 \tag{10}$$

where $C_g$, C, N, and $\phi$ are the parasitic capacitors grounded the JJ, the JJ capacitance, the number of JJ in array, and flux as a quantum coordinate operator, respectively.

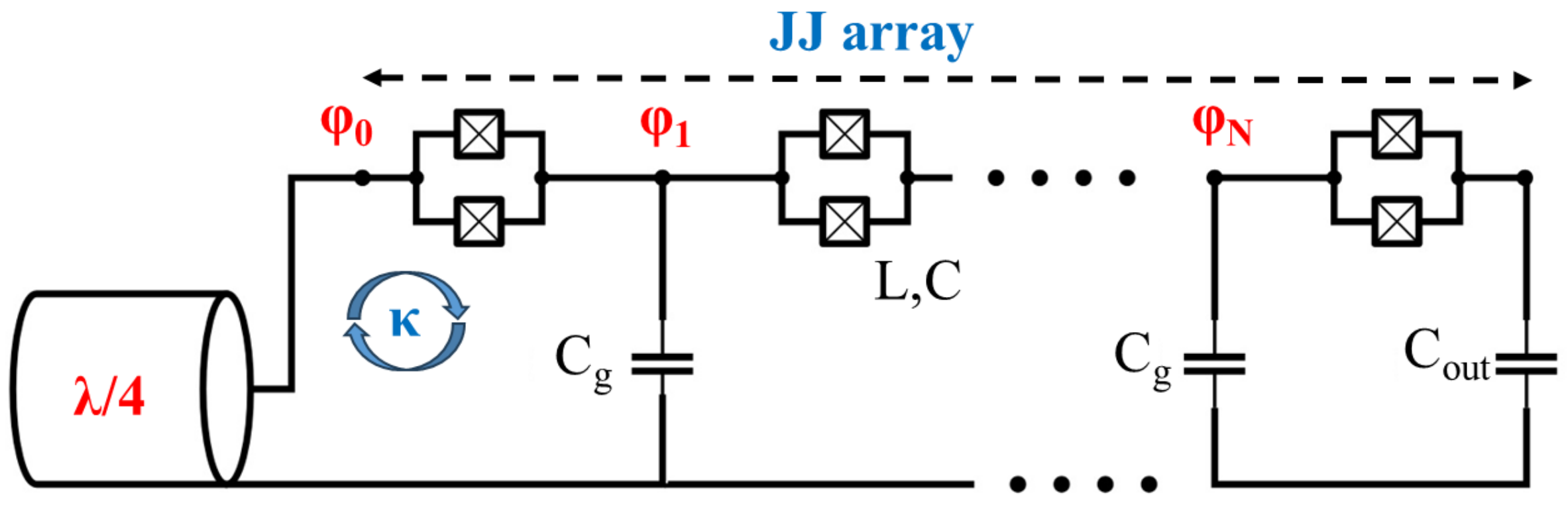

Fig. 12 The schematic of array JJ for JPA engineering: Applying array of JJ to enhance the JPA features.

To simplify the algebra, it is convenient to represent Eq. 1 through N coupled linear equations. Therefore, one can express Eq. 10 as a compact form $L_t = \frac{1}{2}\dot{\phi}^T \hat{C}\dot{\phi} - \frac{1}{2}\phi^T \hat{L}^{-1}\phi$, where $\hat{C}$ and $\hat{L}^{-1}$ are matrixes as:

$$\hat{C} = C\begin{bmatrix} 2 & -1 & 0 & . & . & . & 0 \\ -1 & 2 & -1 & 0 & . & . & 0 \\ 0 & -1 & 2 & -1 & 0 & . & 0 \\ . & 0 & . & . & . & 0 & 0 \\ . & . & 0 & . & . & . & . \\ . & . & . & . & . & . & -1 \\ 0 & 0 & 0 & 0 & 0 & -1 & 1 \end{bmatrix} + C_g\begin{bmatrix} 1 & 0 & 0 & . & . & . & 0 \\ 0 & 1 & 0 & 0 & . & . & 0 \\ 0 & 0 & 1 & 0 & 0 & . & 0 \\ . & 0 & . & . & . & 0 & 0 \\ . & . & 0 & . & . & . & . \\ . & . & . & . & . & . & 0 \\ 0 & 0 & 0 & 0 & 0 & 0 & \frac{C_{out}}{C_g} \end{bmatrix}, \hat{L}^{-1} = L^{-1}\begin{bmatrix} 2 & -1 & 0 & . & . & . & 0 \\ -1 & 2 & -1 & 0 & . & . & 0 \\ 0 & -1 & 2 & -1 & 0 & . & 0 \\ . & 0 & . & . & . & 0 & 0 \\ . & . & 0 & . & . & . & . \\ . & . & . & . & . & . & -1 \\ 0 & 0 & 0 & 0 & 0 & -1 & 1 \end{bmatrix} \tag{11}$$

Using the matrices for capacitors and inductors presented in Eq. 11, one can define matrix $\Omega^2 = \hat{C}^{-1}\hat{L}^{-1}$ [6, 9] to calculate the eigenvalues (dominant frequencies) and eigenvectors ($\Psi_i$ is the wave profile of each mode) of the quantum system under discussion. Thus, the effective capacitance and inductance of the circuits are determined employing the eigenvalues and eigenvectors for each mode as follows: $C_{eff} = \overrightarrow{\Psi_\iota}^T \hat{C}\overrightarrow{\Psi_\iota}$ and $L_{eff}{}^{-1} = \overrightarrow{\Psi_\iota}^T \hat{L}^{-1}\overrightarrow{\Psi_\iota}$. This allows us to map a λ/4 resonator to an equivalent LC circuit [6]. Utilizing the materials mentioned, one can derive the total Hamiltonian of the circuit summarized as:

$$H_t = \omega_{eff}\hat{a}^+\hat{a} - N * E_j\cos\left(\frac{\hat{\varphi}}{N}\right) \simeq \omega_{eff}\hat{a}^+\hat{a} - \frac{E_j}{24N^3}\hat{\varphi}^4 \xrightarrow{\hat{\varphi}=\frac{i}{\sqrt{2}}\left(\frac{8E_c}{E_j}\right)^{0.25}\left(\hat{a}^+-\hat{a}\right)} \omega_{eff}\hat{a}^+\hat{a} - \frac{E_c}{12N}\left(\hat{a}^+ - \hat{a}\right)^4 \tag{12}$$

where a and $a^+$ are the annihilation and creation operators, respectively. In the derivation of the total Hamiltonian, it is supposed that the circuit contains a simple LC oscillator and a nonlinear element. It should be noted that the second term of the Taylor's series of the cosine expansion was absorbed by oscillatory function earlier. Finally, using the quantization for the phase expressed in the equation and applying the small signal fluctuation (some terms such as $a^4$, $a^{+4}$, and so forth have been ignored) the total Hamiltonian can be introduced as:

$$H_t = \omega_{eff} a^+ a - \frac{E_c}{6N} a^{+2} a^2 \tag{13}$$

In this equation, derived nonlinearity terms strongly confined through the number of JJ in the circuit. Using the same approach applied for single JJ JPA, the dynamics equation of motion for the quantum system discussed is examined (relevant quantum Langevin equation) as:

$$\dot{a} = -i\omega_{eff} a - iKa^+ aa - \frac{\kappa}{2} a + \sqrt{\kappa} a_{in} \tag{14}$$

In this equation, K = $E_c$/3N, and "a" determines the intra-cavity signals; to solve the equation, we need to linearize that using a = (α + δa)$\exp^{[-i\omega pt]}$, where α and δa are the DC point and quantum signal fluctuation around the DC point, respectively. Also, $\omega_p$ is the pump frequency. The steady-state solution (replacing a = α and $a_{in} = \alpha_{in}$) and quantum fluctuation dynamics [7, 13] for a coherent pump α = |α|exp(i$\varphi_1$) after some algebra's simplification are introduced as:

$$\begin{gathered} \left(i\left(\omega_{eff} - \omega_p\right) + \frac{\kappa}{2}\right)\alpha + iK\alpha^*\alpha\alpha = \sqrt{\kappa}\alpha_{in} \\ \dot{\delta a} = i\left(\omega_p - \omega_{eff} - 2K\,|\,\alpha\,|^2 + \frac{i\kappa}{2}\right)\delta a - iK\delta a^+ \alpha^2 + \sqrt{\kappa}\delta a_{in} \end{gathered} \tag{15}$$

To calculate the gain of the quantum circuit, it is necessary to handle the quantum fluctuation equation of Eq. 15. However, to do so, it is necessary to find and calculate the DC term of intra-cavity signal (α) from the first part of Eq. 15. To calculate α, one can multiply both side of the steady-state solution with their complex conjugate, by which it is become:

$$\left\{\left(\frac{\omega_{eff} - \omega_p}{\kappa}\right)^2 + \frac{\kappa}{2}\right\}|\,\alpha\,|^2 - \frac{2\left(\omega_{eff} - \omega_p\right)K}{\kappa^2}|\,\alpha\,|^4 + \left(\frac{K}{\kappa}\right)^2|\,\alpha\,|^6 = \frac{1}{\kappa}|\,\alpha_{in}\,|^2 \xrightarrow[\xi \equiv \frac{K}{\kappa}|\alpha_{in}|^2]{n \equiv \frac{|\alpha|^2}{|\alpha_{in}|^2},\ \delta \equiv \frac{\omega_{eff} - \omega_p}{\kappa}} \left(\delta^2 + \frac{1}{4}\right)n^2 - 2\delta\xi n^2 + \xi^2 n^3 - 1 = 0 \tag{16}$$

where δ, n, and ξ are the detuning between pump and effective resonator frequency, average number of pump photons, and relative strength of nonlinearity in the presence of pump field, respectively. The important point here is that the product of drive pump field and JJ nonlinearity determine the total nonlinearity of the quantum system not each quantity itself. That is why the smallest value of K is enough for JPA operation in which we look for a linear amplifier design. We attempt to solve Eq. 16 and analyze the results. Fig. 13a presents the steady-state simulation results, depicting the variation in the intra-cavity average photon number as a function of the relative strength of the nonlinearity. The figure clearly illustrates that an increase in nonlinearity can induce bifurcation [7, 13]. It is crucial to differentiate between K and ζ, where K represents the intrinsic nonlinearity of the JJ, while ζ accounts for the input-driven effect on the system's nonlinearity. This distinction implies that even a weak JJ nonlinearity can be effectively compensated by enhancing the driving power. It is notable to mention that in the design of JPA one has to prefer weak JJ nonlinearity by which the JPA phase noise and $P_{1dB}$ can be enhanced dramatically.

Fig. 13b shows the signal gain of the amplifier as the relative strength of the nonlinearity increases. The increase in signal gain does not fully align with the intra-cavity average number of photons because the signal gain depends on other factors besides *n*. Alongside, increasing ζ leads to change of pump detuning. The subsequent analysis focuses on quantum fluctuations and the gain related to small signals. However, steady state part of Eq. 15 can also be used to calculate the DC point gain (reflection coefficient in a classical sense) $\alpha_{out}/\alpha_{in}$ [7].

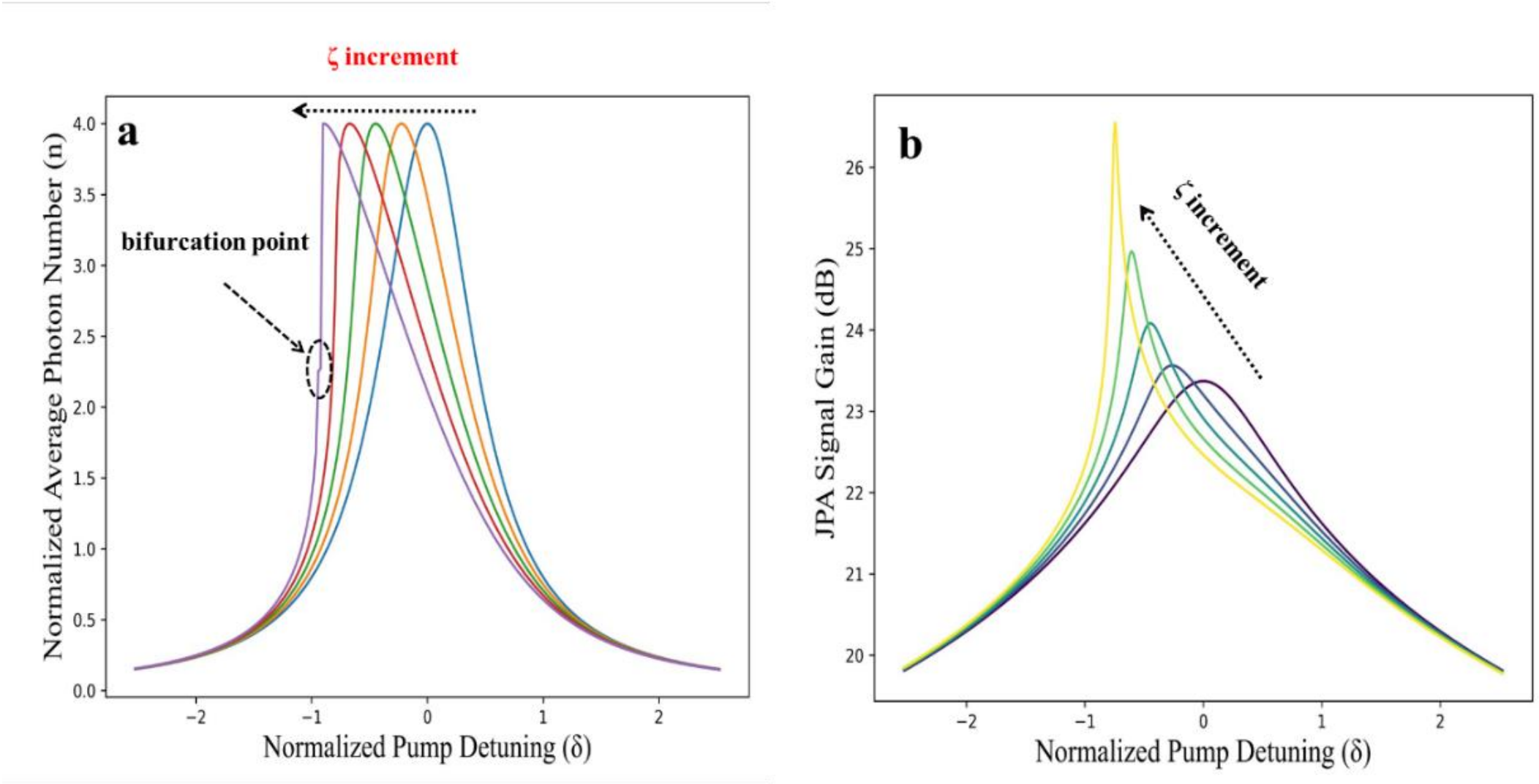

Fig. 13 (a) the effect of the relative strength of the nonlinearity on the normalized average photon number (n) vs normalized pump detuning, (b) JPA signal gain vs normalized pump detuning with $\zeta = \zeta = \zeta_0$ *{0.01, 0.1, 0.4, 0.8, 0.95}, $\zeta_0$=-1/√27 [7].

Finally, to calculate the gain of the quantum signals, the second part of Eq. 15 is transformed into Fourier domain expressed as:

$$-i\left(\omega_s-\omega_p\right)\delta a = i\left[\frac{\left(\omega_p-\omega_{eff}\right)}{\kappa}+i\frac{1}{2}-2\frac{K}{\kappa}|\alpha|^2\right]\delta a - i\frac{K}{\kappa}\delta a^{+}\alpha^2+\frac{\sqrt{\kappa}}{\kappa}\delta a_{in} \xrightarrow{\Delta=\frac{\omega_s-\omega_p}{\kappa}} 0 = i\left[\delta+\Delta+\frac{i}{2}-2\zeta n\right]\delta a - i\zeta n e^{(2i\varphi)}\delta a^{+} + \delta\hat{\alpha}_{in} \tag{17}$$

where Δ is the signal detuning. The last step to calculate the gain is using the conjugate of the Eq. 17 and establish the relating scattering matrix, then employing the input-output formula, the gain of the signal and idler [7, 13, 20] become calculated as follows:

$$\begin{cases}\left[i\left(-\delta-\Delta+2\zeta n\right)+\frac{1}{2}\right]\delta a_{\omega}+i\zeta n e^{(2i\varphi)}\delta a_{\omega}^{+}=\delta\hat{\alpha}_{in\omega}\\ \left[i\left(\delta-\Delta-2\zeta n\right)+\frac{1}{2}\right]\delta a_{\omega}^{+}-i\zeta n e^{(-2i\varphi)}\delta a_{\omega}=\delta\hat{\alpha}_{in\omega}^{+}\end{cases} \longrightarrow \begin{bmatrix} i\left(-\delta-\Delta+2\zeta n\right)+0.5 & i\zeta n e^{(2i\varphi)} \\ -i\zeta n e^{(-2i\varphi)} & i\left(\delta-\Delta-2\zeta n\right)+0.5\end{bmatrix}\begin{bmatrix}\delta a_{\omega}\\ \delta a_{\omega}^{+}\end{bmatrix}=\begin{bmatrix}\delta\hat{\alpha}_{in\omega}\\ \delta\hat{\alpha}_{in\omega}^{+}\end{bmatrix} \tag{18}$$

Finally using a few algebras, the signal gain of the array JJ JPA is calculated as:

$$G_{array_JJ_JPA}=\frac{i\sqrt{\kappa}\left(\delta-\Delta-2\zeta n\right)+0.5}{\left[-\left\{\left(\delta-\Delta-2\zeta n\right)+0.5\right\}\left\{\left(-\delta-\Delta+2\zeta n\right)+0.5\right\}-\zeta^2 n^2\right]}-1 \tag{19}$$

It is clear from the gain relationship in Eq. 19, that it is affected dramatically by some quantities such as signal and pump detuning, the relative strength of the nonlinearity, and also input coupling rate. In the next step, the study focuses on modeling and simulating array-based JJ JPAs with various structural configurations and analyzing the corresponding gain, as theoretically derived in Eq. 19, in detail.

### *Array-based JJ JPAs: Degenerate JPA simulation results*

The schematic depicted in Fig. 14 presents the structure of a degenerate JPA composed of an array of 2048 JJs. The primary function of such a structure is to provide quantum-limited amplification while maintaining phase-sensitive characteristics. The design integrates a resonator to enhance the amplification process and manage impedance matching. The schematic also highlights parasitic capacitances, which are crucial to understanding the performance and stability of the amplifier. Parasitic capacitances arise naturally due to the distributed nature of the JJ array, and their impact must be carefully analyzed to avoid unwanted resonances or gain reductions.

From the schematic, key design parameters such as junction critical current, resonator coupling, and the placement of impedance-matching networks can be inferred. The integration of 2048 JJs suggests a strategy for achieving high gain while preserving the amplifier's bandwidth and high linearity. The number of JJs in the array plays a crucial role in defining the amplification properties, including the gain profile, bandwidth, and nonlinearity effects. Additionally, the resonator's role in shaping the signal path and enhancing selectivity is critical. Depending on its quality factor and coupling strength, the resonator can significantly influence the overall efficiency of the amplifier, making it essential to optimize its design.

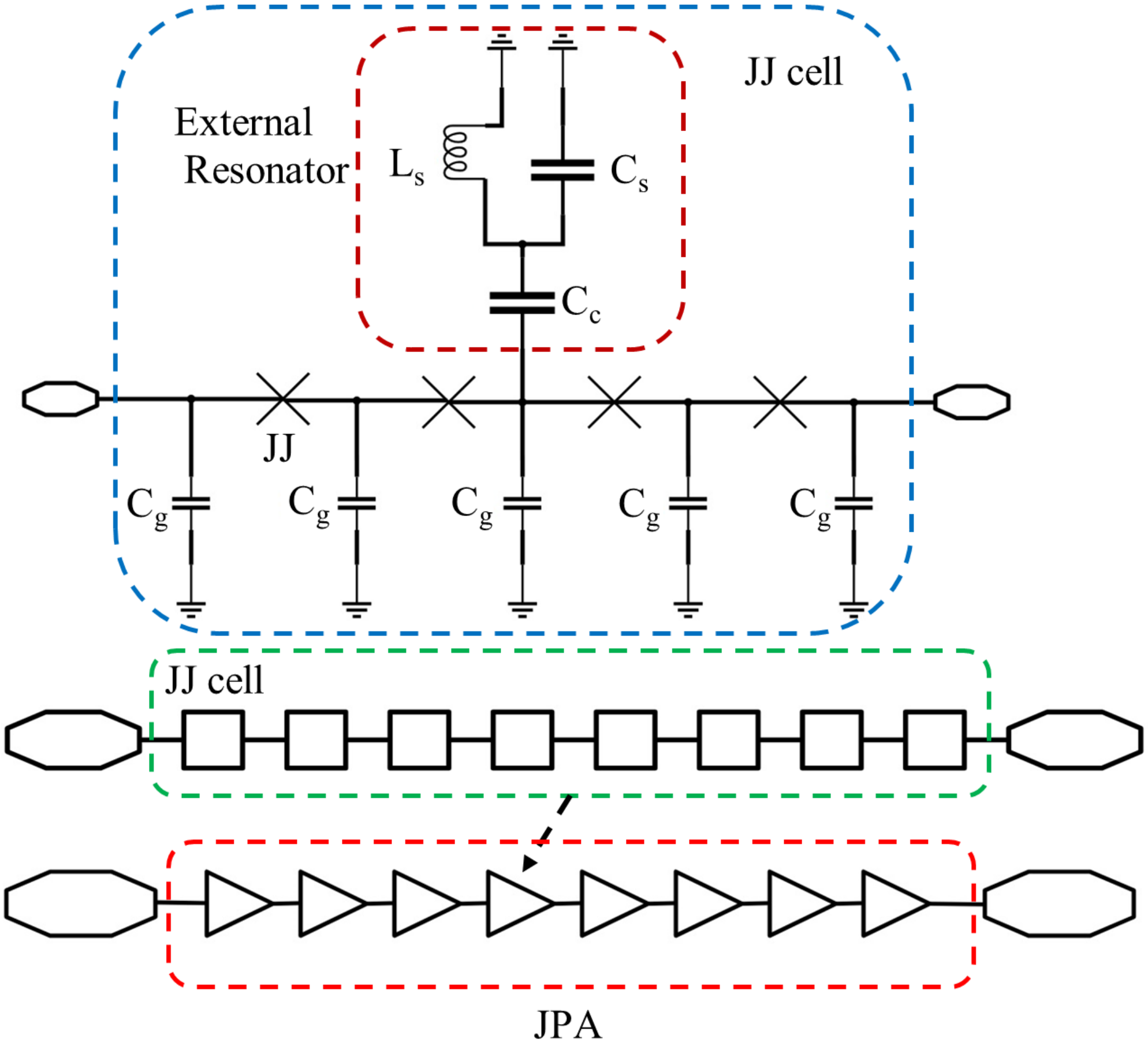

Fig. 14. Schematic of the degenerate JPA using 2048 JJ; information related to the parasitic capacitance and also the resonator connected the array are given.

In the following, we attempt to discuss about the simulation results. The S-parameters illustrated in Fig. 15 provide insights into the reflection and transmission properties of the JPA. At a pump frequency of 7.12 GHz, the S-parameters illustrate how efficiently the device transmits and reflects signals, allowing for an assessment of suitable technical features. A well-designed JPA should exhibit strong transmission at the operational frequency while minimizing unwanted reflections that could degrade performance.

The key parameter in this analysis is $S_{21}$, which represents forward transmission. A significant decreasing in $S_{21}$ around 7.24 GHz would indicate that the most of the power is absorbed by the resonator that has been used in JPA circuit. Around the resonator frequency ~7.24 GHz, the system exhibits strong resonance behavior, which leads to high strong reflection depending on the operational mode. The minimization of $S_{21}$ (i.e., strong dip in transmission) occurs because the resonator effectively absorbs and reflects most of the input power instead of allowing it to transmit. Additionally, $S_{11}$, which denotes reflection at the input port, should be generally minimized to ensure that most of the input signal is utilized effectively. The frequency-dependent behavior of these parameters also helps in understanding the bandwidth limitations of the amplifier. In JPAs, bandwidth is often a trade-off against gain, meaning that a high-gain amplifier typically has a narrower operational bandwidth. The number of JJs in the array influences these parameters, as larger arrays tend to exhibit stronger linearities and modified impedance characteristics.

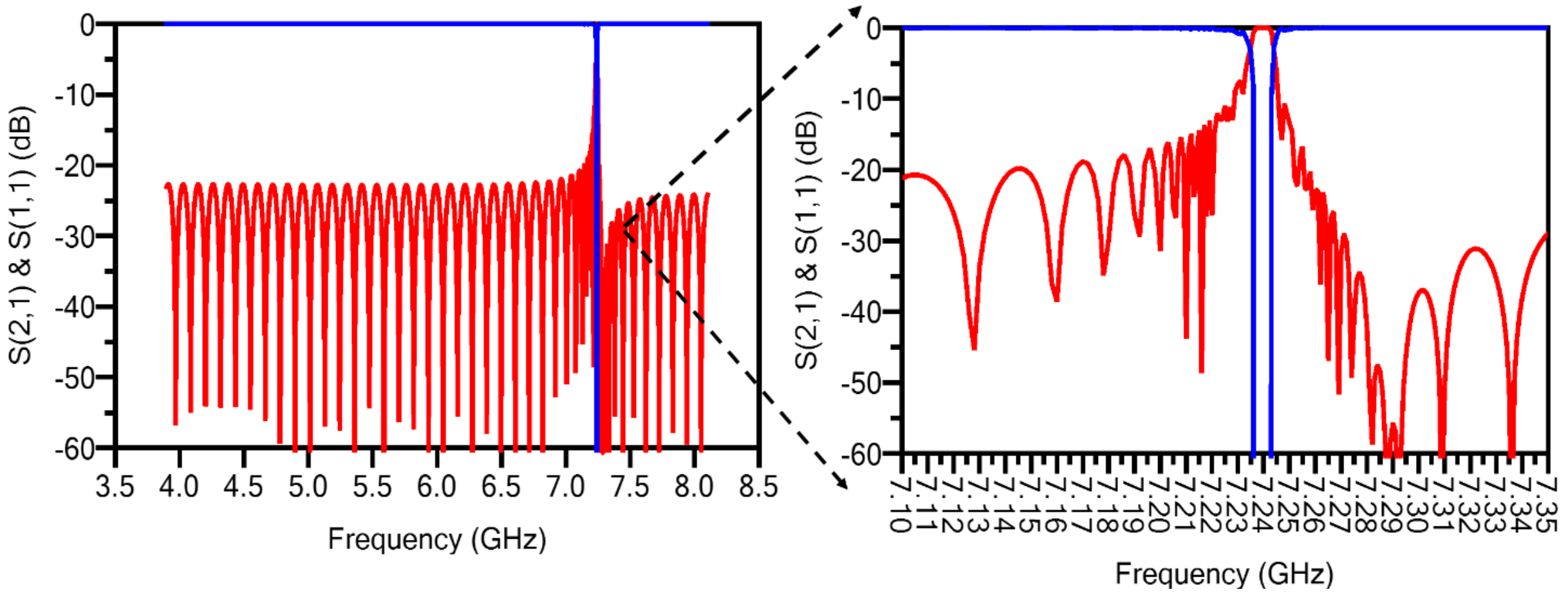

Fig. 15. Scattering Parameters of 2048 JJ as a degenerate JPA, $F_{pump}$ = 7.12 GHz.

The spectrum displayed in Fig. 16 provides critical insights into the signal amplification process for an array JPA with 8×256 JJ; the pump current in this simulation is around 3.68 uA. At an input signal power of -150 dBm, the JPA is expected to operate in its quantum-limited regime, where noise contributions are minimized. The spectrum likely displays a primary amplified signal at the pump frequency and its corresponding sidebands, which result from the parametric mixing process inherent in JPAs. A key characteristic to observe in this spectrum is the presence of gain peaks at specific frequencies. These peaks should align with the expected amplification bands determined by the pump frequency and the device's nonlinear properties. Additionally, the presence of unwanted harmonics or excess noise may indicate the onset of saturation effects or undesired nonlinear interactions. Another crucial aspect is the spectral purity of the amplified signal, which impacts the JPA's usability in quantum applications. A clean spectrum (with no any extra harmonics) with minimal spurious tones ensures that the amplifier can be reliably integrated into quantum measurement setups without introducing excessive noise or distortions.

Herein the design of the array JPA, one of the crucial points is to study the effect of the number of the JJ on the JPA features. In line with, Fig. 17 provides a comparative analysis of the gain profiles for JPAs constructed with varying numbers of JJ arrays including Light Blue (LBlue): LBlue: 5×256 JJ, Pink: 6×256 JJ, Blue: 7×256 JJ, Red: 8×256 JJ. The number of JJs in an array influences the device's gain directly, as larger arrays tend to suppress nonlinear effects, leading to higher parametric amplification. Nonetheless, increasing the number of junctions also introduces additional complexity, such as modified impedance characteristics and increased sensitivity to fabrication variations. The gain profile in this figure likely illustrates how amplification efficiency scales with the number of JJ arrays. For lower numbers of arrays, the gain is moderate, while for higher numbers, the gain may peak at a higher level but possibly at the cost of bandwidth reduction. Understanding this trade-off is essential for optimizing JPA performance in specific applications.

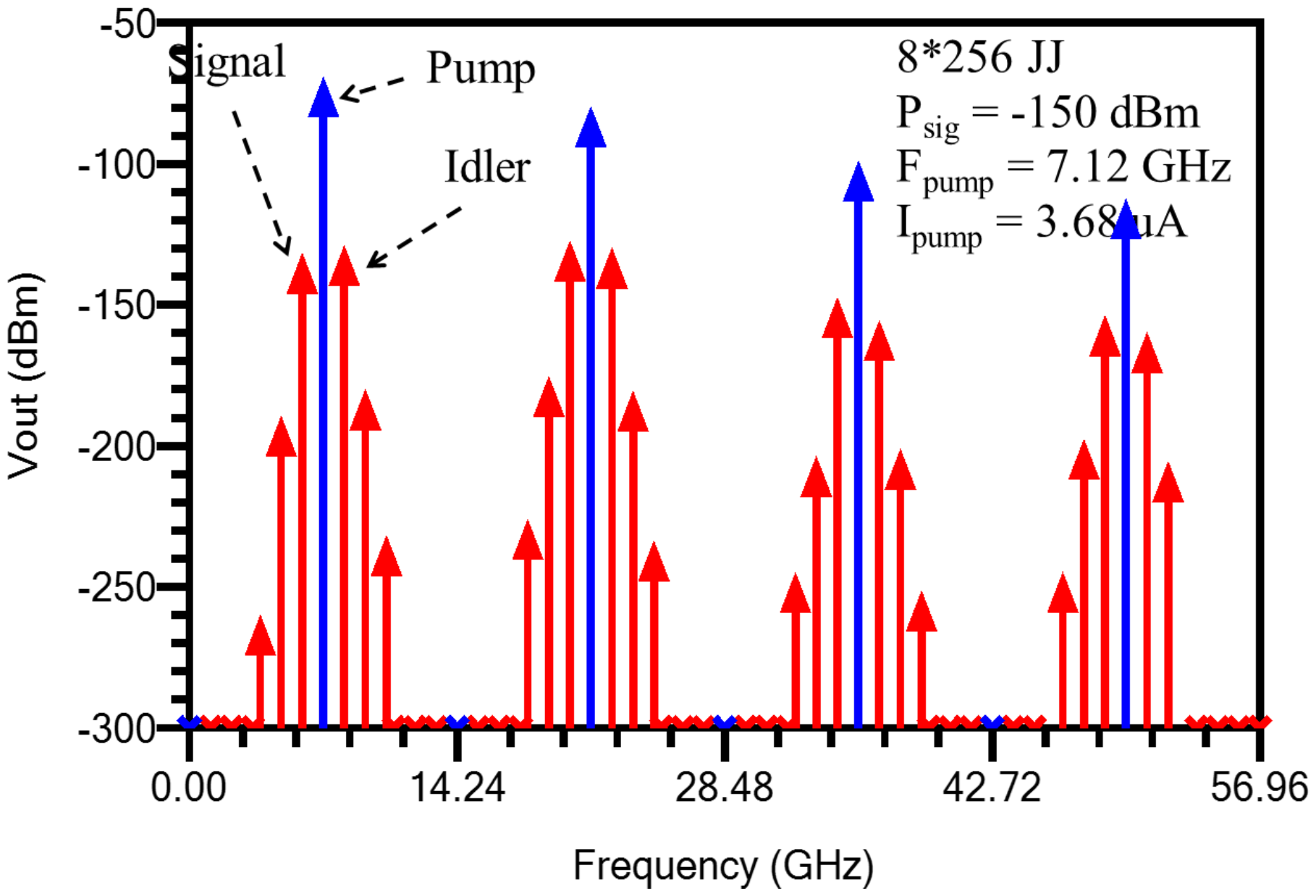

Fig. 16 Spectrum of degenerate JPA; $P_{sig}$ = -150 dBm, $F_{pump}$ = 7.12 GHz
Red: 8×256 JJ.

The other important factor in the JPA design is the $P_{1dB}$, a crucial metric that determines the power level at which the gain drops by 1 dB due to nonlinear saturation. For a JPA with an 8×256 JJ configuration illustrated in Fig. 18, $P_{1dB}$ is analyzed for different pump currents, which affect the amplifier's gain and saturation properties substantially. As $I_{pump}$ increases, the parametric gain of the amplifier also rises due to stronger nonlinear interactions within the Josephson junction array. However, this gain enhancement comes at the cost of reduced linearity, meaning that the amplifier reaches saturation at lower input signal powers. When $I_{pump}$ is small, the JPA operates in a near-linear regime where the gain remains stable over a wider range of input powers. But as $I_{pump}$ increases, the amplifier becomes more susceptible to nonlinear saturation effects, reducing $P_{1dB}$. This occurs because the increased pump strength enhances the device's sensitivity to incoming signals, leading to earlier onset of gain compression.

Another critical factor is the trade-off between high gain and dynamic range. While a higher $I_{pump}$ can boost gain significantly, it also leads to a sharper gain roll-off beyond a certain input power, effectively lowering $P_{1dB}$. This suggests that there exists an optimal $I_{pump}$ where the amplifier achieves maximum gain without excessive compression, ensuring sufficient headroom for handling stronger signals. This balance is crucial for quantum applications, where both high gain and low noise are required. Thus, carefully selecting $I_{pump}$ is essential to maintain a high $P_{1dB}$ while preventing unwanted nonlinear effects, ensuring the JPA operates efficiently within its intended signal range.

In the next step, Non-degenerate JPA as a versatile JPA will be designed simply through engineering the circuit and find that the JPA characteristics can be changed completely.

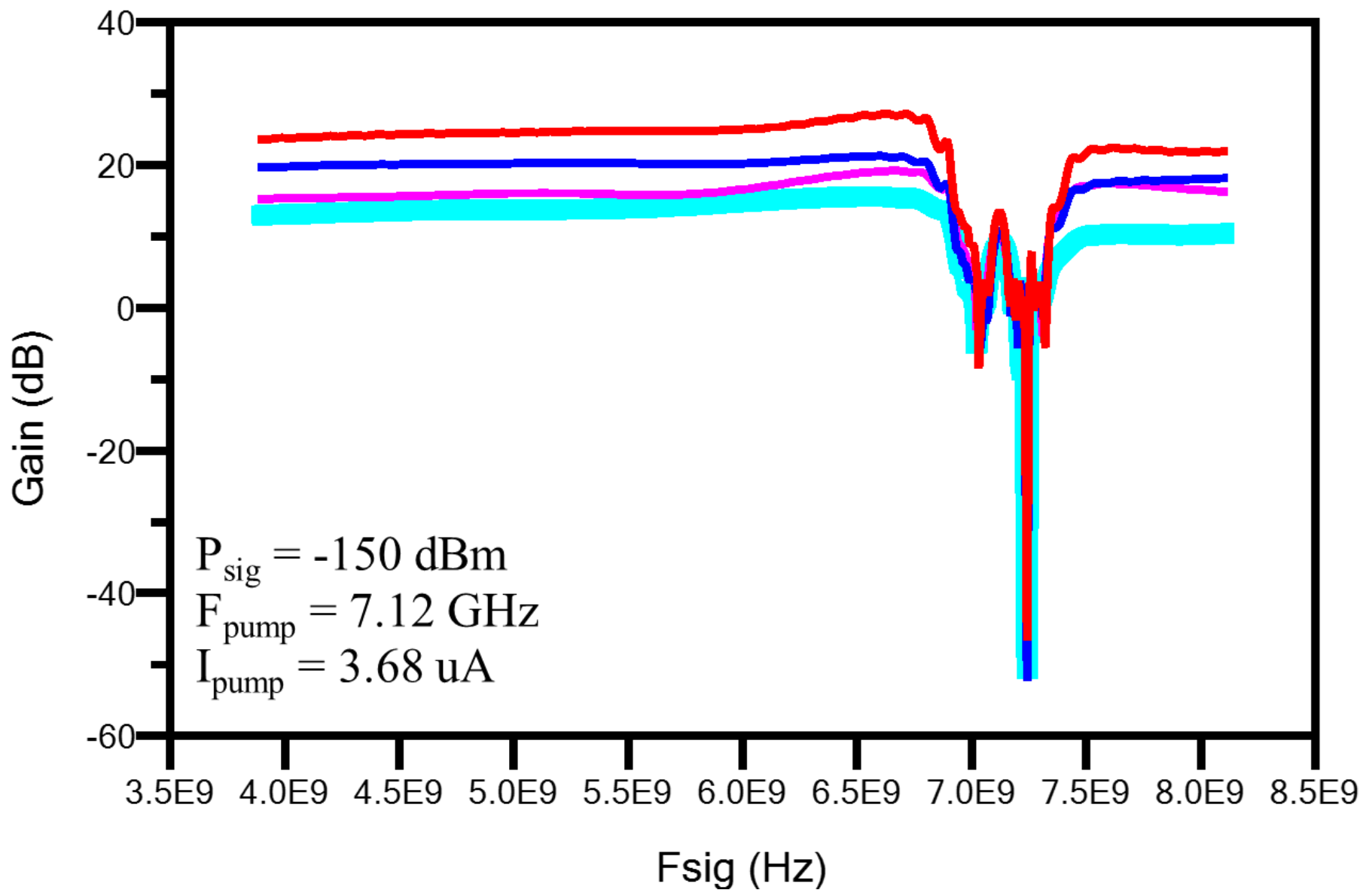

Fig. 17 Gain of degenerate JPA with different number of arrays, LBlue: 5×256 JJ, Pink: 6×256 JJ, Blue: 7× 256 JJ, and Red: 8× 256 JJ.

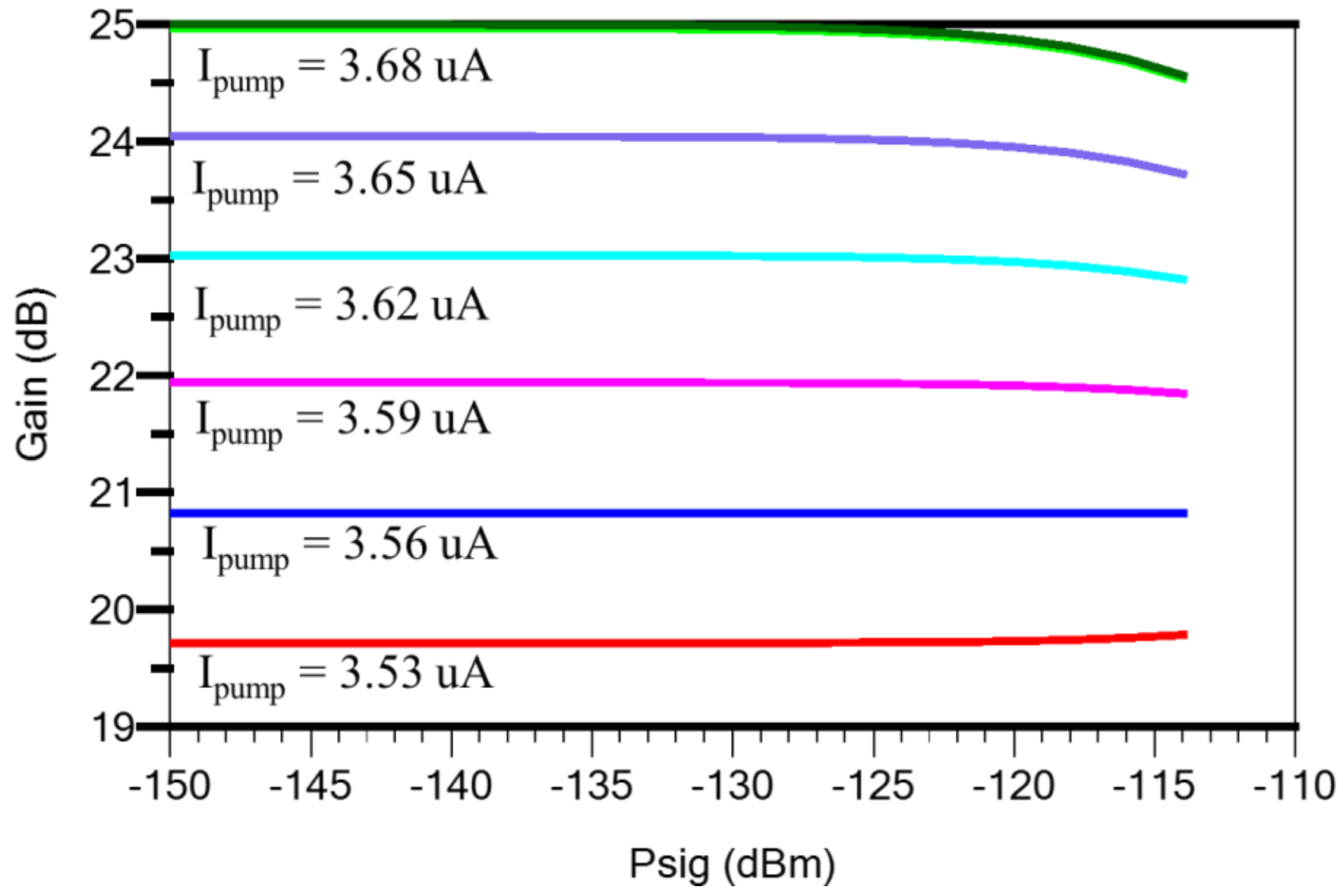

Fig. 18 $P_{1dB}$ of degenerate JPA with 8×256 JJ for different pump current, $F_{sig}$ = 6.0 GHz, $F_{pump}$ = 7.12 GHz.

*Array-based JJ JPAs: Non-degenerate JPA simulation results*

The schematic shown in Fig. 19 presents the non-degenerate JPA, which, similar to its degenerate counterpart, consists of an array of 2048 JJs arranged to provide parametric amplification. Still, a key distinction in this architecture is the inclusion of an interstage capacitance ($C_c$: combination of $C_1$, $C_2$, and $C_3$), which introduces non-degeneracy into the system. The role of $C_c$ is crucial as it modifies the phase-matching conditions and alters the parametric interactions within the amplifier. From a circuit design perspective, the interstage capacitance introduces additional degrees of freedom in tailoring the amplifier's response, enabling better impedance matching and gain control. The schematic also highlights the presence of parasitic capacitances and a coupled resonator, which are integral to defining the bandwidth and stability of the JPA. Compared to the degenerate JPA, all of the important parameters relating to a JPA such as gain, bandwidth, and also the nonlinearity will be discussed.

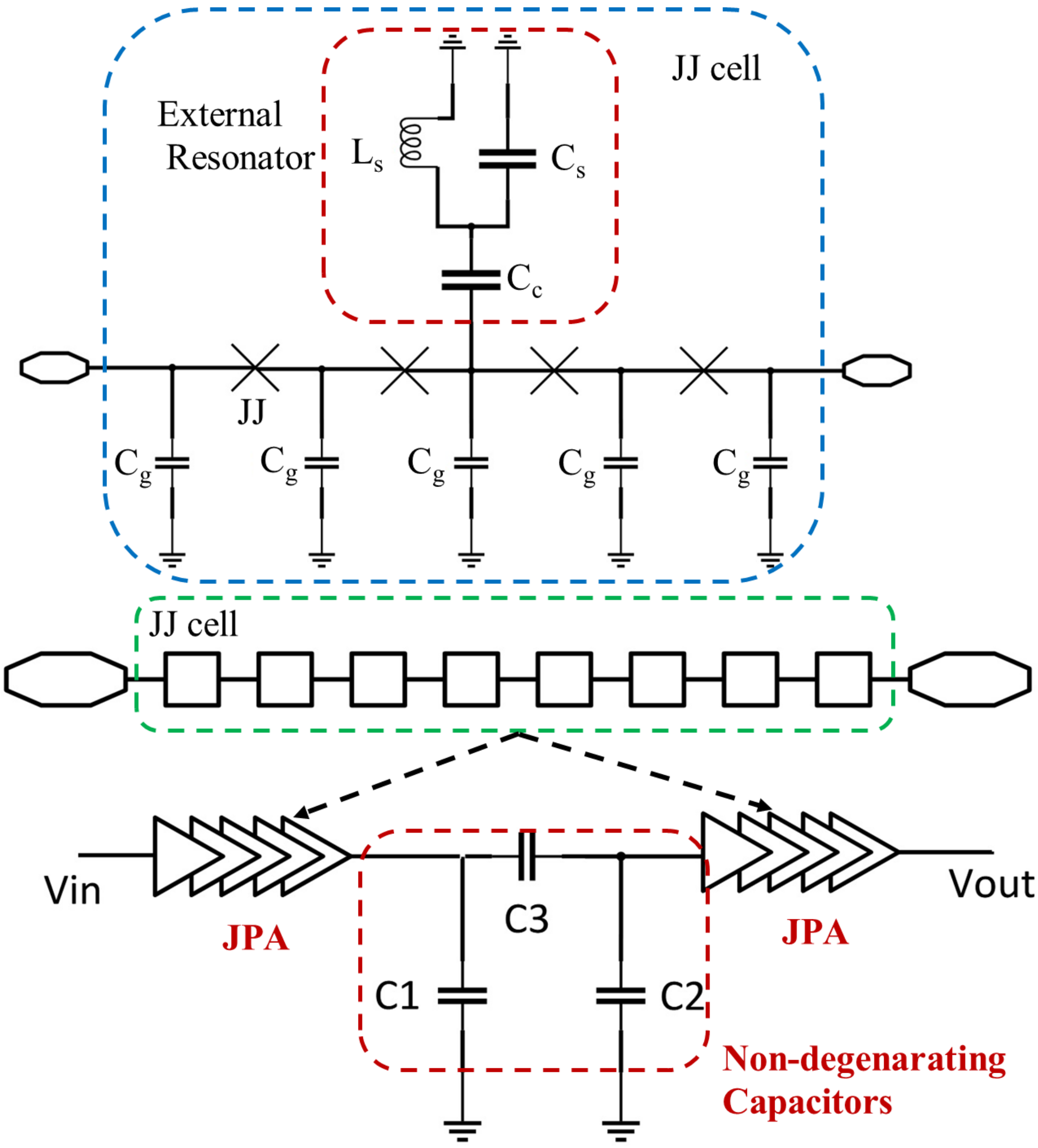

Fig. 19 Schematic of the degenerate JPA using 2048 JJ; $C_1$, $C_2$, and $C_3$ herein are the interstage capacitance that generate nondegeneracy in the circuit. Also, information related to the parasitic capacitance and also the resonator connected the array are given.

The scattering parameters illustrating the transmission and reflection properties of the non-degenerate JPA with an interstage capacitance $C_3$ = 8 pF is shown in Fig. 20. In comparison to the degenerate JPA, where gain profiles were strictly tied to phase-sensitive parametric amplification, the non-degenerate design exhibits a broader and more stable gain response due to the phase-insensitive nature introduced by $C_c$.

The parameter $S_{21}$ (forward transmission) is a critical metric in assessing amplifier gain. A well-defined peak in $S_{21}$ suggests strong amplification, while $S_{11}$ (input reflection coefficient) should ideally be minimized to ensure efficient signal coupling. The introduction of $C_c$ in the non-degenerate design leads to introduce the nondegeneracy which is shown in the zoom out figure depicted. Hence, one can arrange the frequency of the pump and also signal and idler based on regarding the degeneracy shown.

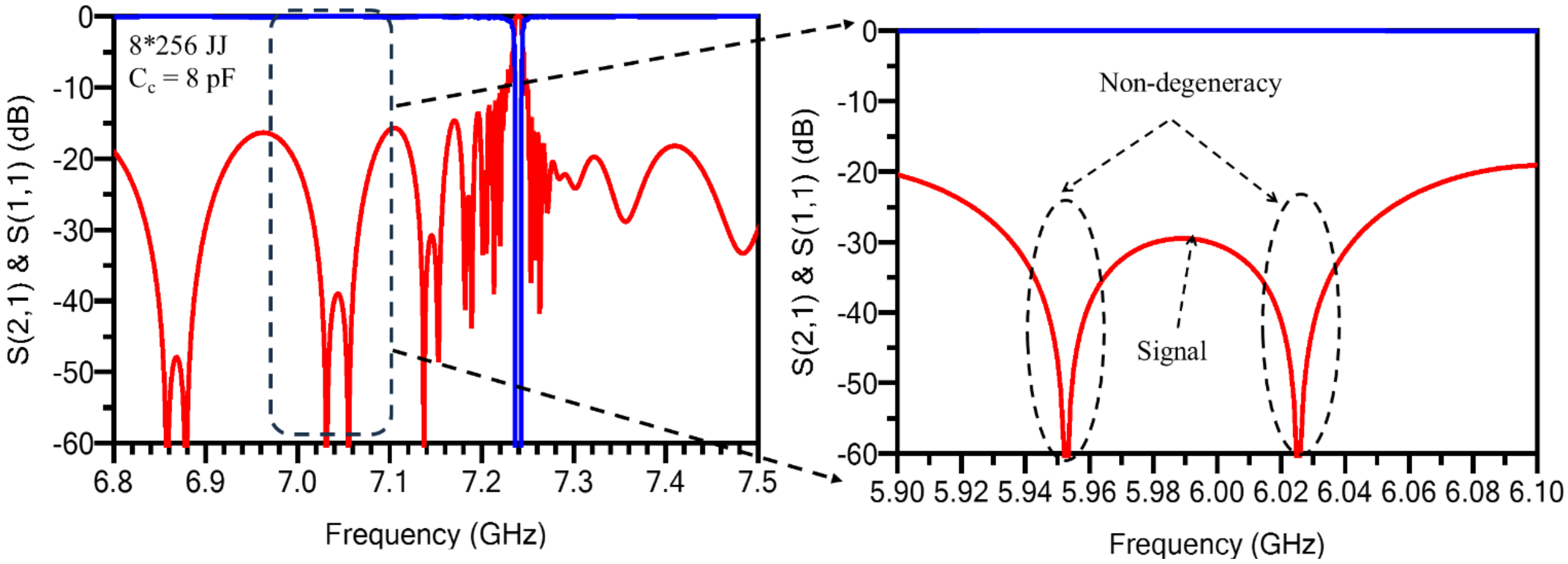

Fig. 20 Scattering Parameters of 8×256 JJ, $F_{pump}$ = 7.12 GHz, with $C_3$ = 8pF.

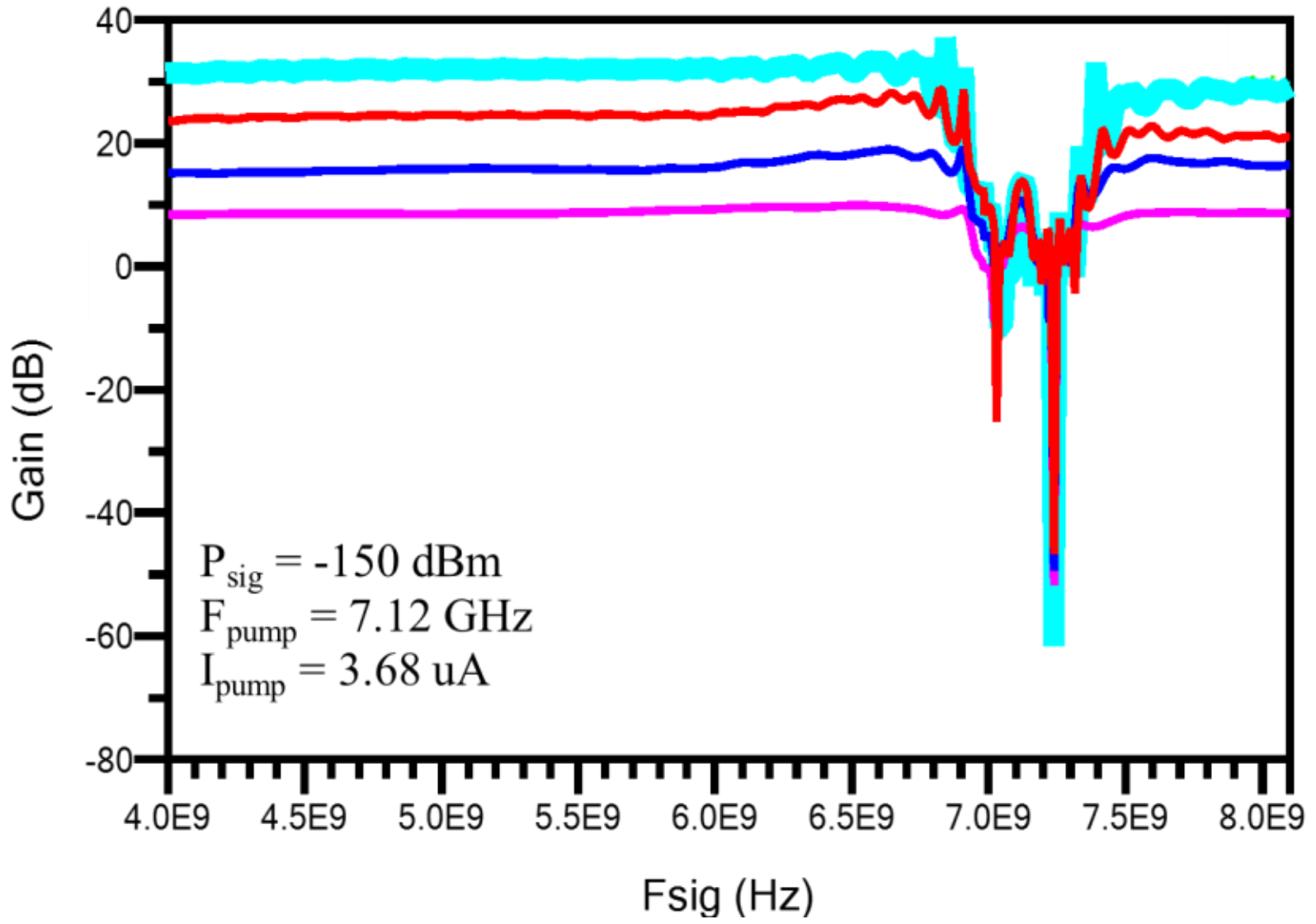

Fig. 21 Gain of JPA with different number of arrays, with $C_3$ = 8pF, Pink: 6×256 JJ, Blue: 7×256 JJ, Red: 8×256 JJ, LBlue: 9×256 JJ.

In Fig. 21, the gain performance of the non-degenerate JPA for different numbers of JJ arrays is presented demonstrating how the amplifier's gain scales with increased array's JJ number. Unlike the degenerate JPA, where gain scaling is primarily dictated by the nonlinear parametric response, the non-degenerate JPA benefits from an additional degree of freedom introduced by $C_c$, allowing for more controlled and tunable gain characteristics. One key observation in this comparison is that while both degenerate and non-degenerate JPAs exhibit increased gain with larger arrays, the non-degenerate configuration tends to show a more stable gain curve. This stability may arise due to the suppression of phase-sensitive fluctuations that can occur in degenerate parametric amplification. Furthermore, the gain-bandwidth product in the non-degenerate JPA is generally superior, meaning that it can sustain relatively high gain without sacrificing bandwidth as significantly as the degenerate counterpart. The other important parameter is the $P_{1dB}$.

Fig. 22 examines the $P_{1dB}$ for a JPA with 8×256 JJs, analyzing how it varies with different pump currents. In degenerate JPAs, $P_{1dB}$ tends to decrease more rapidly with increasing $I_{pump}$ due to stronger nonlinearities and gain saturation. In contrast, the non-degenerate JPA benefits from a more gradual compression onset, allowing for a higher dynamic range. This suggests that for applications requiring strong signal handling capabilities, the non-degenerate JPA is a better candidate, as it maintains higher $P_{1dB}$ values over a wider range of pump currents; this point becomes clearer in the next figure in which the effect of the $C_c$ is studied on the 1 dB compression point. Additionally, excessive pumping in degenerate JPAs can introduce instability, whereas in non-degenerate JPAs, the added capacitance moderates these effects, leading to smoother compression characteristics.

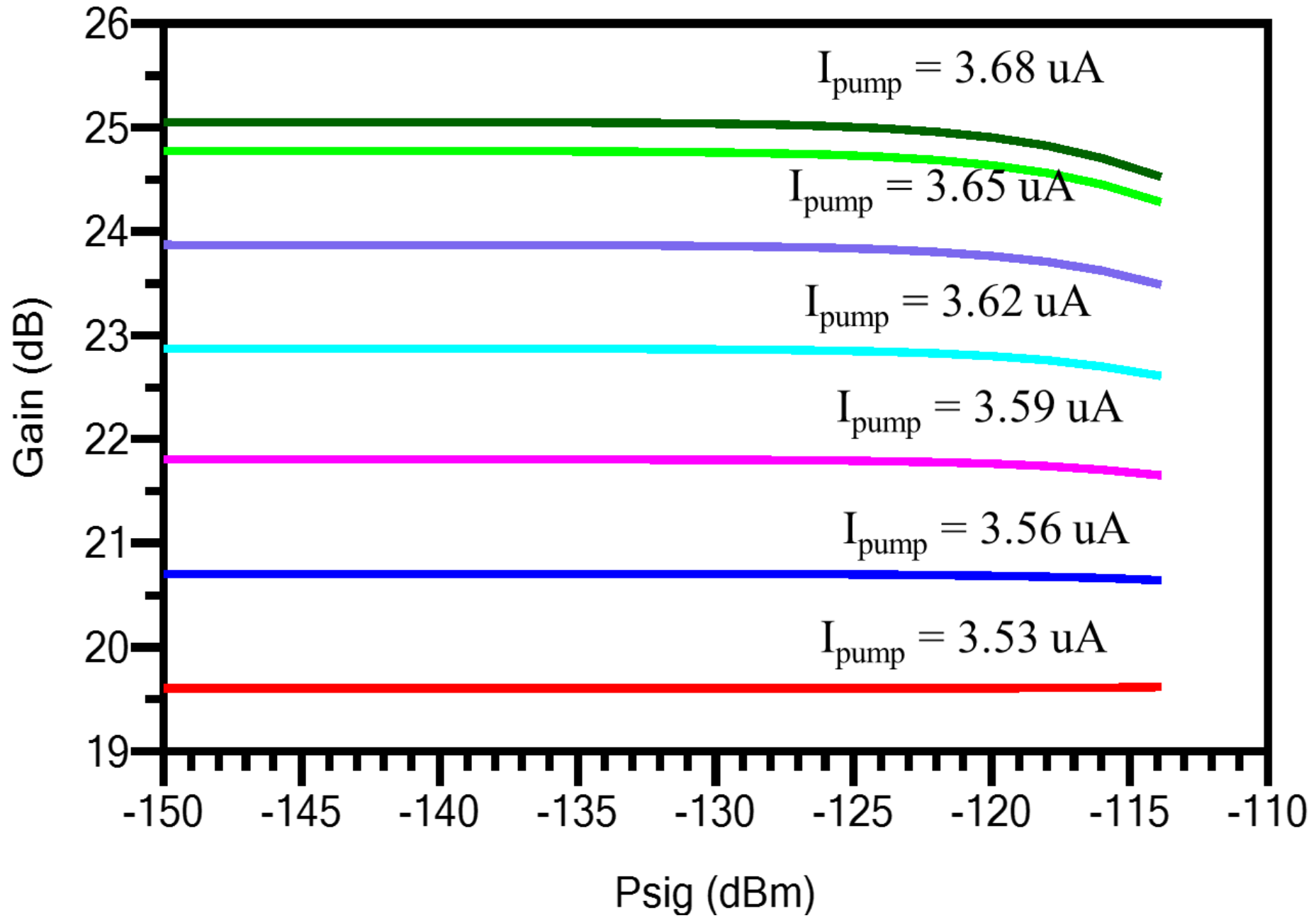

Fig. 22 $P_{1dB}$ of a non-degenerate JPA with 8×256 JJ for different $I_{pump}$, with $C_3$ = 8pF, $F_{sig}$ = 6.0 GHz, $F_{pump}$ = 7.12 GHz, $I_{pump}$ = 3.68 uA.

Finally, Fig. 23 analyzes how $P_{1dB}$ varies with different values of $C_c$, highlighting its impact on amplifier linearity and saturation performance. Increasing $C_3$ generally enhances phase-insensitive amplification, leading to improved compression point performance. Unlike the degenerate JPA, where $P_{1dB}$ is tightly constrained by intrinsic nonlinearities, the presence of $C_c$ in the non-degenerate design enables greater control over the gain and bandwidth. The graphs in the figure imply that increasing $C_c$ can be advantageous not only for achieving higher $P_{1dB}$ values, but also for higher gain. However, excessively large $C_c$ values can also introduce bandwidth trade-offs, meaning that an optimal value must be determined for a given application. Compared to the degenerate JPA, the non-degenerate version with a properly tuned $C_c$ provides a superior balance between high gain, broad bandwidth, and improved linearity, making it more suitable for real-world quantum applications where robustness and stability are critical.

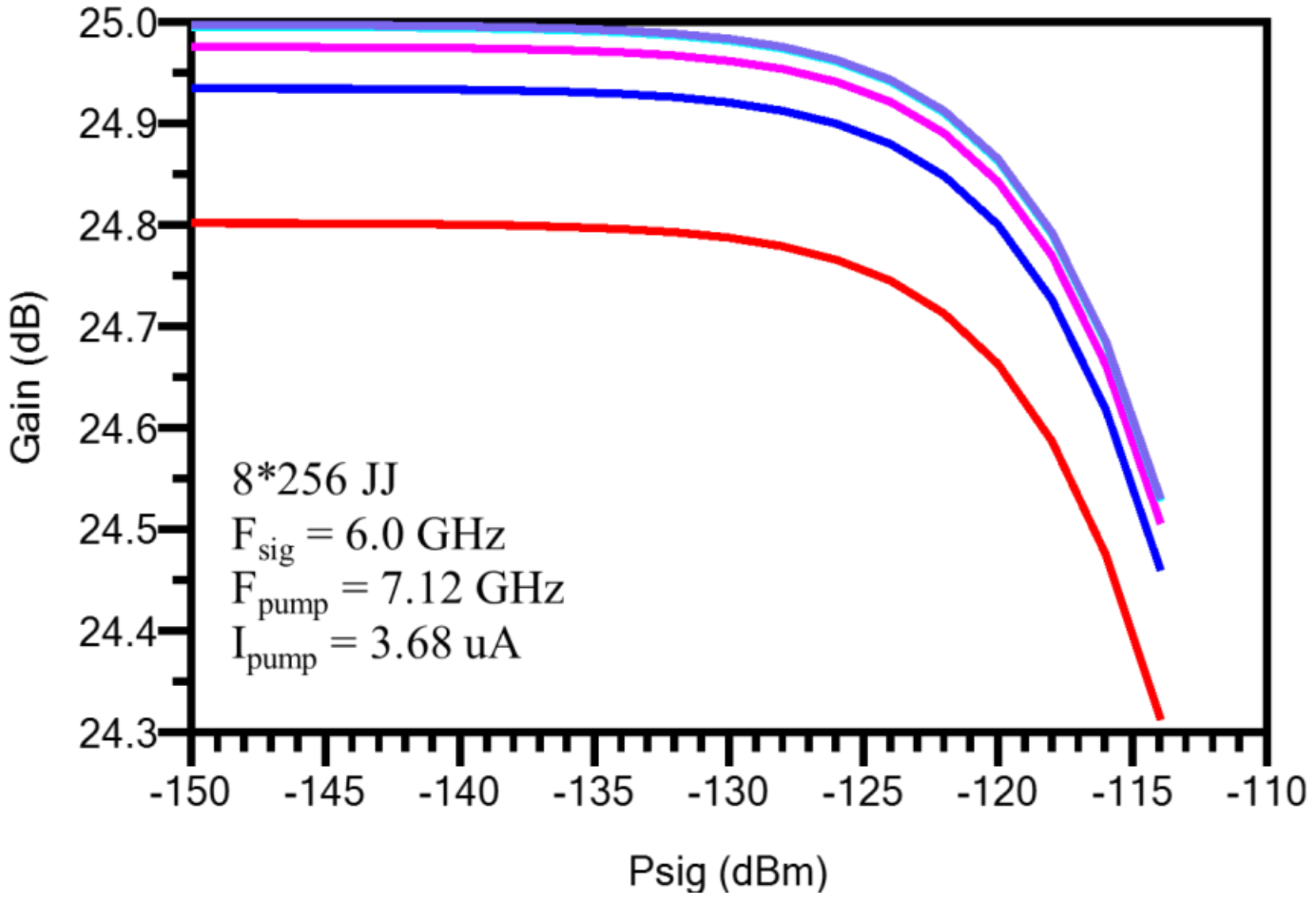

Fig. 23 $P_{1dB}$ of a non-degenerate JPA with 8×256 JJ for different interstage capacitance $C_3$, Red: $C_3$ = 8 pF, Blue: $C_3$ = 16 pF, Pink: $C_3$ = 24 pF, LBlue: $C_3$ = 32 pF.

Up to this point, we consider three different structures for JPA including single JJ JPA, degenerate array JPA, and Non-degenerate array JPA. In a JPA, discussed in theoretical section, the Kerr nonlinearity arises due to the Josephson junction's inductance, which introduces a quartic ($\phi^4$) term in the Hamiltonian. This nonlinearity causes effects like Stark shifts, which limit the amplifier's dynamic range. One intuitive approach to mitigate this issue is **"arraying"**—replacing a single junction with an array of N junctions in series, effectively distributing the nonlinearity across multiple elements.

Although different array configurations [16] derive scaling laws for the Kerr effect, in our presented modeling, an array of N junctions replaces a single junction, each junction carries only a fraction (1/N) of the total superconducting phase. This suggests that the Kerr nonlinearity should decrease by a factor of **$1/N^2$**. However, this assumes that each junction's Josephson energy scales accordingly, which is challenging in practical fabrication. Nonetheless, a more practical scenario where junctions have a fixed Josephson energy ($E_J$), increasing N requires a simultaneous change in capacitance to maintain the same resonance frequency [16]. This leads to a dilution of the Kerr effect, but only as **1/M**, not $1/M^2$.

The key conclusion is that while increasing M can reduce the Kerr nonlinearity, the scaling is weak and does not strictly follow the $1/M^2$ dependence expected in an idealized model. Instead, due to practical constraints like stray inductances, mode impedance variations, and capacitance adjustments, the effective Kerr suppression follows a much slower decay, often just 1/M. This explains why simply increasing the number of junctions does not dramatically improve the amplifier's dynamic range. This insight is significant because it challenges the assumption that making larger arrays of Josephson junctions will always lead to better performance in parametric amplifiers. Instead, careful engineering is required to balance Kerr suppression while maintaining sufficient tunability and gain for practical applications. To study the point mentioned, in the following a new version of the JPA called Blochnium based JPA (BJPA) is introduced.

*New Architecture for JPA:* Blochnium based JPA

The Blochnium-based Josephson Parametric Amplifier (BJPA) marks a major advancement in superconducting quantum circuits, utilizing the distinctive properties of Blochnium to significantly improve signal amplification and operational stability [13]. Blochnium is a recently introduced superconducting element known for its exceptional coherence and strong resistance to quasiparticle poisoning, making it particularly suitable for quantum technologies [10]. Conventional JPAs are often constrained by limited power handling and elevated noise levels due to imperfections in the Josephson nonlinearity. By integrating Blochnium, the BJPA design aims to overcome these challenges, delivering enhanced gain, improved linearity, and lower noise performance at cryogenic temperatures.

The configuration shown in Fig. 24 features an array of N Quarton structures [10–13], where each consists of M secondary SQUIDs with identical Josephson energy $E_{Js}$, and one primary SQUID with Josephson energy $E_{Jm}$. An important parasitic element impacting the system's performance is the capacitance $C_g$, which is difficult to control in the depicted quantum setup. This capacitance influences the effective impedance of the quantum circuit, which must be matched to the intrinsic impedance $Z_0$ of a $\lambda/4$ resonator for optimal performance. It is assumed that the phase drop $\varphi$ from the flux node is uniformly distributed across the array, ensuring that each SQUID experiences the same phase variation [6–7, 13]. The flux variable $\phi$ represents the magnetic flux threading each SQUID loop and its relationship to the externally applied flux $\phi_{ext}$ introduces tunability, typically expressed through terms like $\cos(2\pi(\phi-\phi_{ext})/\Phi_0)$, where $\Phi_0$ is the magnetic flux quantum. Maintaining impedance matching is crucial, as any mismatch reduces the coupling rate $\kappa$ between the system and its environment. A quantum mechanical analysis shows that the total Hamiltonian of this structure is fundamentally distinct from that of traditional Fluxonium devices or conventional SQUID arrays [6–7]. This difference arises not only from the physical design of Blochnium but also from its more effective control over Josephson junction nonlinearity. Moreover, the BJPA exhibits unique characteristics that are unattainable with earlier designs.

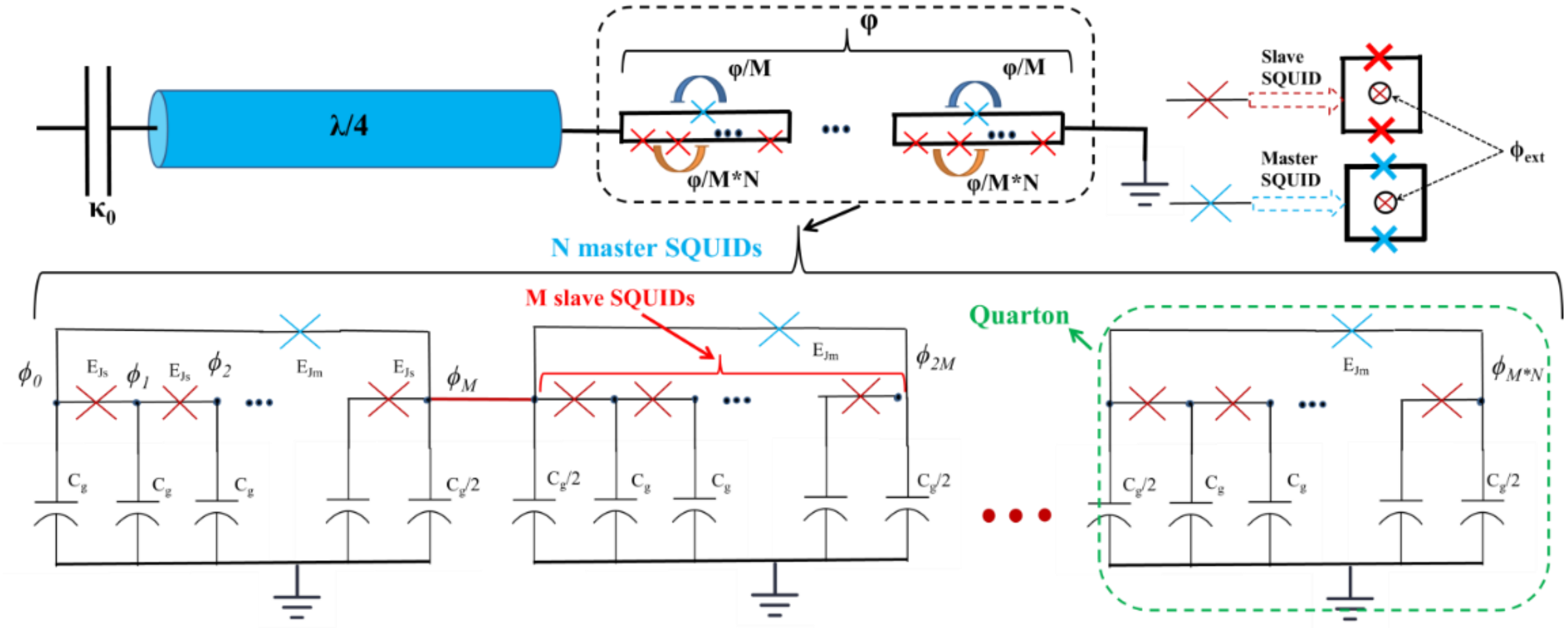

Fig 24. The electrical schematic of a Blochnium structure (as a $\lambda/4$ resonator) containing N Quarton, meaning N master SQUIDs and each Quarton contains M slave SQUIDs [13].

In the following sections, the quantum mechanical aspects of the BJPA are explored in detail, deriving key parameters such as gain and nonlinearity. The theoretical results highlight Blochnium's advantages in tunability and linearity, emphasizing its potential for next-generation quantum technologies. This comprehensive analysis illustrates the clear superiority of the BJPA over traditional designs, positioning it as a promising solution for advancing the performance and scalability of superconducting quantum systems.

*Theoretical background: BJPA gain calculation using quantum theory*

To gain a complete understanding of the structure presented in Fig. 24, the approach begins with the theoretical derivation of the system's Lagrangian and total Hamiltonian, followed by a dynamic analysis using the quantum Langevin equation. The total Lagrangian [13, 69] of the Blochnium structure shown in Fig. 24 is given by:

$$
\begin{aligned}
L_t =\; & \frac{C_g}{2}\sum_{k=0}^{M*N}\dot{\phi}_k^{\,2} + \frac{C_{Js}}{2}\sum_{k=0}^{M*N-1}\left(\dot{\phi}_k - \dot{\phi}_{k+1}\right)^2 + \frac{C_{Jm}}{2}\sum_{k=0}^{N-1}\left(\dot{\phi}_{M*k} - \dot{\phi}_{M*(k+1)}\right)^2 \\
& - \frac{1}{2L_{Js}}\sum_{k=0}^{M*N-1}\left(\phi_k - \phi_{k+1}\right)^2 - \frac{1}{2L_{Jm}}\sum_{k=0}^{N-1}\left(\phi_{M*k} - \phi_{M*(k+1)}\right)^2 + M*N*E_{Js}\cos\left(\frac{\varphi}{M*N}\right) + N*E_{Jm}\cos\left(\frac{\varphi}{N}\right)
\end{aligned}
\qquad (20)
$$

Here, $C_g$ represents the parasitic capacitance grounding the SQUIDs; ($C_{Js}$, $L_{Js}$, $E_{Js}$) denote the capacitance, inductance, and Josephson energy of the slave junctions; ($C_{Jm}$, $L_{Jm}$, $E_{Jm}$) correspond to the capacitance, inductance, and Josephson energy of the master junctions; N is the number of master junctions; M is the number of slave SQUIDs within each B-cell; and φ is the flux, treated as a quantum coordinate operator. To simplify the algebra, it is convenient to reformulate Eq. (20) as a set of coupled equations. In this simplification process, the last two terms in the Lagrangian are temporarily neglected but will be reintroduced when calculating the total Hamiltonian. Thus, Eq. (20) can be expressed in a more compact form as $L_t = \frac{1}{2}\dot{\phi}^T\hat{C}\dot{\phi} - \frac{1}{2}\phi^T\hat{L}^{-1}\phi$, where $\hat{C}$ and $\hat{L}^{-1}$ are tri-angular matrixes as [13]:

$$
\hat{C} = \begin{bmatrix}
\underbrace{\begin{matrix}
C_g+C_j+C_M & -C_j & 0 & 0 & \bullet & -C_M \\
-C_j & C_g+2C_j & -C_j & 0 & \bullet & 0 \\
0 & -C_j & \bullet & -C_j & \bullet & 0 \\
0 & 0 & -C_j & \bullet & \bullet & \bullet \\
\bullet & \bullet & \bullet & \bullet & C_g+2C_j & -C_j \\
-C_M & 0 & \bullet & 0 & -C_j & C_g+2C_j+2C_M
\end{matrix}}_{\text{The first Quarton}} & \bullet\;\bullet & \begin{matrix}
0 & 0 & 0 & \bullet & \bullet & 0 \\
0 & 0 & 0 & \bullet & \bullet & 0 \\
0 & 0 & 0 & \bullet & \bullet & 0 \\
\bullet & \bullet & \bullet & \bullet & \bullet & 0 \\
\bullet & \bullet & \bullet & \bullet & 0 & 0 \\
0 & 0 & 0 & \bullet & \bullet & 0
\end{matrix} \\
\vdots & \ddots & \vdots \\
\begin{matrix}
0 & 0 & 0 & \bullet & \bullet & 0 \\
0 & 0 & 0 & \bullet & \bullet & 0 \\
0 & 0 & 0 & \bullet & \bullet & 0 \\
\bullet & \bullet & \bullet & \bullet & \bullet & 0 \\
\bullet & \bullet & \bullet & \bullet & 0 & 0 \\
0 & 0 & 0 & \bullet & \bullet & 0
\end{matrix} & \bullet\;\bullet & \underbrace{\begin{matrix}
C_g+2C_j+2C_M & -C_j & 0 & 0 & \bullet & -C_M \\
-C_j & C_g+2C_j & -C_j & 0 & \bullet & 0 \\
0 & -C_j & \bullet & -C_j & \bullet & 0 \\
0 & 0 & -C_j & \bullet & \bullet & \bullet \\
\bullet & \bullet & 0 & \bullet & C_g+2C_j & -C_j \\
-C_M & 0 & \bullet & 0 & -C_j & C_g+2C_j+C_M
\end{matrix}}_{\text{The last Quarton}}
\end{bmatrix}
$$

$$
\hat{L}^{-1} = L^{-1}\begin{bmatrix}
\begin{matrix}
2 & -1 & 0 & 0 & \bullet & -1 \\
-1 & 2 & -1 & 0 & 0 & 0 \\
0 & -1 & \bullet & -1 & 0 & 0 \\
0 & 0 & -1 & \bullet & \bullet & \bullet \\
\bullet & 0 & 0 & \bullet & 2 & -1 \\
-1 & 0 & \bullet & 0 & -1 & 4
\end{matrix} & \bullet\;\bullet & \begin{matrix}
0 & 0 & 0 & \bullet & \bullet & 0 \\
0 & 0 & 0 & \bullet & \bullet & 0 \\
0 & 0 & 0 & \bullet & \bullet & 0 \\
\bullet & \bullet & \bullet & \bullet & \bullet & 0 \\
\bullet & \bullet & \bullet & \bullet & 0 & 0 \\
0 & 0 & 0 & \bullet & \bullet & 0
\end{matrix} \\
\vdots & \ddots & \vdots \\
\begin{matrix}
0 & 0 & 0 & \bullet & \bullet & 0 \\
0 & 0 & 0 & \bullet & \bullet & 0 \\
0 & 0 & 0 & \bullet & \bullet & 0 \\
\bullet & \bullet & \bullet & \bullet & \bullet & 0 \\
\bullet & \bullet & \bullet & \bullet & 0 & 0 \\
0 & 0 & 0 & \bullet & \bullet & 0
\end{matrix} & \bullet\;\bullet & \begin{matrix}
4 & -1 & 0 & 0 & \bullet & -1 \\
-1 & 2 & -1 & 0 & \bullet & 0 \\
0 & -1 & \bullet & -1 & 0 & 0 \\
0 & 0 & -1 & \bullet & \bullet & \bullet \\
\bullet & 0 & 0 & \bullet & 2 & -1 \\
-1 & 0 & \bullet & 0 & -1 & 2
\end{matrix}
\end{bmatrix}
\qquad (21)
$$

By employing the capacitance and inductance matrices given in Eq. (21), the matrix $\Omega^2 = \hat{C}^{-1}\hat{L}^{-1}$ [6, 9] can be defined to determine the eigenvalues (corresponding to the dominant frequencies) and eigenvectors (where $\Psi_i$ represents the wave profile of each mode) of the quantum system under consideration. Based on these results, the effective capacitance and inductance associated with the structure are calculated. Specifically, the effective circuit parameters for each mode can be derived from the eigenvalues and eigenvectors as follows: $C_{eff} = \overrightarrow{\Psi_i}^T \hat{C} \overrightarrow{\Psi_i}$ and $L_{eff}{}^{-1} = \overrightarrow{\Psi_i}^T \hat{L}^{-1} \overrightarrow{\Psi_i}$. This approach allows the mapping of a λ/4 resonator to an equivalent LC circuit [6]. As a result, the Blochnium structure developed in this work can be modeled as a series combination of effective inductors and capacitors, along with nonlinear elements arising from the terms isolated in Eq. (20). Analyzing this simplified circuit model greatly facilitates the evaluation of the quantum circuit's impedance $Z_{eff} = \sqrt{L_{eff}/C_{eff}}$ and subsequently obtain the coupling rate of the quantum system to the environment $\kappa_{eff} = \omega_{eff}/Q_{eff}$, where $Q_{eff}$ is the circuit quality factor related to mismatching. Therefore, the total Hamiltonian [13, 69] of the effective circuit is summarized as:

$$H_t = \omega_{eff} a^+ a - M * N * E_{Js} \cos\left(\frac{\varphi}{M * N}\right) - N * E_{Jm} \cos\left(\frac{\varphi}{N}\right) \tag{22}$$

where a and $a^+$ denote the annihilation and creation operators, respectively. In deriving the total Hamiltonian, it is assumed that the circuit consists of a simple LC oscillator coupled to a nonlinear element. The second and third terms can be conveniently expressed in terms of ladder operators by applying a Taylor expansion of the cosine function [6, 7], truncated at the fourth order, as follows:

$$H_t = \omega_{eff} a^+ a - M * N * E_{Js} \left(\frac{\varphi^4}{4! M^4 * N^4}\right) - N * E_{Jm} \left(\frac{\varphi^4}{4! N^4}\right) \tag{23}$$

It is important to note that the second term in the Taylor expansion of the cosine function was previously absorbed into the oscillatory component of the system. To simplify the analytical calculations, the assumption $E_{Jm} = \alpha_c^* E_{Js}$ is employed, where $\alpha_c^*$ represents the ratio of the junction areas. Precise control of junction areas is critical in superconducting circuits, as the Josephson energy $E_J$ depends directly on the critical current $I_c$, which in turn scales with the junction area. However, fabrication imperfections—such as variations arising from lithography and deposition processes—can lead to deviations from the intended energy ratio. These deviations can result in shifts in energy levels, reduced coupling efficiency, and increased sensitivity to noise. To mitigate these effects, strategies such as dynamic tuning of $E_J$ via magnetic flux, post-fabrication characterization and compensation, and the use of advanced fabrication techniques like atomic-layer deposition are employed to enhance junction uniformity and device reliability. Finally, by applying the phase quantization described earlier and assuming that the resonance frequency of each Quarton in the structure remains constant under the aforementioned conditions, the total Hamiltonian can be expressed as [13]:

$$H_t = \omega_{eff} a^+ a - \frac{E_c}{6N}\left(\frac{1}{M} - \alpha_c^*\right) a^{+2} a^2 \tag{24}$$

In this equation, the derived nonlinearity terms explicitly depend on the number of Quartons in the circuit, the number of SQUIDs within each Quarton, and the ratio between the master and slave Josephson junctions. With the complete Hamiltonian established, the system dynamics can be analyzed using the quantum Langevin equation [18–19], ultimately allowing for the calculation of signal gain and other relevant parameters. The dynamics equation of motion for this system is examined (quantum Langevin equation) [69] as:

$$\dot{a} = -i\omega_{eff} a - iKa^+ aa - \frac{\kappa}{2} a + \sqrt{\kappa} a_{in} \tag{25}$$

In this equation, $K = -E_c/6N\left(1/M - \alpha_c^*\right)$, $\alpha_c^* \equiv \alpha_c/M$, where $\alpha_c$ is the nonlinearity factor, and "a" determines the intra-cavity signals expressed as a = α + δa, where α and δa are the DC point and quantum signal fluctuation around the DC point, respectively. The steady-state solution (replacing a = α and $a_{in}$ =

$\alpha_{in}$) [19] for a coherent pump $\alpha = |\alpha|\exp(i\varphi_1)$ after some algebra's simplification is introduced as:

$$\left[\frac{1}{4}+\left(\frac{\omega_p-\omega_{eff}}{\kappa}\right)^2\right]|\alpha|^2-\frac{\left(\omega_p-\omega_{eff}\right)K}{\kappa^2}|\alpha|^4+\left(\frac{K}{\kappa}\right)^2|\alpha|^6-\frac{K}{\kappa^2}|\alpha_{in}|^2=0\xrightarrow[\zeta=\left(\frac{K}{\kappa}\right)|\alpha_{in}|^2,\ \hat{\alpha}_{in}=\alpha_{in}\frac{1}{\sqrt{\kappa}}]{n=\frac{|\alpha|^2}{|\alpha_{in}|^2},\ \delta=\frac{\omega_p-\omega_{eff}}{\kappa}}\left[\frac{1}{4}+\delta^2\right]n-2\delta\zeta n^2+\zeta^2n^3-1=0 \tag{26}$$

Using Eq. 26, the normalized intra-cavity average photon number n, generated by the pump effect, can be determined by solving the roots of the equation. In this expression, δ and ζ denote the normalized pump detuning and the effective nonlinearity in the presence of the pump field, respectively. However, it is crucial to distinguish between K and ζ [7, 13]: K refers to the intrinsic nonlinearity of JJ or an array of JJs, while ζ reflects the influence of the input signal on the overall nonlinear response. This implies that the inherently weak nonlinearity of a JJ can be effectively compensated by increasing the pump power. The analysis that follows will address quantum fluctuations and the small-signal gain. Nonetheless, Eq. 26 is also applicable for estimating of DC gain, interpreted as the classical reflection coefficient as $\alpha_{out}/\alpha_{in}$ [7]. Finally, the linearized response for weak quantum signals can be expressed as:

$$\dot{\delta a}=\left[i\left(\omega_p-\omega_{eff}\right)-\frac{\kappa}{2}-i2K|\alpha|^2\right]\delta a-iK\delta a^{+}\alpha^2+\sqrt{\kappa}\delta a_{in} \tag{27}$$

Since the equation derived and its conjugate is linear, so one can transform it to the Fourier domain and decompose all the modes relevant as:

$$-i\left(\omega_s-\omega_p\right)\delta a=i\left[\frac{\left(\omega_p-\omega_{eff}\right)}{\kappa}+i\frac{1}{2}-2\frac{K}{\kappa}|\alpha|^2\right]\delta a-i\frac{K}{\kappa}\delta a^{+}\alpha^2+\frac{\sqrt{\kappa}}{\kappa}\delta a_{in}\xrightarrow{\Delta=\frac{\omega_s-\omega_p}{\kappa}}0=i\left[\delta+\Delta+\frac{i}{2}-2\zeta n\right]\delta a-i\zeta ne^{(2i\varphi)}\delta a^{+}+\delta\hat{\alpha}_{in} \tag{28}$$

where Δ denotes the signal detuning. To compute the gain, the final step involves taking the complex conjugate of Eq. 28 and constructing the corresponding scattering matrix. By applying the input-output formalism, the gains of both the signal and idler can then be derived [7, 19, 20], as shown below:

$$\begin{cases}\left[i\left(-\delta-\Delta+2\zeta n\right)+\frac{1}{2}\right]\delta a_{\omega}+i\zeta ne^{(2i\varphi)}\delta a_{\omega}^{+}=\delta\hat{\alpha}_{in\omega}\\ \left[i\left(\delta-\Delta-2\zeta n\right)+\frac{1}{2}\right]\delta a_{\omega}^{+}-i\zeta ne^{(-2i\varphi)}\delta a_{\omega}=\delta\hat{\alpha}_{in\omega}^{+}\end{cases}\longrightarrow\begin{bmatrix}i\left(-\delta-\Delta+2\zeta n\right)+0.5 & i\zeta ne^{(2i\varphi)}\\ -i\zeta ne^{(-2i\varphi)} & i\left(\delta-\Delta-2\zeta n\right)+0.5\end{bmatrix}\begin{bmatrix}\delta a_{\omega}\\ \delta a_{\omega}^{+}\end{bmatrix}=\begin{bmatrix}\delta\hat{\alpha}_{in\omega}\\ \delta\hat{\alpha}_{in\omega}^{+}\end{bmatrix} \tag{29}$$

Thus, BJPA's gain using a few algebras and after some simplifications is calculated as [13]:

$$G_{BJPA}=\left\{\frac{i\sqrt{\kappa}\left(\delta-\Delta-2\zeta n\right)+0.5}{\left[-\left\{\left(\delta-\Delta-2\zeta n\right)+0.5\right\}\left\{\left(-\delta-\Delta+2\zeta n\right)+0.5\right\}-\zeta^2n^2\right]}-1\right\}-\left\{\frac{i\sqrt{\kappa}\zeta ne^{(2i\varphi)}}{\left[-\left\{\left(\delta-\Delta-2\zeta n\right)+0.5\right\}\left\{\left(-\delta-\Delta+2\zeta n\right)+0.5\right\}-\zeta^2n^2\right]}\right\} \tag{30}$$

In this expression, the first term corresponds to the signal gain, while the second represents the idler gain. As evident from the gain relation in Eq. 30, the overall gain is significantly influenced by several factors, including the signal and pump detuning, the effective nonlinearity strength, and the input coupling rate. In the following, we highlight several technical characteristics of the BJPA that underscore its distinct advantages over conventional parametric amplifiers. The simulation results reveal the significant benefits of BJPAs, particularly their enhanced output $P_{1dB}$—an indicator of the amplifier's linearity limit—and their frequency tunability, including operation within the C-band. Notably, improved linearity, which is critical for quantum applications, is achieved through precise manipulation of the system's inherent nonlinearity. Fig. 25a illustrates simulation results for $P_{1dB}$ across various BJPA configurations, demonstrating that parameters M, N, and $\alpha_c$ significantly influence linearity performance. These effects are rooted in the modulation of system nonlinearity enabled by the Quarton architecture. The data show that increasing M (the number of secondary JJs per Quarton) has a more substantial impact on enhancing $P_{1dB}$ than increasing

N (the number of Quartons), likely due to the critical dependence on the interplay between M and $\alpha_c$. Simulations confirm that BJPAs can deliver signal gains above 25 dB with $P_{1dB}$ values exceeding –92 dBm. A configuration with N=70 and M=8, highlighted with dashed lines in Fig. 25a, achieves this performance. Remarkably, this surpasses the output of a conventional straight JJ array reported in [9], despite utilizing significantly fewer junctions—a key advantage that simplifies fabrication compared to architectures such as Fluxonium, thereby improving scalability and integration potential.

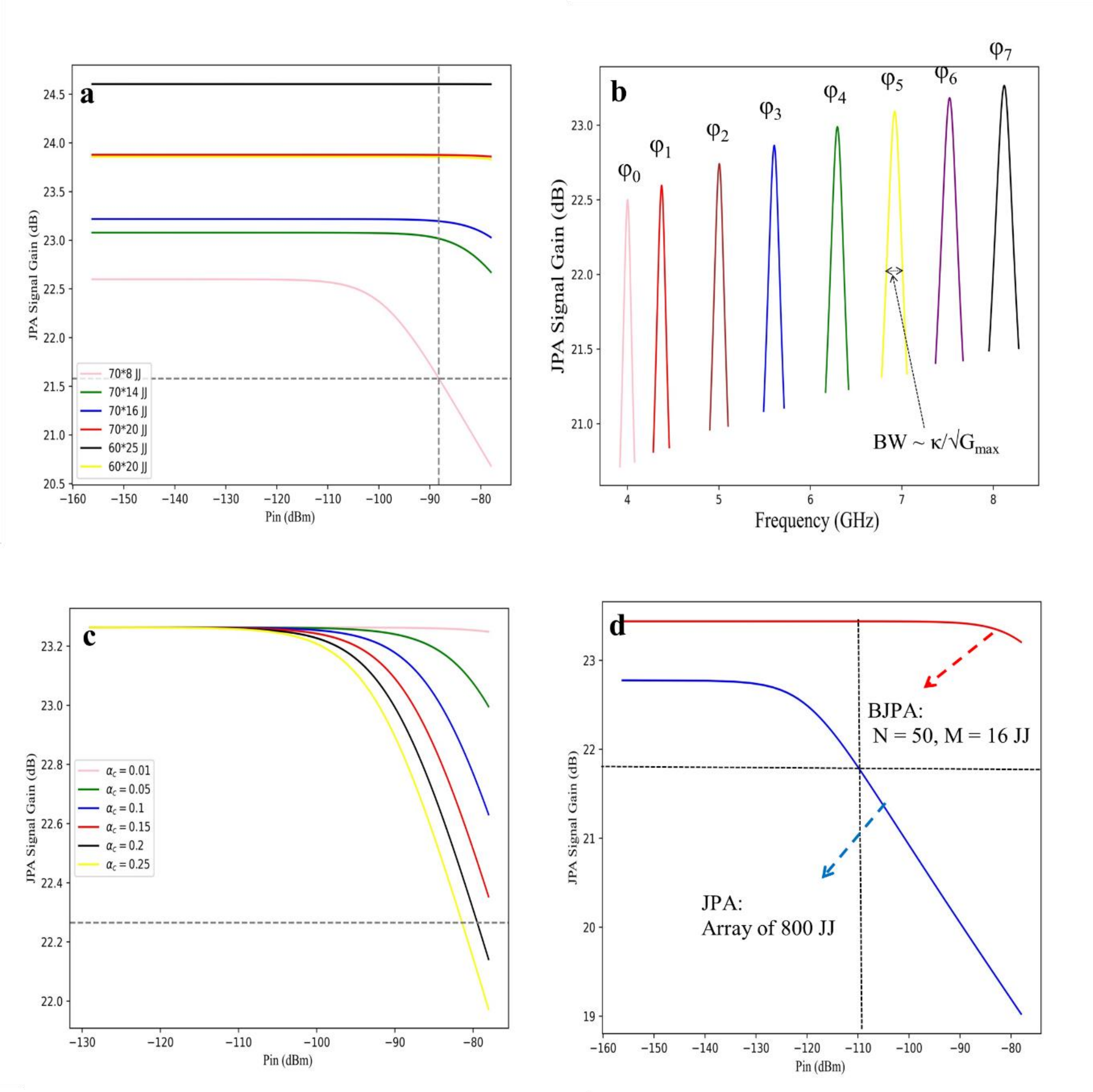

Fig 25. (a) the effect of the number of Quarton and slave SQUIDs on the compression points $P_{1dB}$ vs the pump power, $P_{in}$ (dBm), $\alpha_c = 0.1$, b) the effect of the JJ's nonlinear inductor on the BJPA's signal gain and sweeping throughout the C-band for N = 40 and M = 14, $\alpha_c = 0.1$, BW: bandwidth, (c) the effect of the nonlinearity factor $\alpha_c$ on $P_{1dB}$ for N = 40 and M = 14, (d) the comparison of the JPA linearity between a straight array of 800 JJ and a BJPA with N = 50 and M = 16; For all figures the signal and pump detuning are considered to be zero $\Delta = \delta = 0$, and $\zeta = 0.01\zeta_0$ [13].

Fig. 25b presents results for a BJPA with N=40 and M=16, highlighting the tunability of the junction's effective inductance via modulation of the Josephson phase $\phi$, where $L_J=L_{J0}/\cos\phi$. This tunability facilitates dynamic adjustment of the resonance frequency across the operational bandwidth, enabling performance optimization. The parameter $\alpha_c$*, defined as the ratio of Josephson energies between the primary and secondary junctions within a Quarton, plays a critical role in shaping key amplifier characteristics, including $P_{1dB}$. As shown in Fig. 25c, tuning $\alpha_c$ allows for control over the system's nonlinearity: a reduction in $\alpha_c$ leads to a notable improvement in linearity. This behavior aligns with the expression for the nonlinearity coefficient K in Eq. 25, which explicitly incorporates $\alpha_c$. The linearity performance of a BJPA with N=50, M=16 is compared to a traditional JPA comprising an array of 800 JJs and shown in Fig. 25d. This comparison is particularly relevant given the extensive study of large-array JPAs in both theory and experiment [5, 9]. While BJPA delivers comparable gain, it achieves a significantly higher compression point, confirming its superior linearity. Although the BJPA demonstrates strong potential—offering improved $P_{1dB}$ and full C-band tunability—real-world implementation must consider possible variations stemming from fabrication tolerances, material inconsistencies, and environmental effects, which may impact performance stability. Nonetheless, alignment with prior theoretical and experimental results [5, 6, 9] reinforces the validity of the simulations. Beyond electrical advantages, the BJPA's compact architecture—with fewer components—supports more efficient layout, improved reliability, and streamlined integration into modern quantum hardware [10].

Crucially, if BJPA performance consistently exceeds –92 dBm in $P_{1dB}$, as theoretical results suggest, it opens the door to a two-stage JPA configuration, potentially replacing the need for follow-up amplification stages such as HEMTs [13, 14]. This would reduce system noise and enhance the overall fidelity of quantum readout chains. The BJPA's extensive tunability and precise control over key parameters make it a versatile and powerful component for quantum experiments, where fine-grained control of qubit interactions and readout is essential. Coupled with its high gain, enhanced linearity, and compact design, the BJPA represents a scalable and efficient solution poised to support the next generation of quantum technologies.

*New architecture of BJPA: Design for selectivity in gain*

As discussed above, a key feature of the BJPA is its ability to operate with high coherence while maintaining strong nonlinear interactions necessary for parametric amplification. Unlike traditional JPAs, which rely on a single JJ or a long array, the BJPA utilizes an N×M JJ array with a low N and M. The BJPA architecture also provides enhanced tunability, enabling the amplifier to cover a broader range of operating frequencies without significant performance degradation. The incorporation of Blochnium also introduces new challenges, particularly in fabrication and biasing control. Maintaining phase uniformity across such a large array of JJs requires precise control over fabrication parameters, such as junction critical currents and capacitance values. Furthermore, the increased complexity of the array structure demands careful tuning of the bias current and pump power to avoid unwanted frequency shifts or gain saturation effects. Despite these challenges, the theoretical and modeling results indicate [13] that the Blochnium-based JPA outperforms conventional JPAs in several key metrics, including gain and $P_{1dB}$. These improvements suggest that BJPA could play a crucial role in the next generation of quantum computing and superconducting microwave technologies, providing robust amplification with minimal added noise. However, in this section, we propose a modified design of the BJPA, not the traditional BJPA addressed in [13], which produces a distinctive and interesting gain profile tailored for quantum applications. This design innovation enables the modified BJPA to not only amplify quantum signals but also suppress qubit energy leakage into the readout chain, thereby enhancing system coherence and readout efficiency [88]. The modified BJPA schematic shown in Fig. 26 illustrates the architecture relevant, designed for quantum-limited signal amplification. The system is composed of multiple cells, each realized as a composite structure consisting of Quarton A, a central resonator (Res.), and Quarton B. Each Quarton includes a primary JJ or SQUID (with Josephson energy $E_{Jm}$), which plays a dominant role in defining the nonlinearity and the effective potential landscape of the cell. Also, the Quarton hosts the secondary JJ or SQUIDs (with the same Josephson energy of $E_{Js}$), indicating a weaker junction that assists in controlling the parametric gain and dynamic behavior of the amplifier. The central section, labeled as "Res.," acts as a coupling resonator between two Quartons, facilitating energy exchange and phase-sensitive amplification through the Josephson nonlinearity. The upper part of the schematic connected to λ/4 shows a linear cascade of cells, demonstrating how multiple unit cells are interconnected to form a multi-stage amplifier to enhance amplifier features. This modular approach allows for flexible design and scalability. In the same way with BJPA, one of the important elements, parasitic element, that can affect the performance of the system is $C_g$, which is hardly a controllable parameter in the quantum system depicted. It is also assuming that the phase drop (φ) from the flux node distributed homogenously over the array (displayed in the figure); hence each SQUID senses the same phase drop [6-7, 13]. The flux variable ϕ represents the magnetic flux threading the SQUID loop and its relationship with an applied flux $\phi_{ext}$ can be interpreted as phase shift in energy, in which an applied flux $\phi_{ext}$ typically manifests in terms like $\cos(2\pi(\phi-\phi_{ext})/\Phi_0)$, where $\Phi_0$ is the flux quanta. This modifies the energy landscape and introduces tunability to the system.

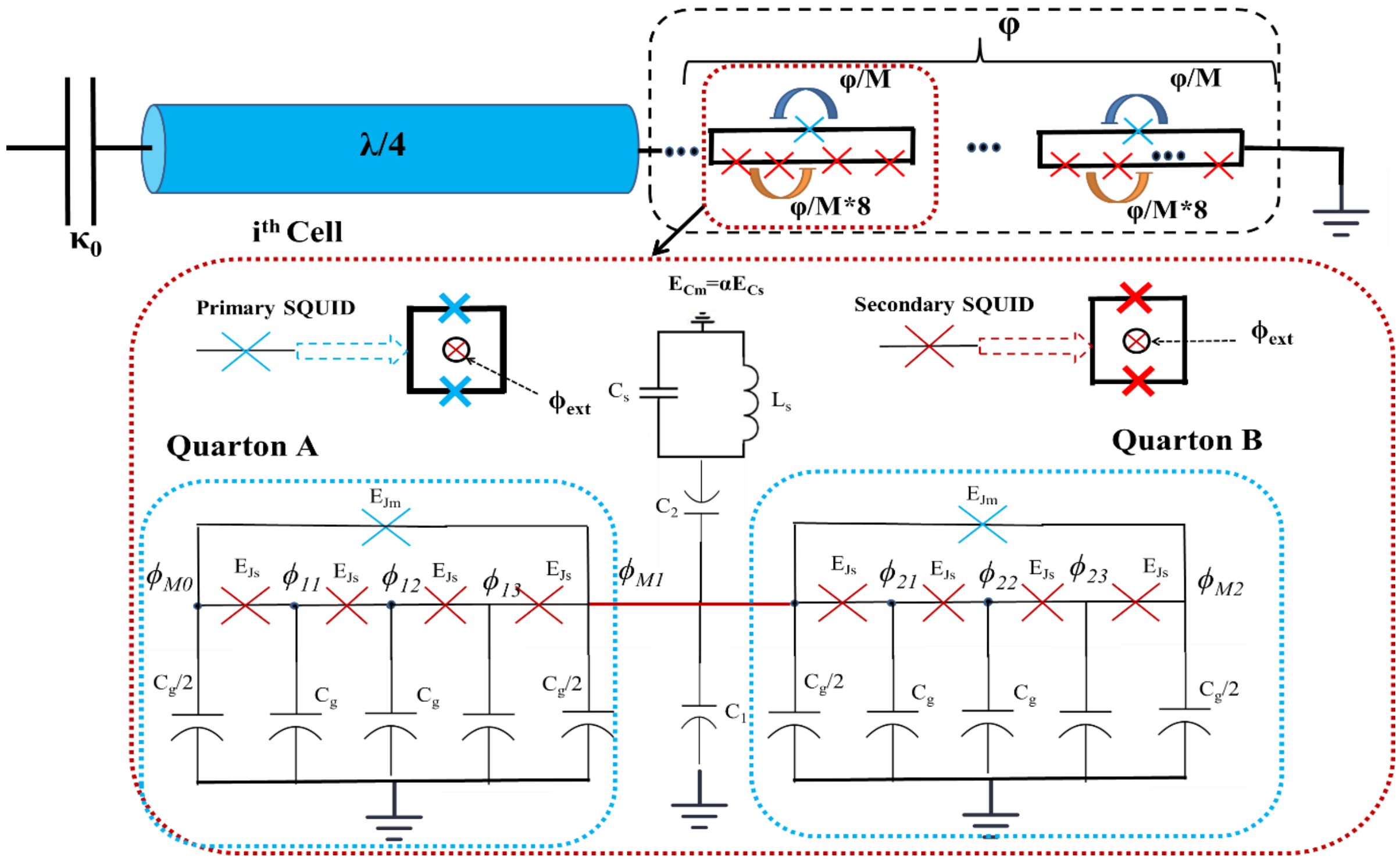

Fig. 26 Schematic of the proposed modified BJPA architecture, consisting of two coupled Quartons (A and B), each composed of an array of secondary JJ with shared parasitic capacitance $C_g$. Each Quarton is controlled via a primary SQUID through flux-biased via $\phi_{ext}$. The two Quartons are interconnected through a superconducting resonator (λ/4) to mediate energy transfer and enforce nonreciprocal dynamics. Each Quarton includes a primary JJ with energy $E_{Jm}$ and four secondary JJ with energy $E_{Js}$, couples to each other through the external resonator [88].

To thoroughly analyze the system, the configuration shown in Fig. 26 undergoes a complete quantum mechanical evaluation. The process begins with the formal derivation of the system's Lagrangian, followed by the computation of its total Hamiltonian. The system's dynamics are then investigated using the quantum Langevin approach, which offers valuable insights into the mechanisms of signal amplification and energy dissipation. The total Lagrangian [6-7, 13, 69] corresponding to the quantum circuit—specifically for the $i^{\text{th}}$ unit cell comprising two consecutive Quartons coupled via a standard resonator, as illustrated in Fig. 26—is given by [88]:

$$\begin{aligned} L_t \simeq\ & \frac{C_g}{2}\sum_{\substack{k=0\\k\neq 4}}^{8}\dot{\phi}_k^{\,2} + \frac{C_J}{2}\sum_{k=0}^{7}\left(\dot{\phi}_k-\dot{\phi}_{k+1}\right)^2 + \frac{C_m}{2}\left\{\left(\dot{\phi}_0-\dot{\phi}_4\right)^2+\left(\dot{\phi}_4-\dot{\phi}_8\right)^2\right\} + \left(\frac{C_1+C_2+C_s}{2}\right)\dot{\phi}_4^{\,2} - \frac{1}{2L_s}\phi_4^{\,2} \\ & -\frac{1}{2L_{Js}}\sum_{k=0}^{7}\left(\phi_k-\phi_{k+1}\right)^2 - \frac{1}{2L_{Jm}}\left\{\left(\phi_0-\phi_4\right)^2+\left(\phi_4-\phi_8\right)^2\right\} + 8N*E_{Js}\cos\left(\frac{\varphi}{8N}\right) + N*E_{Jm}\cos\left(\frac{\varphi}{N}\right) \end{aligned} \tag{31}$$

where $C_g$ denotes the parasitic capacitance grounding the SQUIDs; ($C_{Js}$,$L_{Js}$,$E_{Js}$) represent the capacitance, inductance, and Josephson energy of the secondary junctions; ($C_{Jm}$,$L_{Jm}$,$E_{Jm}$) correspond to those of the primary junctions; N is the number of primary SQUIDs within each unit cell; and $\phi$ refers to the flux, treated as a quantum coordinate operator. To streamline the algebraic manipulation, it is advantageous to reformulate Eq. 31 into a set of coupled equations. As part of this simplification, the last two terms of the Lagrangian are temporarily excluded, though they will be reintroduced during the derivation of the total Hamiltonian. With this approach, Eq. 31 can be concisely expressed in the following compact form: $L_t = \frac{1}{2}\dot{\phi}^T\hat{C}\dot{\phi} - \frac{1}{2}\phi^T\hat{L}^{-1}\phi$, [6-11, 13] where $\hat{C}$ and $\hat{L}^{-1}$ are tri-angular matrixes represented as:

$$\hat{C} = \begin{bmatrix} C_g+C_j+C_M & -C_j & 0 & 0 & -C_m & 0 & 0 & 0 & 0 \\ -C_j & C_g+2C_j & -C_j & 0 & 0 & 0 & 0 & 0 & 0 \\ 0 & -C_j & C_g+2C_j & -C_j & 0 & 0 & 0 & 0 & 0 \\ 0 & 0 & -C_j & C_g+2C_j & -C_j & 0 & 0 & 0 & 0 \\ -C_m & 0 & 0 & -C_j & C_X & -C_j & 0 & 0 & -C_m \\ 0 & 0 & 0 & 0 & -C_j & C_g+2C_j & -C_j & 0 & 0 \\ 0 & 0 & 0 & 0 & 0 & -C_j & C_g+2C_j & -C_j & 0 \\ 0 & 0 & 0 & 0 & 0 & 0 & -C_j & C_g+2C_j & -C_j \\ 0 & 0 & 0 & 0 & -C_m & 0 & 0 & -C_j & C_g+C_j+C_M \end{bmatrix}$$

$$\hat{L}^{-1} = L^{-1}\begin{bmatrix} 2 & -1 & 0 & 0 & -1 & 0 & 0 & 0 & 0 \\ -1 & 2 & -1 & 0 & 0 & 0 & 0 & 0 & 0 \\ 0 & -1 & 2 & -1 & 0 & 0 & 0 & 0 & 0 \\ 0 & 0 & -1 & 2 & -1 & 0 & 0 & 0 & 0 \\ -1 & 0 & 0 & -1 & L_X & -1 & 0 & 0 & -1 \\ 0 & 0 & 0 & 0 & -1 & 2 & -1 & 0 & 0 \\ 0 & 0 & 0 & 0 & 0 & -1 & 2 & -1 & 0 \\ 0 & 0 & 0 & 0 & 0 & 0 & -1 & 2 & -1 \\ 0 & 0 & 0 & 0 & -1 & 0 & 0 & -1 & 2 \end{bmatrix} \tag{32}$$

where, $C_X$=$C_1$+$C_2$+$C_s$+2$C_m$+2$C_j$ and $L_X$=4+L/$L_s$. As shown in Eq. 32, the configuration of the capacitance and inductance matrices in this structure differs notably from that of the BJPA as discussed in [13], with $C_X$ and $L_X$ playing key roles in defining the system's behavior. This distinction indicates that the dynamical response of this newly proposed structure—when functioning as a parametric amplifier—should fundamentally differ from that of the BJPA. Moreover, the revised design introduces several additional degrees of freedom, offering enhanced flexibility and tunability. These added parameters contribute to a broader design space, facilitating the realization of a more adaptable and high-performance JPA suited for diverse quantum applications. The gain of the parametric amplifier can, in principle, be calculated using the same analytical framework applied to the BJPA as outlined in [13]. Nevertheless, in this section, we place greater emphasis on the technical outcomes derived from simulation to better understand the impact of modifying the amplifier's architecture. By examining how the structure evolves in the modified BJPA design, we gain deeper insight into the resulting changes in performance metrics—such as gain, bandwidth, and dynamic range—highlighting the advantages and trade-offs introduced by this novel configuration.
The spectral response of the modified BJPA depicted in Fig. 27 under degenerate pumping conditions reveals the fundamental nonlinear processes driving parametric amplification. The JPA configuration is characterized by a strong pump signal at $F_{pump}$=7.12 GHz in the same way with latter sections. Unlike conventional single-JJ JPAs, which typically display notable variations between the reflection coefficient ($S_{11}$) and the transmission coefficient ($S_{21}$) across the frequency spectrum, the modified BJPA structure exhibits only a minimal difference between these parameters. This observation suggests a significant mismatch between the structure and the terminated loads at both the input and output ports. In other words, the impedance matching between the amplifier and its external circuitry is not well-optimized in this configuration, potentially leading to inefficient power transfer and reflection-related losses. While the internal dynamics of the BJPA may still support strong parametric amplification, the external coupling efficiency is adversely affected by this mismatch. Addressing this issue through improved impedance engineering—such as incorporating matching networks or modifying the interface components—would be essential to fully leverage the amplifier's internal gain potential and ensure seamless integration into quantum systems.

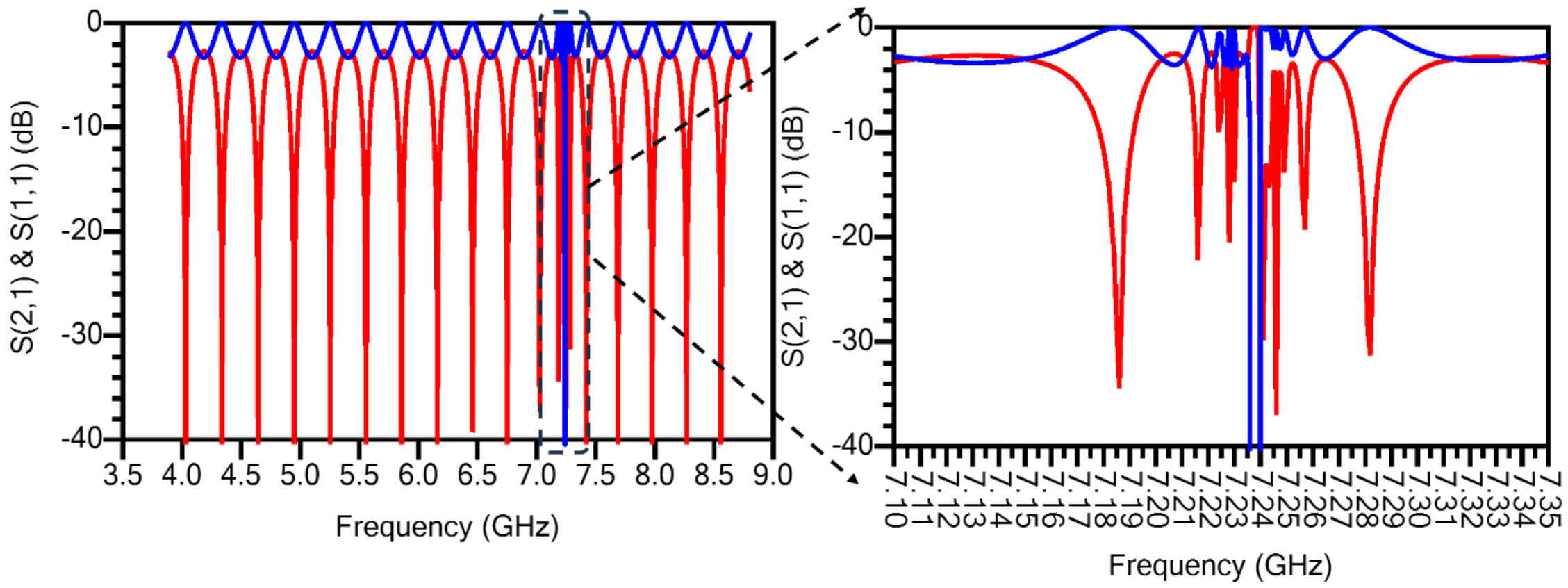

Fig. 27 Spectrum of the designed JPA; $F_{pump}$ = 7.12 GHz, with 8*192 JJ.

The gain performance of the revised BJPA shown in Fig. 28 provides critical insights into its amplification efficiency and nonlinearity characteristics. In this configuration, the JPA is driven by a pump current of 3.96 μA while amplifying a weak signal at $F_{sig}$=6.18 GHz. The amplifier demonstrates a peak gain of approximately 25 dB, closely matching the performance reported for the BJPA in [13]. This substantial gain arises from the effective nonlinear inductance modulation achieved by the Blochnium array, which enables strong and efficient energy transfer from the pump signal to the target signal frequency. The mechanism behind this gain enhancement lies in the periodic structure of the amplifier, composed of repeating nonlinear cells—each formed by two coupled Quartons—with energy oscillating back and forth through the resonator positioned between them. These periodic interactions give rise to resonance conditions that reinforce signal amplification at specific frequencies. What makes this figure particularly significant is its depiction of a nonuniform, yet highly structured, gain profile extending across the C-band (approximately 4–8 GHz). The gain spectrum is marked by sharp and distinct peaks reaching up to ~24–25 dB, interspersed with deep minima approaching 0 dB. This comb-like structure indicates frequency-selective behavior, where the amplifier provides strong gain only at certain resonant modes. While this could be seen as a limitation in applications requiring broadband gain, it presents a valuable opportunity in frequency-multiplexed quantum systems, where each qubit operates at a distinct frequency. By aligning the readout frequencies with the gain peaks, one can maximize amplification efficiency while minimizing qubit's frequency at non-target frequencies to kill any qubit's energy leakage to readout circuit [88]. Furthermore, this gain landscape can be strategically engineered by tuning the inter-Quarton coupling and resonator characteristics, effectively allowing chip designers to tailor the amplifier response to match specific quantum circuit requirements. However, it also implies that outside of the designed gain peaks, signal amplification is minimal or nonexistent, which necessitates careful spectral planning in system-level integration. Thus, the figure reveals not only the power of the modified BJPA structure in achieving high gain, but also its unique potential for frequency-domain optimization in scalable quantum architectures.

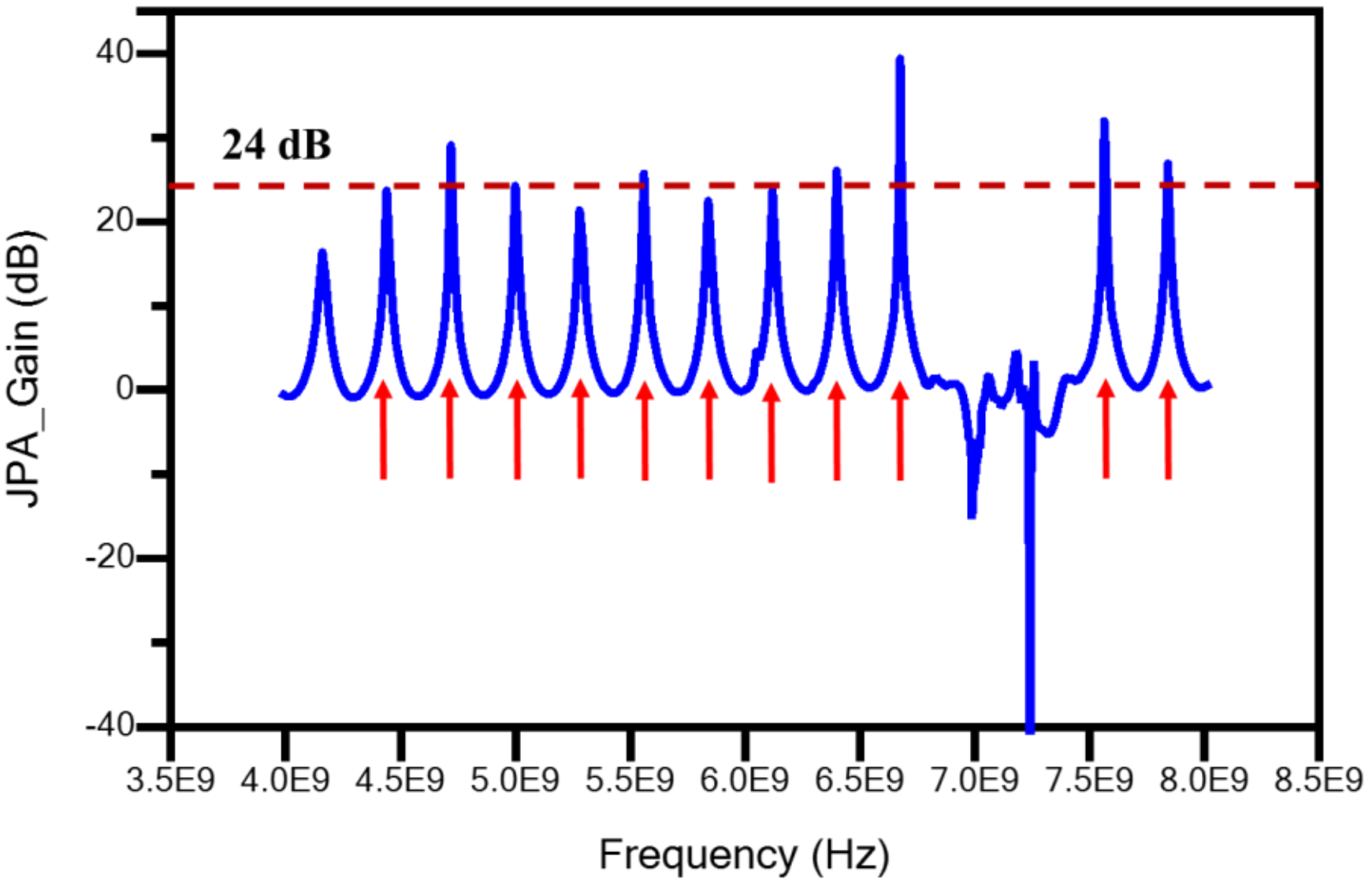

Fig. 28 JPA gain spectrum; $F_{pump}$ = 7.12 GHz, with 8*192 JJ, $P_{sig}$ = -150 dBm, $F_{sig}$ = 6.18 GHz, and $I_{pump}$ = 3.96 uA [88].

In the same way with the latter designs, it is necessary to study the 1-dB compression point as a critical figure of merit that characterizes the power-handling capacity and linearity of the amplifier. As illustrated in Fig. 29, $P_{1dB}$ is plotted as a function of varying pump current, providing direct insight into how the amplifier's performance evolves under different pumping conditions. At lower pump currents, the amplifier maintains a relatively high $P_{1dB}$, signifying its ability to process stronger input signals before the onset of gain compression. This reflects the amplifier's operation in a quasi-linear regime, where the nonlinearities of the Josephson junctions are moderate and well-managed, thus allowing the device to maintain consistent gain across a broader range of input powers. As the pump current is gradually increased, the amplifier transitions into a regime of stronger parametric interaction. While this enhances the overall gain, it also accelerates the onset of gain saturation, leading to a gradual decrease in $P_{1dB}$. This trade-off is a common characteristic in parametric amplifiers: increased pumping boosts the energy available for signal amplification but simultaneously pushes the system closer to its nonlinear limits. At this point, the amplifier becomes increasingly sensitive to fluctuations in signal amplitude, and small increases in input power result in disproportionately large distortions in gain. When the pump current approaches or surpasses a critical threshold, the system enters a highly nonlinear regime where the phase-space dynamics of the junctions become unstable. In this region, even minimal increases in the input power cause the amplifier's gain to collapse rapidly—hence the sharp drop in $P_{1dB}$. Additionally, strong pumping can initiate unwanted higher-order parametric processes, such as third-order harmonics or intermodulation distortions, which further erode linearity and reduce the usable dynamic range. Therefore, although stronger pumping can initially improve gain, it must be carefully optimized to avoid compromising linear performance.

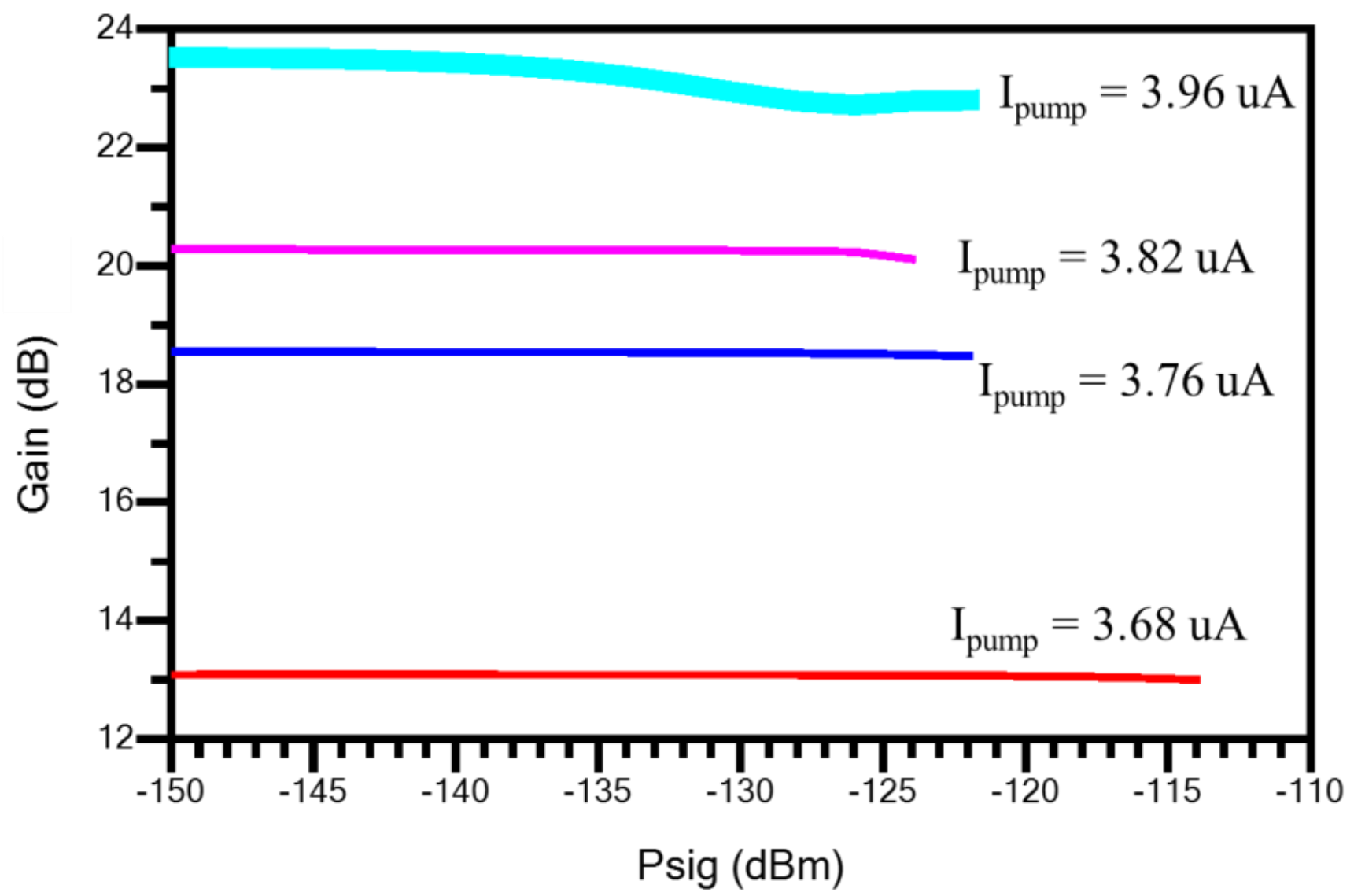

Fig. 29 $P_{1dB}$ for a JPA with 8*192 JJ for different $I_{pump}$, $F_{sig}$ = 6.0 GHz, $F_{pump}$ = 7.12 GHz [88].

Finally, Fig. 30 examines the distribution of power among different output modes of the modified BJPA, particularly the first-order and third-order output modes. The analysis of output modes is crucial for understanding the amplifier's spectral purity and harmonic content. Ideally, a JPA should concentrate most of its amplified power within the first-order mode while minimizing higher-order harmonics, as excess power in higher harmonics can lead to spectral contamination and signal distortion. The results indicate that the first-order output mode retains the majority of the amplified signal power, confirming that the proposed JPA efficiently concentrates parametric gain into the intended frequency band. This behavior is a direct consequence of the Blochnium array's improved coherence properties, which suppress higher-order nonlinear interactions that typically arise in conventional JPA designs. By minimizing third-order harmonic generation, the JPA ensures that the amplified signal remains clean and free from unwanted spectral artifacts, which is essential for applications such as quantum state readout and microwave signal processing. However, while third-order mode suppression is beneficial for maintaining spectral purity, it also suggests that the amplifier operates with reduced nonlinearity in certain regimes. This can be advantageous for applications requiring stable and predictable gain, but it may limit the JPA's ability to engage in more complex multi-mode parametric interactions. In contrast, traditional JJ-array JPAs often exhibit significant third-order harmonic power due to their stronger nonlinear interactions, which can be useful in applications such as frequency conversion and non-degenerate amplification.

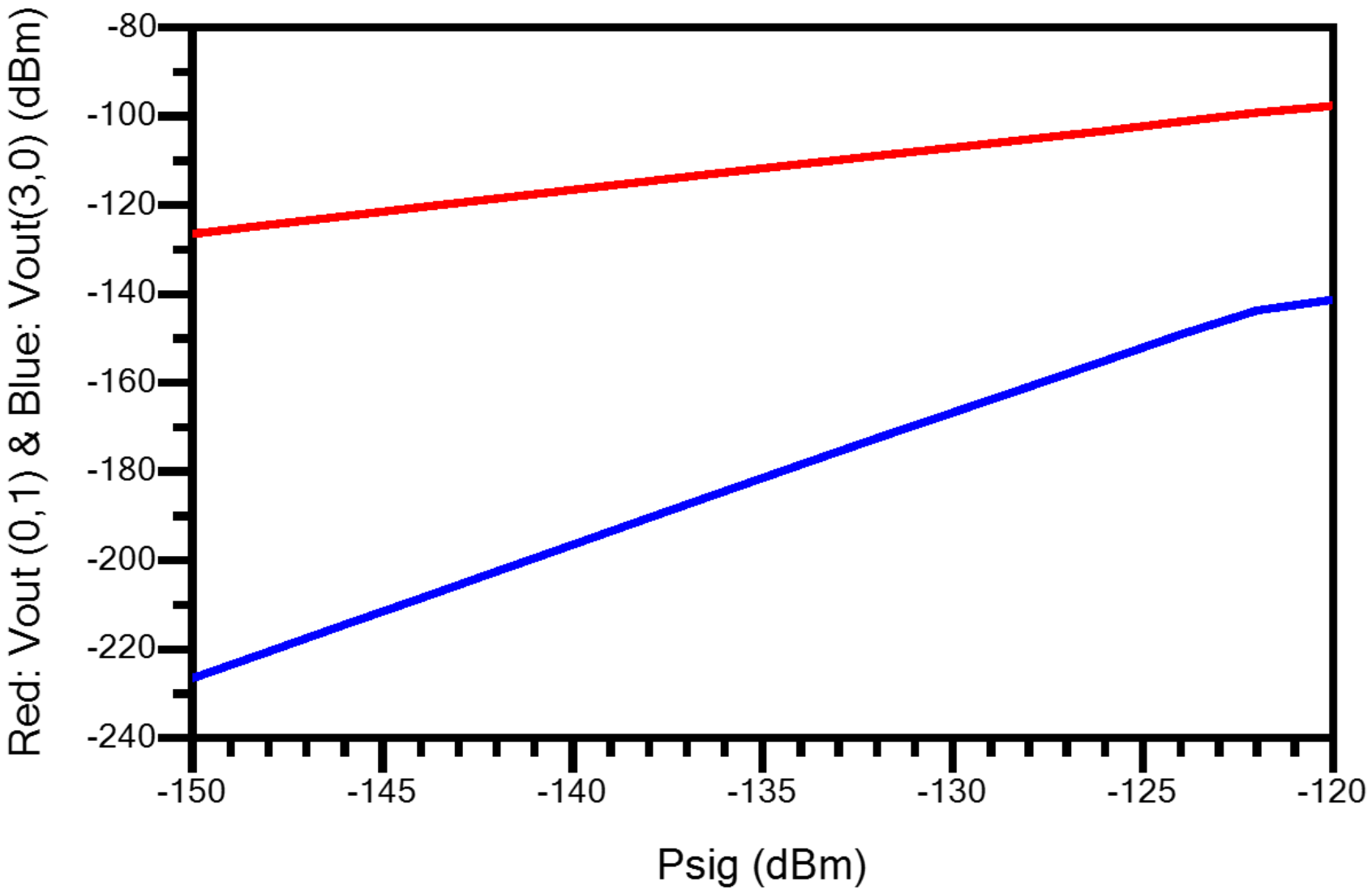

Fig. 30 JPA first and third output mode power, $F_{pump}$ = 7.12 GHz, with 8*192 JJ, $P_{sig}$ = -150 dBm, $F_{sig}$ = 6.18 GHz, $I_{pump}$ = 3.96 uA [88].

## Technical Comparison between different JPA designs

As a results, Table. 2 presents a comparative analysis of four JPA architectures: the Single JJ JPA, Arrayed JJ JPA, BJPA, and a newly proposed Modified BJPA. Each configuration is evaluated based on its signal gain, bandwidth, $P_{1dB}$ compression point, noise figure, and inherent design advantages and limitations. Single JJ JPAs [69-80] are well-established devices offering near-quantum-limited noise performance and gains typically ranging between 20–25 dB. However, they suffer from a limited power handling capability ($P_{1dB}$ often below –115 dBm) and relatively narrow bandwidths (<50 MHz), restricting their utility in multi-qubit systems or broadband measurements. Their simplicity and compatibility with traditional fabrication methods make them appealing for low-noise, single-channel readout tasks. Arrayed JJ JPAs [81-87] improves upon this by utilizing multiple junctions or SQUIDs to broaden bandwidth (up to several GHz in JTWPA architectures) and increase gain. These configurations often demonstrate enhanced linearity and tunability but at the cost of increased design complexity and power consumption. They require careful phase matching and impedance engineering, particularly in traveling-wave formats. The BJPA [13] introduces Blochnium elements, providing high coherence, improved nonlinearity control, and stronger resilience to fabrication imperfections. With simulated gain above 25 dB, bandwidth coverage across the C-band, and compression points exceeding –92 dBm, the BJPA offers a compact, scalable alternative to large JJ arrays, suitable for next-generation quantum systems. The Modified BJPA [88] further advances this architecture by introducing a spectrally engineered gain profile. While maintaining high gain, its comb-like response is especially advantageous in frequency-multiplexed qubit readout schemes. Each gain peak can be aligned with a qubit's readout frequency, enhancing selectivity and reducing leakage without the need for external filters. However, its complex impedance matching and spectral gaps demand careful system-level planning. Overall, the table underscores the trade-offs between gain, bandwidth, and linearity, and highlights how innovations like Blochnium and spectral engineering are reshaping JPA design for scalable quantum technologies.

**Table 2:** Comparison of key performance parameters for four JPA architectures. Data are based on theoretical modeling, simulation, and experimental literature. The modified BJPA demonstrates a frequency-selective gain profile suitable for multiplexed quantum readout, while the BJPA shows superior linearity and gain efficiency compared to traditional designs.

| **Parameter** | **Single JJ JPA [71-79]** | **Arrayed JJ JPA [80-87]** | **BJPA [13]** | **Modified BJPA [88]** |
|---|---|---|---|---|
| **Gain** | 20–25 dB typical | 11–31 dB depending on design | ~25 dB simulated | ~25 dB with frequency-selective peaks across C-band (comb-like structure) |
| **Bandwidth** | Narrow (typically 10–50 MHz) | Very broad for JTWPAs | Tunable across C-band (~4–8 GHz) | Tunable; structured comb profile across C-band (~4–8 GHz); not flat but engineered for selective frequency gain |
| **$P_{1dB}$** | –115 to –133 dBm | ~-125 to –95 dBm (distributed 1000-JJ JPA) | ~ –92 dBm (N = 70, M = 8); improved by Quarton architecture | Variable; better than ~–115 dBm |
| **Noise Figure** | Near quantum limit (~1–2× SQL) | Near quantum limit | Near quantum limit due to high coherence of Blochnium | Low noise; harmonic purity maintained with low 3rd-order output modes |
| **Advantages** | Simple fabrication, well understood, low noise | Higher gain-bandwidth product, more robust power handling, tunability via array design | High gain with fewer junctions, improved linearity and power handling, compact design, tunable nonlinearity | Selective gain structure ideal for frequency-multiplexed qubit readout; suppresses qubit leakage into readout chain |
| **Disadvantages** | Narrow bandwidth, limited power handling, pump leakage challenges | Fabrication complexity, phase matching sensitivity, higher power demand in large arrays | Requires precise control over junction parameters, new fabrication challenges with Blochnium | Complex impedance matching; sharp gain dips between peaks require careful frequency planning |

It is important to note that the data provided in the tables and throughout the article are derived from an extensive literature review conducted as part of this study. While every effort has been made to ensure accuracy and consistency, variations in measurement conditions, fabrication processes, and experimental setups across different sources may lead to minor discrepancies. Nonetheless, the compiled data offer a reliable and comprehensive comparison of key amplifier characteristics, supporting the analysis and insights presented in this work.

**Conclusions:**

In quantum applications, particularly at cryogenic temperatures, Josephson Parametric Amplifiers (JPAs) are often preferred over CMOS and HEMT amplifiers due to their unique ability to operate at the quantum limit of noise. JPAs are designed with superconducting Josephson junctions, which allow them to achieve ultra-low noise figures, typically below 0.5 dB. This characteristic is crucial in quantum systems, where preserving the quantum nature of signals is paramount. The noise performance of JPAs is superior to both CMOS and HEMT technologies, which, although they offer low noise figures at cryogenic temperatures (around 2-5 dB for CMOS and <1 dB for HEMTs), cannot match the quantum-limited performance of JPAs. This low noise is essential for accurately reading out quantum states and minimizing decoherence, making JPAs the preferred choice in highly sensitive quantum experiments and quantum computing systems. Moreover, JPAs operate at much lower temperatures (around 10 mK) than CMOS and HEMT amplifiers, which typically function at around 4.2 K. The operation at mK temperatures aligns perfectly with the environment required for superconducting qubits and other quantum devices, ensuring that the amplifier does not introduce thermal noise or other disturbances that could degrade the quantum state. In contrast, CMOS and HEMT amplifiers, although capable of operating at cryogenic temperatures, are not designed to work at the extreme low temperatures where quantum systems operate most effectively. This makes them less suitable for integration into quantum systems that require amplification without compromising the delicate quantum states.

Another critical factor is the power consumption of JPAs, which is significantly lower than that of CMOS and HEMT amplifiers. Operating in the microwatt range, JPAs are extremely power-efficient, which is particularly advantageous in cryogenic environments where minimizing heat generation is crucial to maintaining the low-temperature conditions necessary for quantum systems. In contrast, CMOS and HEMT amplifiers consume more power (typically in the milliwatt range for CMOS and tens of milliwatts for HEMTs), which can introduce unwanted heating and complicate the cooling requirements of the system. While CMOS and HEMT technologies offer advantages in terms of integration, scalability, and bandwidth, their limitations in noise performance, temperature compatibility, and power efficiency make them less ideal for quantum applications. JPAs, despite their narrower bandwidth and more challenging scalability, provide unparalleled noise performance and compatibility with quantum systems, making them the amplifier of choice in applications where maintaining quantum coherence and minimizing noise are of utmost importance. This makes JPAs indispensable in advancing quantum technologies, where precision, noise reduction, and compatibility with cryogenic environments are critical. Nonetheless, this review focuses on the JPA engineering and reports some different structures by which a more suitable JPA can be designed. Accordingly, single-JJ JPAs provide compact and analytically tractable designs but are limited by nonlinear saturation and early gain compression. Our simulations confirm that while these devices can achieve substantial gain, their performance degrades rapidly under high pump powers. The use of JJ arrays mitigates these limitations by distributing nonlinearity, enhancing the 1 dB compression point (up to –107 dBm for traditional structure and ~-92 dB for a special structure), and improving the linearity and tunability of the amplifier. Quantum theoretical models based on Hamiltonians and input-output formalism have enabled accurate gain and noise predictions, matching well with experimental observations and guiding future designs.

As a main conclusion, the study underscores that JPAs are indispensable for quantum computing, metrology, and sensing, particularly in systems requiring low power, ultra-low noise, and cryogenic operation. Continued efforts in array optimization, on-chip integration, and nonreciprocal configurations are likely to address current limitations and pave the way for next-generation quantum readout and communication technologies. As quantum systems scale up, JPAs will remain at the forefront of enabling high-fidelity, noise-resilient signal processing.

**Acknowledgement:** The authors would like to express their sincere gratitude to the Iranian Quantum Technologies Research Center (IQTEC) for their full support of this work.